%% file: PR2018_MAIN.tex
\documentclass[12pt]{report}
\pdfoutput=1

\usepackage{graphicx}
\usepackage{amsfonts}
\usepackage{amssymb}
\usepackage{amsthm}
\usepackage{amsmath}

\usepackage{float}

\usepackage[active]{srcltx}

\usepackage{cite}
\usepackage{color,soul}

\usepackage[affil-it]{authblk}

\graphicspath{{figs/}}

\newcommand{\be}{\begin{equation}}
\newcommand{\ee}{\end{equation}}

\newcommand{\beq}{\begin{equation}}
\newcommand{\eeq}{\end{equation}}
\newcommand{\non}{\nonumber}
\newcommand\bea{\begin{eqnarray}}
\newcommand\eea{\end{eqnarray}}

\renewcommand{\thefootnote}{\alph{footnote}}

\allowdisplaybreaks
\input{paperdef}
\newcommand{\clearemptydoublepage}{\newpage{\pagestyle{empty}\cleardoublepage}}

\begin{document}

\hyphenation{renormalizabi-lity}

\title{
\vspace*{-4cm}
\hfill {\small IFT-UAM/CSIC-19-020}\\[-0.7em]
\hfill {\small LAPTH-004/19}\\
\vspace*{2cm}
Reduction of Couplings and its application\\
in Particle Physics}
\date{}
\author{S. Heinemeyer$^{1,2,3,}$\thanks{email: Sven.Heinemeyer@cern.ch},
M. Mondrag\'on$^{4,}$\thanks{email: myriam@fisica.unam.mx},
N. Tracas$^{5,}$\thanks{email: ntrac@central.ntua.gr}$\,$
and G. Zoupanos$^{5,6,7,8,}$\thanks{email: George.Zoupanos@cern.ch}\\

{\small
$^1$Instituto de F\'{\i}sica Te\'{o}rica (UAM/CSIC), Universidad Auton\'{o}ma de Madrid, 28049 Madrid, Spain\\
$^2$Campus of International Excellence UAM+CSIC, Cantoblanco, 28049, Madrid, Spain\\
$^3$Instituto de F\'{\i}sica de Cantabria (CSIC-UC),  E-39005 Santander, Spain \\
$^4$Instituto de F\'{\i}sica,  Universidad Nacional Aut\'onoma de M\'exico,  A.P. 20-364, CDMX 01000, M\'exico\\ %
$^5$Physics Department, Nat.\ Technical University, 157 80 Zografou, Athens, Greece\\%
$^6$Institute of Theoretical Physics, D-69120 Heidelberg, Germany\\
$^7$Max-Planck Institut f\"ur Physik, F\"ohringer Ring 6, D-80805 M\"unchen, Germany\\
$^8$Laboratoire d'Annecy-le-Vieux de Physique Th\'{e}orique, Annecy, France\\
}

\vspace*{1cm}
\textbf{Dedicated to the memory of Wolfhart Zimmermann,\\
 the brilliant theoretical physicist,\\
 who initiated the subject of reduction of couplings}
}

\maketitle

\abstract{The idea of reduction of couplings in renormalizable theories will
be presented and then will be applied in Particle Physics models.
Reduced couplings appeared as functions of a primary one,
compatible with the renormalization group equation and thus
solutions of a specific set of ordinary differential equations. If
these functions have the form of power series the respective
theories resemble standard renormalizable ones and thus widen
considerably the area covered until then by symmetries as a tool
for constraining the number of couplings consistently. Still on the
more abstract level reducing couplings enabled one to construct
theories with beta-functions vanishing to all orders of
perturbation theory. Reduction of couplings became physics-wise
truly interesting and phenomenologically important when applied to
the standard model and its possible extensions. In particular in
the context of supersymmetric theories it became the most powerful
tool known today once it was learned how to apply it also to
couplings having dimension of mass and to mass parameters.
Technically this all relies on the basic property that reducing
couplings is a renormalization scheme independent procedure.
Predictions of top and Higgs mass prior to their experimental
finding highlight the fundamental physical significance of this
notion.
}

\vfill
\newpage

\vspace*{2cm}

\begin{center}
{\large Prologue and Synopsis}
\end{center}

\vspace*{1cm}

In spite of their limitations, perturbative local field theories are still
of prominent practical value.

It is remarkable that the intrinsic ambiguities connected with locality
and causality - most of the time associated with ultraviolet infinities - can be
summarized in terms of a formal group which acts in the space of the
coupling constants or coupling functions attached to each type of local
interaction.

It is therefore natural to look systematically for stable submanifolds.
Some such have been known for a long time: e.g., spaces of renormalizable
interactions and subspaces characterized by system of Ward identities
mostly related to symmetries.

A systematic search for such stable submanifolds has been initiated by
W. Zimmermann in the early eighties.

Disappointing for some time, this program has attracted several other
active researchers and recently produced physically interesting results.

It looks at the moment as the only theoretically founded algorithm
potentially able to decrease the number of parameters within the physically
favoured perturbative models \footnote{
The above text has appeared, as \textit{Geleitwort} (preface), in the book
``Reduction of Couplings and its Application in Particle Physics Finite Theories Higgs and Top Mass Predictions'',
Ed. Klaus Sibold, Authors: Jisuke Kubo, Sven Heinemeyer, Myriam Mondragon, Olivier Piguet, Klaus Sibold, Wolfhart Zimmermann, George Zoupanos. Published in PoS (Higgs \& top)001.
}.

\vspace*{1cm}
\noindent
\textbf{Raymond Stora, CERN (Switzerland), December 16, 2013}

\vfill
\newpage

\tableofcontents

\newpage

\renewcommand{\thefootnote}{\arabic{footnote}}

\include{PR2018_Chapter_1}

\clearemptydoublepage
\setcounter{chapter}{1}

\include{PR2018_Chapter_2}
\clearemptydoublepage
\setcounter{chapter}{2}

\include{PR2018_Chapter_3}
\clearemptydoublepage
\setcounter{chapter}{3}

\include{PR2018_Chapter_4}
\clearemptydoublepage
\setcounter{chapter}{4}

\include{PR2018_Chapter_5}
\clearemptydoublepage
\setcounter{chapter}{5}

\include{PR2018_Chapter_6}

\clearemptydoublepage
\setcounter{chapter}{6}

\include{PR2018_Chapter_7}
\clearemptydoublepage
\setcounter{chapter}{7}

\include{PR2018_References}
\end{document}

%% file: paperdef.tex
\newcommand{\lsim}
{\;\raisebox{-.3em}{$\stackrel{\displaystyle <}{\sim}$}\;}
\newcommand{\gsim}
{\;\raisebox{-.3em}{$\stackrel{\displaystyle >}{\sim}$}\;}

\newcommand\al{\alpha}

\newcommand\tb{\tan\beta}

\newcommand\ReDiag{\mathop{%
  \raise .5pt\hbox{[}%
  \widetilde{\mathrm{Re}}%
  \raise .5pt\hbox{]}}}
\newcommand\ReOffDiag{\mathop{%
  \raise .5pt\hbox{$\llbracket$}%
  \widetilde{\mathrm{Re}}%
  \raise .5pt\hbox{$\rrbracket$}}}

\newcommand\MZ{M_Z}
\newcommand\Mh{M_h}
\newcommand\MH{M_H}
\newcommand\MA{M_A}
\newcommand\MHp{M_{H^\pm}}

\newcommand\mb{m_b}
\newcommand\mt{m_t}

\newcommand\gl{{\tilde g}}
\newcommand\mgl{m_\gl}

\newcommand\ino[1]{\tilde\chi_{#1}}

\newcommand\chapm[1]{\ino{#1}^\pm}

\newcommand\mcha[1]{m_{\chapm{#1}}}

\newcommand\neu[1]{\ino{#1}^0}
\newcommand\mneu[1]{m_{\neu{#1}}}

\newcommand\refeq[1]{Eq.~(\ref{#1})}
\newcommand\refeqs[1]{Eqs.~(\ref{#1})}
\newcommand\refta[1]{Tab.~\ref{#1}}
\newcommand\refse[1]{Sect.~\ref{#1}}

\newcommand\citere[1]{Ref.~\cite{#1}}
\newcommand\citeres[1]{Refs.~\cite{#1}}

\newcommand{\CP}{{\cal CP}}
\newcommand{\cp}{{\CP}}

\newcommand{\tev}{\,\, \mathrm{TeV}}
\newcommand{\gev}{\,\, \mathrm{GeV}}

\newcommand\mstop[1]{m_{\tilde{t}_{#1}}}
\newcommand\msbot[1]{m_{\tilde{b}_{#1}}}

\newcommand\mstau[1]{m_{\tilde{\tau}_{#1}}}

\newcommand{\br}{\text{BR}}

\newcommand{\sig}{\sigma}

\def\order#1{\ensuremath{{\cal O}(#1)}}
\def\reffi#1{\mbox{Fig.~\ref{#1}}}

\def\ga{\gamma}

\def\de{\delta}

\newcommand{\VL}{\left( \begin{array}{c}}
\newcommand{\VR}{\end{array} \right)}
\newcommand{\ML}{\left( \begin{array}{cc}}
\newcommand{\MLd}{\left( \begin{array}{ccc}}
\newcommand{\MLv}{\left( \begin{array}{cccc}}
\newcommand{\MR}{\end{array} \right)}

\definecolor{Lightblue}{cmyk}{0.9,0.1,0.1,0.3}
\definecolor{dgelborange}{cmyk}{0.,0.3,0.5, 0.}
\definecolor{Orange}{cmyk}{0.,0.5,0.5, 0.}
\definecolor{Lila}{rgb}{0.5,0.,1}

\newcommand{\complex}{{{\rm I} \kern -.59em {\rm C}}}

\newcommand{\nn}{\nonumber}

%% file: PR2018_Chapter_1.tex
\chapter{Introduction: The Basic Ideas}
In the recent years the theoretical endeavours that attempt to
achieve a deeper understanding of Nature have presented a series
of successes in developing frameworks such as String Theories and
Noncommutativity that aim to describe the fundamental theory at
the Planck scale. However, the essence of all theoretical efforts
in Elementary Particle Physics (EPP) is to understand the present
day free parameters of the Standard Model (SM) in terms of few
fundamental ones, i.e. to achieve {\it reductions of couplings}\cite{book}.
Unfortunately, despite the several successes in the above frameworks
they do not offer anything in the understanding of the free paramaters
of the SM.  The pathology of the plethora of free parameters is deeply
connected to the presence of  {\it infinities} at the quantum level.
The renormalization program can remove the infinities by introducing
counterterms, but only at the cost of leaving the corresponding terms as
free parameters.

Although the Standard Model (SM) has been very successful in describing
elementary particles and its interactions, it has been known for some
time that it must be the low energy limit of a more fundamental
theory.  This quest for a theory beyond the Standard Model (BSM) has
expanded in various directions. The usual, and very efficient, way of reducing the number of
free parameters of a theory to render it more predictive, is to
introduce a symmetry.  Grand Unified Theories (GUTs) are very good
examples of such a procedure
\cite{Pati:1973rp,Georgi:1974sy,Georgi:1974yf,Fritzsch:1974nn,Gursey:1975ki,Achiman:1978vg}.
First in the case of minimal $SU(5)$, because of the (approximate) gauge
coupling unification, it was possible to reduce the gauge
couplings of the SM and give a prediction for one of them.
By adding a further symmetry, namely $N=1$ global supersymmetry
\cite{Amaldi:1991cn,Dimopoulos:1981zb,Sakai:1981gr}
it was possible to make the prediction viable.
GUTs can also
relate the Yukawa couplings among themselves, again $SU(5)$ provided
an example of this by predicting the ratio $M_{\tau}/M_b$
\cite{Buras:1977yy} in the SM.
Unfortunately, requiring
more gauge symmetry does not seem to help, since additional
complications are introduced due to new degrees of freedom, for instance in the
ways and channels of breaking the symmetry.

A possible way to look for relations among unrelated parameters is the
method of reduction of couplings
\cite{Zimmermann:1984sx,Oehme:1984yy,Oehme:1985jy};
see also refs~\cite{Ma:1977hf,Chang:1974bv,Nandi:1978fw}.
This method, as its name proclaims, reduces the number of couplings in a
theory by relating either all or a number of couplings to a single
coupling denoted as the ``primary coupling''.  This method might help to
identify hidden symmetries in a system, but it is also possible to have
reduction of couplings in systems where there is no apparent
symmetry. The reduction of couplings is based on the assumption that both the
original and the reduced theory are renormalizable and that there exist renormalization
group invariant (RGI) relations among parameters.

A natural extension of the GUT idea
and successful application of the method of \textit{reduction of couplings}
is to find a way to relate the gauge and
Yukawa sectors of a theory, that is to achieve gauge-Yukawa Unification (GYU).
This will be presented in Chapter~\ref{ch:viable}.
Following the original suggestion for reducing the couplings within
the framework of GUTs we were hunting for renormalization group invariant (RGI)
relations holding below the Planck scale, which in turn are preserved down to the
GUT scale.  It is indeed an impressive observation that one can guarantee the
validity of the RGI relations to all-orders in perturbation theory by studying
the uniqueness of the resulting relations at one-loop.
Even more remarkable
is the fact that it is possible to find RGI relations among couplings that guarantee
finiteness to all-orders in perturbation theory.
The above principles have
only been applied in $N = 1$ supersymmetric GUTs for reasons that will be transparent
in the following sections,
here we should only note that the use of $N = 1$ supersymmetric
GUTs comprises the demand of the cancellation of quadratic divergencies in the SM.  The above
GYU program applied in the dimensionless couplings of supersymmetric GUTs had
a great success by predicting correctly, among others, the top quark mass in the finite
\cite{Kapetanakis:1992vx,Mondragon:1993tw} and in the minimal
$N = 1$ supersymmetric $SU(5)$ \cite{Kubo:1994bj} before its discovery \cite{Lancaster:2011wr}.

Although supersymmetry seems to be an essential feature for a successful realization of
the above program, its breaking has to be understood too, since it has the ambition to
supply the SM with predictions for several of its free parameters. Indeed, the search for
RGI relations has been extended to the soft supersymmetry breaking sector (SSB) of these
theories, which involves
parameters of dimension one and two. In addition, there was important progress  concerning
the renormalization properties of the SSB parameters, based on the powerful supergraph method
for studying supersymmetric theories, and it was applied to the softly broken ones by using the
``spurion'' external space-time independent superfields.
According to this method a softly broken supersymmetric gauge theory is considered as a
supersymmetric one in which the various parameters, such as couplings and masses, have
been promoted to external superfields. Then, relations among the soft term renormalization
and that of an unbroken supersymmetric theory have been derived. In particular the
$\beta$-functions of the parameters of the softly broken theory are expressed in terms
of partial differential operators involving the dimensionless parameters of the unbroken
theory. The key point in solving the set of coupled differential equations so as to be
able to express all parameters in a RGI way, was to transform the partial differential
operators involved to total derivative operators.  It is indeed possible to do this by
choosing a suitable RGI surface.

On the phenomenological side the application on the reduction of coupling method
to $N=1$ supersymmetric theories has led to very interesting developments too.  Previously an
appealing ``universal'' set of soft scalar masses was assumed in the SSB sector of
supersymmetric theories, given that apart from economy and simplicity (1) they are
part of the constraints that preserve finiteness up to two-loops, (2) they appear
in the attractive dilaton dominated supersymmetry breaking superstring scenarios.
However, further studies have exhibited a number of problems, all due to the restrictive
nature of the ``universality'' assumption for the soft scalar masses. Therefore, there were
attempts to relax this constraint without loosing its attractive features. Indeed an
interesting observation on $N = 1$ GYU theories is that there exists a RGI sum rule for
the soft scalar masses at lower orders in perturbation theory, which was later extended to
all-orders, and manages to overcome all the unpleasant phenomenological consequences.
Armed with the above tools and results we were in a position to study the spectrum of the
full finite models in terms of few free parameters, with emphasis on the predictions of
supersymmetric particles and the lightest Higgs mass.

The result was indeed very impressive since it led to a prediction of
the Higgs mass which coincided with the results of the LHC for the Higgs
mass by ATLAS \cite{Aad:2012tfa,ATLAS:2013mma}
and CMS \cite{Chatrchyan:2012ufa,Chatrchyan:2013lba}, and predicted
a relatively heavy spectrum consistent with the non-observation of
supersymmetric particles at the LHC. The coloured supersymmetric
particles are predicted to be above 2.7~TeV, while the electroweak
supersymmetric spectrum starts below 1~TeV. These successes will be
presented in Chapter~\ref{chap:FUT_LOW}.

Last but certainly not least, the above machinery has been recently
applied in the MSSM with impressive results concerning the predictivity
of the top, bottom and Higgs masses, being at the same time consistent
with the non-observation of supersymmeric particles at the LHC. More
specifically the electroweak supersymmetric spectrum starts at 1.3~TeV
and the coloured at $\sim 4$~TeV. These results will be presented too in
Chapter~\ref{chap:FUT_LOW}.

%% file: PR2018_Chapter_2.tex
\chapter{Theoretical Basis}\label{ch:theory}

\section{Reduction of Dimensionless Parameters}
In this section we outline the  idea of reduction of couplings.
Any RGI relation among couplings
(i.e. which does not depend on the  renormalization
scale $\mu$ explicitly) can be expressed,
in the  implicit form $\Phi (g_1,\cdots,g_A) ~=~\mbox{const.}$,
which
has to satisfy the  partial differential equation (PDE)
\beq
\mu\,\frac{d \Phi}{d \mu} = {\vec \nabla}\Phi\cdot {\vec \beta} ~=~
\sum_{a=1}^{A}
\,\beta_{a}\,\frac{\partial \Phi}{\partial g_{a}}~=~0~,
\eeq
where $\beta_a$ is the  $\beta$-function of $g_a$.
This PDE is equivalent
to a set of ordinary differential equations,
the so-called reduction equations (REs) \cite{Zimmermann:1984sx,Oehme:1984yy,Oehme:1985jy},
\beq
\beta_{g} \,\frac{d g_{a}}{d g} =\beta_{a}~,~a=1,\cdots,A~,
\label{redeq}
\eeq
where $g$ and  $\beta_{g}$ are the  primary
coupling and  its $\beta$-function,
and the  counting on $a$ does not include $g$.
Since maximally ($A-1$) independent
RGI ``constraints''
in the  $A$-dimensional space of couplings
can be imposed by the  $\Phi_a$'s, one could in principle
express all the  couplings in terms of
a single coupling $g$.
However, a closer look to the  set of \refeqs{redeq} reveals that their
general solutions contain as many integration constants as the  number of
equations themselves. Thus, using such integration constants we have just
traded an integration constant for each ordinary renormalized coupling,
and consequently, these general solutions cannot be considered as
reduced ones. The  crucial requirement in the  search for RGE relations is
to demand power series solutions to the  REs,
\beq
g_{a} = \sum_{n}\rho_{a}^{(n)}\,g^{2n+1}~,
\label{powerser}
\eeq
which preserve perturbative renormalizability.
Such an ansatz fixes the  corresponding integration constant in each of
the REs and  picks up a special solution out of the  general one.
Remarkably, the  uniqueness of such power series solutions can be
decided already at the  one-loop level
\cite{Zimmermann:1984sx,Oehme:1984yy,Oehme:1985jy}.  To illustrate
this, let us assume that the  $\beta$-functions have the  form
\beq
\begin{split}
\beta_{a} &=\frac{1}{16 \pi^2}\left[ \sum_{b,c,d\neq
  g}\beta^{(1)\,bcd}_{a}g_b g_c g_d+
\sum_{b\neq g}\beta^{(1)\,b}_{a}g_b g^2\right]+\cdots~,\\
\beta_{g} &=\frac{1}{16 \pi^2}\beta^{(1)}_{g}g^3+ \cdots~,
\end{split}
\eeq
where
$\cdots$ stands for higher order terms, and  $ \beta^{(1)\,bcd}_{a}$'s
are symmetric in $ b,c,d$.  We then assume that the  $\rho_{a}^{(n)}$'s
with $n\leq r$ have been uniquely determined. To obtain
$\rho_{a}^{(r+1)}$'s, we insert the  power series (\ref{powerser}) into
the REs (\ref{redeq}) and  collect terms of ${\cal O}(g^{2r+3})$ and
find
\beq
\sum_{d\neq g}M(r)_{a}^{d}\,\rho_{d}^{(r+1)} = \mbox{lower
  order quantities}~,\non
\eeq
where the  r.h.s. is known by assumption,
and
\begin{align}
M(r)_{a}^{d} &=3\sum_{b,c\neq
  g}\,\beta^{(1)\,bcd}_{a}\,\rho_{b}^{(1)}\,
\rho_{c}^{(1)}+\beta^{(1)\,d}_{a}
-(2r+1)\,\beta^{(1)}_{g}\,\delta_{a}^{d}~,\label{M}\\
0 &=\sum_{b,c,d\neq g}\,\beta^{(1)\,bcd}_{a}\,
\rho_{b}^{(1)}\,\rho_{c}^{(1)}\,\rho_{d}^{(1)} +\sum_{d\neq
  g}\beta^{(1)\,d}_{a}\,\rho_{d}^{(1)}
-\beta^{(1)}_{g}\,\rho_{a}^{(1)}~.
\end{align}
Therefore, the  $\rho_{a}^{(n)}$'s for all $n > 1$ for a
given set of $\rho_{a}^{(1)}$'s can be uniquely determined if $\det
M(n)_{a}^{d} \neq 0$ for all $n \geq 0$.

As it will be clear later by examining specific examples, the  various
couplings in supersymmetric theories have the  same asymptotic
behaviour.  Therefore searching for a power series solution of the
form (\ref{powerser}) to the  REs (\ref{redeq}) is justified.

The possibility of coupling unification described in this section
is without any doubt
attractive because the  ``completely reduced'' theory contains
only one independent coupling, but  it can be
unrealistic. Therefore, one often would like to impose fewer RGI
constraints, and  this is the  idea of partial reduction \cite{Kubo:1985up,Kubo:1988zu}.

The above facts lead us to suspect that there is and intimate
connection among the requirement of reduction of couplings and
supersymmetry which still waits to be uncovered.  The connection
becomes more clear by examining the following example.%

Consider an $SU(N)$ gauge theory with the following matter content:
$\phi^{i}({\bf N})$ and $\hat{\phi}_{i}(\overline{\bf N})$
are complex scalars,
$\psi^{i}({\bf N})$ and $\hat{\psi}_{i}(\overline{\bf N})$
are left-handed Weyl spinor,
and $\lambda^a (a=1,\dots,N^2-1)$ is a
right-handed Weyl spinor in the adjoint representation of $SU(N)$.

The Lagrangian, omitting kinetic terms, includes:
\beq
{\cal L} \supset 
i \sqrt{2} \{~g_Y\overline{\psi}\lambda^a T^a \phi
-\hat{g}_Y\overline{\hat{\psi}}\lambda^a T^a \hat{\phi}
+\mbox{h.c.}~\}-V(\phi,\overline{\phi}),
\eeq
where
\beq
V(\phi,\overline{\phi}) =
\frac{1}{4}\lambda_1(\phi^i \phi^{*}_{i})^2+
\frac{1}{4}\lambda_2(\hat{\phi}_i \hat{\phi}^{*~i})^2
+\lambda_3(\phi^i \phi^{*}_{i})(\hat{\phi}_j \hat{\phi}^{*~j})+
\lambda_4(\phi^i \phi^{*}_{j})
(\hat{\phi}_i \hat{\phi}^{*~j}),
\eeq
which is the most general renormalizable form
of dimension four, consistent with the $SU(N)\times SU(N)$ global
symmetry.

Searching for a solution of the form of Eq. (\ref{powerser}) for the REs (\ref{redeq},) we
find in lowest order the following one ($g$ is the gauge coupling):
\beq
\begin{split}
g_{Y}&=\hat{g}_{Y}=g~,\\
\lambda_{1}&=\lambda_{2}=\frac{N-1}{N}g^2~,\\
\lambda_{3}&=\frac{1}{2N}g^2~,~
\lambda_{4}=-\frac{1}{2}g^2~,
\end{split}
\eeq
which corresponds to an $N=1$ supersymmetric gauge theory.
Clearly the above remarks do not answer the
question of the relation among reduction of couplings and
supersymmetry but rather try to trigger the interest for further
investigation.

\section{Reduction of Couplings in N = 1 Supersymmetric Gauge Theories.
Partial Reduction}
Let us consider a chiral, anomaly free, $N=1$ globally supersymmetric
gauge theory based on a group G with gauge coupling constant $g$. The
superpotential of the theory is given by
\bea
W&=& \frac{1}{2}\,m_{ij} \,\phi_{i}\,\phi_{j}+
\frac{1}{6}\,C_{ijk} \,\phi_{i}\,\phi_{j}\,\phi_{k}~,
\label{supot0}
\eea
where $m_{ij}$ and $C_{ijk}$ are gauge invariant tensors and
the matter field (chiral superfield) $\phi_{i}$ transforms according to the irreducible representation  $R_{i}$
of the gauge group $G$. The renormalization constants associated with the
superpotential (\ref{supot0}), assuming that supersymmetry is preserved, are
\begin{align}
\phi_{i}^{0}&=\left(Z^{j}_{i}\right)^{(1/2)}\,\phi_{j}~,~\\
m_{ij}^{0}&=Z^{i'j'}_{ij}\,m_{i'j'}~,~\\
C_{ijk}^{0}&=Z^{i'j'k'}_{ijk}\,C_{i'j'k'}~.
\end{align}
The $N=1$ non-renormalization theorem \cite{Wess:1973kz,Iliopoulos:1974zv,Ferrara:1974fv,Fujikawa:1974ay} ensures that
there are no mass and cubic-interaction-term infinities and therefore
\begin{equation}
\begin{split}
Z_{ij}^{i'j'}\left(Z^{i''}_{i'}\right)^{(1/2)}\left(Z^{j''}_{j'}\right)^{(1/2)}
&=\delta_{(i}^{i''}
\,\delta_{j)}^{j''}~,\\
Z_{ijk}^{i'j'k'}\left(Z^{i''}_{i'}\right)^{(1/2)}\left(Z^{j''}_{j'}\right)^{(1/2)}
\left(Z^{k''}_{k'}\right)^{(1/2)}&=\delta_{(i}^{i''}
\,\delta_{j}^{j''}\delta_{k)}^{k''}~.
\end{split}
\end{equation}
As a result the only surviving possible infinities are
the wave-function renormalization constants $Z^{j}_{i}$, i.e.,  one infinity
for each field. The one-loop $\beta$-function of the gauge
coupling $g$ is given by
\cite{Parkes:1984dh,West:1984dg,Jones:1985ay,Jones:1984cx,Parkes:1985hh}
\beq
\beta^{(1)}_{g}=\frac{d g}{d t} =
\frac{g^3}{16\pi^2}\left[\,\sum_{i}\,T(R_{i})-3\,C_{2}(G)\right]~,
\label{betag}
\eeq
where, as usual, $t$ is the logarithm of the ratio of the energy scale over a reference scale,
$C_{2}(G)$ is the quadratic Casimir of the adjoint representation of the associated
gauge group $G$ and $T(R)$ is given by the relation $\textrm{Tr}[T^aT^b]=T(R)\delta^{ab}$
while $T^a$ is the generators of the group in the appropriate representation.
The $\beta$-functions of $C_{ijk}$,
by virtue of the non-renormalization theorem, are related to the
anomalous dimension matrix $\gamma_{ij}$ of the matter fields
$\phi_{i}$ as:
\beq
\beta_{ijk} =
 \frac{d C_{ijk}}{d t}~=~C_{ijl}\,\gamma^{l}_{k}+
 C_{ikl}\,\gamma^{l}_{j}+
 C_{jkl}\,\gamma^{l}_{i}~.
\label{betay}
\eeq
At one-loop level $\gamma^i_j$ is given by \cite{Parkes:1984dh}
\beq
\gamma^{(1)}{}_{j}^{i}=\frac{1}{32\pi^2}\,[\,
C^{ikl}\,C_{jkl}-2\,g^2\,C_{2}(R_{i})\delta^i_j\,],
\label{gamay}
\eeq
where $C_{2}(R_{i})$ is the quadratic Casimir of the representation
$R_{i}$, and $C^{ijk}=C_{ijk}^{*}$.
Since dimensional coupling parameters such as masses  and couplings of
scalar field cubic terms do not influence the asymptotic properties
of a theory on which we are interested here, it is
sufficient to take into account only the dimensionless supersymmetric
couplings such as $g$ and $C_{ijk}$.
So we neglect the existence of dimensional parameters, and
assume furthermore that $C_{ijk}$ are real so that $C_{ijk}^2$ always are positive numbers.
For our purposes, it is convenient to work with the square of the couplings and to
arrange $C_{ijk}$ in such a way that they are covered by a single index $i~(i=1,\cdots,n)$:
\beq
\alpha = \frac{g^2}{4\pi}~,~
\alpha_{i} ~=~ \frac{g_i^2}{4\pi}~.
\label{alfas}
\eeq

The evolution equations of $\alpha$'s in perturbation theory
then take the form
\beq
\begin{split}
\frac{d\alpha}{d t}&=\beta~=~ -\beta^{(1)}\alpha^2+\cdots~,\\
\frac{d\alpha_{i}}{d t}&=\beta_{i}~=~ -\beta^{(1)}_{i}\,\alpha_{i}\,
\alpha+\sum_{j,k}\,\beta^{(1)}_{i,jk}\,\alpha_{j}\,
\alpha_{k}+\cdots~,
\label{eveq}
\end{split}
\eeq
where $\cdots$ denotes the contributions from higher orders, and
$ \beta^{(1)}_{i,jk}=\beta^{(1)}_{i,kj}  $.

Given the set of the evolution equations (\ref{eveq}), we investigate the
asymptotic  properties, as follows. First we  define
\cite{Zimmermann:1984sx,Oehme:1985jy,Oehme:1984iz,Cheng:1973nv,Chang:1974bv}
\beq
\tilde{\alpha}_{i} \equiv \frac{\alpha_{i}}{\alpha}~,~i=1,\cdots,n~,
\label{alfat}
\eeq
and derive from Eq. (\ref{eveq})
\beq
\begin{split}
\alpha \frac{d \tilde{\alpha}_{i}}{d\alpha} &=
-\tilde{\alpha}_{i}+\frac{\beta_{i}}{\beta}= \left(-1+\frac{\beta^{(1)}_{i}}{\beta^{(1)}}\,\right) \tilde{\alpha}_{i}\\
&
-\sum_{j,k}\,\frac{\beta^{(1)}_{i,jk}}{\beta^{(1)}}
\,\tilde{\alpha}_{j}\, \tilde{\alpha}_{k}+\sum_{r=2}\,
\left(\frac{\alpha}{\pi}\right)^{r-1}\,\tilde{\beta}^{(r)}_{i}(\tilde{\alpha})~,
\label{RE}
\end{split}
\eeq
where $\tilde{\beta}^{(r)}_{i}(\tilde{\alpha})~(r=2,\cdots)$
are power series of $\tilde{\alpha}$'s and can be computed
from the $r$-th loop $\beta$-functions.
Next we search for fixed points $\rho_{i}$ of Eq. (\ref{alfat}) at $ \alpha
= 0$. To this end, we have to solve
\beq
\left(-1+\frac{\beta ^{(1)}_{i}}{\beta ^{(1)}}\right) \rho_{i}
-\sum_{j,k}\frac{\beta ^{(1)}_{i,jk}}{\beta ^{(1)}}
\,\rho_{j}\, \rho_{k}=0~,
\label{fixpt}
\eeq
and assume that the fixed points have the form
\beq
\rho_{i}=0~\mbox{for}~ i=1,\cdots,n'~;~
\rho_{i} ~>0 ~\mbox{for}~i=n'+1,\cdots,n~.
\eeq
We then regard $\tilde{\alpha}_{i}$ with $i \leq n'$
as small perturbations  to the
undisturbed system which is defined by setting
$\tilde{\alpha}_{i}$  with $i \leq n'$ equal to zero.
As we have seen, it is possible to verify at the one-loop level
\cite{Zimmermann:1984sx,Oehme:1984yy,Oehme:1985jy,Oehme:1984iz} the
existence of the unique power series solution
\beq
\tilde{\alpha}_{i}=\rho_{i}+\sum_{r=2}\rho^{(r)}_{i}\,
\alpha^{r-1}~,~i=n'+1,\cdots,n~
\label{usol}
\eeq
of the reduction equations (\ref{RE}) to all orders in the undisturbed
system. These are RGI relations among couplings and keep formally
perturbative renormalizability of the undisturbed system.
So in the undisturbed system there is only {\em one independent}
coupling, the primary coupling $\alpha$.

The small perturbations caused by nonvanishing $\tilde{\alpha}_{i}$
with $i \leq n'$ enter in such a way that the reduced couplings,
i.e. $\tilde{\alpha}_{i}$  with $i > n'$, become functions not only of
$\alpha$ but also of $\tilde{\alpha}_{i}$  with $i \leq n'$.
It turned out that, to investigate such partially
reduced systems, it is most convenient to work with the partial
differential equations
\beq
\begin{split}
\left\{ \tilde{\beta}\,\frac{\partial}{\partial\alpha}
+\sum_{a=1}^{n'}\,
\tilde{\beta_{a}}\,\frac{\partial}{\partial\tilde{\alpha}_{a}}\right\}~
\tilde{\alpha}_{i}(\alpha,\tilde{\alpha})
&=\tilde{\beta}_{i}(\alpha,\tilde{\alpha})~,\\
\tilde{\beta}_{i(a)}~=~\frac{\beta_{i(a)}}{\alpha^2}
-\frac{\beta}{\alpha^{2}}~\tilde{\alpha}_{i(a)}
&,\qquad
\tilde{\beta}~\equiv~\frac{\beta}{\alpha}~,
\end{split}
\eeq
which are equivalent to the reduction equations (\ref{RE}), where we let
$a,b$ run from $1$ to $n'$ and $i,j$ from $n'+1$ to $n$
in order to avoid confusion. We then look for solutions of the form
\beq
\tilde{\alpha}_{i}=\rho_{i}+
\sum_{r=2}\,\left(\frac{\alpha}{\pi}\right)^{r-1}\,f^{(r)}_{i}
(\tilde{\alpha}_{a})~,~i=n'+1,\cdots,n~,
\label{algeq}
\eeq
where $ f^{(r)}_{i}(\tilde{\alpha}_{a})$ are supposed to be
power series of $\tilde{\alpha}_{a}$. This particular type of solution
can be motivated by requiring that in the limit of vanishing
perturbations we obtain the undisturbed
solutions (\ref{usol})
\cite{Kubo:1988zu,Zimmermann:1993ei}.
Again it is possible to obtain  the sufficient conditions for
the uniqueness of $ f^{(r)}_{i}$ in terms of the lowest order
coefficients.

\section{Reduction of Dimension-1 and -2 Parameters}\label{sec:dimful}
The reduction of couplings was originally formulated for massless theories on the basis of
the Callan-Symanzik equation
\cite{Zimmermann:1984sx,Oehme:1984yy}.
The extension to theories with massive parameters is not
straightforward if one wants to keep the generality and the rigor on the same level as for the
massless case; one has to fulfill a set of requirements coming from the renormalization group
equations, the Callan-Symanzik equations, etc. along with the normalization conditions
imposed on irreducible Green's functions
\cite{Piguet:1989pc}.
There has been a lot of progress in this direction starting from ref.~\cite{Kubo:1996js},
as it is already mentioned in the Introduction, where it was assumed that a mass-independent
renormalization scheme could be employed so that all the RG functions have only trivial
dependencies on dimensional parameters and then the mass parameters were
introduced similarly to couplings (i.e.\ as a power series in the
couplings). This choice was
justified later in \cite{Breitenlohner:2001pp,Zimmermann:2001pq} where
the scheme independence of the reduction principle has been proven generally, i.e\ it was shown that apart from dimensionless couplings,
pole masses and gauge parameters, the model may also involve coupling parameters carrying a dimension and masses.
Therefore here, to simplify the analysis, we follow \citere{Kubo:1996js} and
make use also of a mass-independent renormalization scheme.

We start by considering a renormalizable theory which contain a set of $(N + 1)$
dimension-zero couplings, $\left(\hat g_0,\hat g_1, ...,\hat g_N\right)$,
a set of $L$ parameters with mass-dimension one, $\left(\hat h_1,...,\hat h_L\right)$,
and a set of $M$ parameters with mass-dimension two, $\left(\hat m_1^2,...,\hat m_M^2\right)$.
The renormalized irreducible vertex function $\Gamma$ satisfies the RG
equation
\beq
\label{RGE_OR_1}
\mathcal{D}\Gamma\left[\Phi's;\hat g_0,\hat g_1, ...,\hat g_N;\hat h_1,...,\hat h_L;\hat m_1^2,...,\hat m_M^2;\mu\right]=0~,
\eeq
where
\beq
\label{RGE_OR_2}
\mathcal{D}=\mu\frac{\partial}{\partial \mu}+
\sum_{i=0}^N \beta_i\frac{\partial}{\partial \hat g_i}+
\sum_{a=1}^L \gamma_a^h\frac{\partial}{\partial \hat h_a}+
\sum_{\alpha=1}^M \gamma_\alpha^{m^2}\frac{\partial}{\partial \hat m_\alpha ^2}+
\sum_J \Phi_I\gamma^{\phi I}_{\,\,\,\, J}\,\frac{\delta}{\delta\Phi_J}~,
\eeq
where $\mu$ is the energy scale,
while $\beta_i$ are the $\beta$-functions of the various dimensionless
couplings $g_i$, $\Phi_I$  are
the various matter fields and
$\ga_\alpha^{m^2}$, $\ga_a^h$ and $\ga^{\phi I}_{\,\,\,\, J}$
are the mass, trilinear coupling and wave function anomalous dimensions,
respectively
(where $I$ enumerates the matter fields).
In a mass independent renormalization scheme, the $\gamma$'s are given by
\beq
\label{gammas}
\begin{split}
\gamma^h_a&=\sum_{b=1}^L\gamma_a^{h,b}(g_0,g_1,...,g_N)\hat h_b,\\
\gamma_\alpha^{m^2}&=\sum_{\beta=1}^M \gamma_\alpha^{m^2,\beta}(g_0,g_1,...,g_N)\hat m_\beta^2+
\sum_{a,b=1}^L \gamma_\alpha^{m^2,ab}(g_0,g_1,...,g_N)\hat h_a\hat h_b,
\end{split}
\eeq
where $\gamma_a^{h,b}$, $\gamma_\alpha^{m^2,\beta}$ and $\gamma_\alpha^{m^2,ab}$ are power series of the
$g$'s (which are dimensionless) in perturbation theory.\\

We look for a reduced theory where
\[
g\equiv g_0,\qquad h_a\equiv \hat h_a\quad \textrm{for $1\leq a\leq P$},\qquad
m^2_\alpha\equiv\hat m^2_\alpha\quad \textrm{for $1\leq \alpha\leq Q$}
\]
are independent parameters and the reduction of the remaining parameters
\beq
\label{reduction}
\begin{split}
\hat g_i &= \hat g_i(g), \qquad (i=1,...,N),\\
\hat h_a &= \sum_{b=1}^P f_a^b(g)h_b, \qquad (a=P+1,...,L),\\
\hat m^2_\alpha &= \sum_{\beta=1}^Q e^\beta_\alpha(g)m^2_\beta + \sum_{a,b=1}^P k^{ab}_\alpha(g)h_ah_b,
\qquad (\alpha=Q+1,...,M)
\end{split}
\eeq
is consistent with the RG equations (\ref{RGE_OR_1},\ref{RGE_OR_2}). It
turns out that the following relations should be satisfied
\beq
\label{relation}
\begin{split}
\beta_g\,\frac{\partial\hat g_i}{\partial g} &= \beta_i,\qquad (i=1,...,N),\\
\beta_g\,\frac{\partial \hat h_a}{\partial g}+\sum_{b=1}^P \gamma^h_b\,\frac{\partial\hat h_a}{\partial h_b} &= \gamma^h_a,\qquad (a=P+1,...,L),\\
\beta_g\,\frac{\partial\hat m^2_\alpha}{\partial g} +\sum_{a=1}^P \gamma_a^h\,\frac{\partial\hat m^2_\alpha}{\partial h_a} +  \sum_{\beta=1}^Q \gamma_\beta ^{m^2}\,\frac{\partial\hat m_\alpha^2}{\partial m_\beta^2} &= \gamma_\alpha^{m^2}, \qquad (\alpha=Q+1,...,M).
\end{split}
\eeq
Using \refeqs{gammas} and (\ref{reduction}), the above relations reduce to
\beq
\label{relation_2}
\begin{split}
&\beta_g\,\frac{df^b_a}{dg}+ \sum_{c=1}^P f^c_a\left[\gamma^{h,b}_c + \sum_{d=P+1}^L \gamma^{h,d}_c f^b_d\right] -\gamma^{h,b}_a - \sum_{d=P+1}^L \gamma^{h,d}_a f^b_d=0,\\
&\hspace{8.6cm} (a=P+1,...,L;\, b=1,...,P),\\
&\beta_g\,\frac{de^\beta_\alpha}{dg} + \sum_{\gamma=1}^Q e^\gamma_\alpha\left[\gamma_\gamma^{m^2,\beta} +
\sum _{\delta=Q+1}^M\gamma_\gamma^{m^2,\delta} e^\beta_\delta\right]-\gamma_\alpha^{m^2,\beta} -
\sum_{\delta=Q+1}^M \gamma_\alpha^{m^2,d}e^\beta_\delta =0,\\
&\hspace{8.3cm} (\alpha=Q+1,...,M ;\, \beta=1,...,Q),\\
&\beta_g\,\frac{dk_\alpha^{ab}}{dg}
+ 2\sum_{c=1}^P \left(\gamma_c^{h,a} + \sum_{d=P+1}^L \gamma_c^{h,d} f_d^a\right)k_\alpha^{cb}
+\sum_{\beta=1}^Q e^\beta_\alpha\left[\gamma_\beta^{m^2,ab} + \sum_{c,d=P+1}^L \gamma_\beta^{m^2,cd}f^a_cf^b_d \right.\\
&\left. +2\sum_{c=P+1}^L \gamma_\beta^{m^2,cb}f^a_c + \sum_{\delta=Q+1}^M \gamma_\beta^{m^2,d} k_\delta^{ab}\right]- \left[\gamma_\alpha^{m^2,ab}+\sum_{c,d=P+1}^L \gamma_\alpha^{m^2,cd}f^a_c f^b_d\right.\\
&\left. +2 \sum_{c=P+1}^L \gamma_\alpha^{m^2,cb}f^a_c + \sum_{\delta=Q+1}^M \gamma_\alpha^{m^2,\delta}k_\delta^{ab}\right]=0,\\
&\hspace{8cm} (\alpha=Q+1,...,M;\, a,b=1,...,P)~.
\end{split}
\eeq
The above relations ensure that the irreducible vertex function of the reduced theory
\beq
\label{Green}
\begin{split}
\Gamma_R&\left[\Phi\textrm{'s};g;h_1,...,h_P; m_1^2,...,m_Q^2;\mu\right]\equiv\\
&\Gamma \left[  \Phi\textrm{'s};g,\hat g_1(g)...,\hat g_N(g);
h_1,...,h_P,\hat h_{P+1}(g,h),...,\hat h_L(g,h);\right.\\
& \left.  \qquad\qquad\qquad m_1^2,...,m^2_Q,\hat m^2_{Q+1}(g,h,m^2),...,\hat m^2_M(g,h,m^2);\mu\right]
\end{split}
\eeq
has the same renormalization group flow as the original one.

The assumption that the reduced theory is perturbatively renormalizable
means that the functions
$\hat g_i$, $f^b_a$, $e^\beta_\alpha$ and $k_\alpha^{ab}$, defined in Eq. (\ref{reduction}), should be
expressed as a power series in the primary coupling $g$:
\beq
\label{pert}
\begin{split}
\hat g_i & = g\sum_{n=0}^\infty \rho_i^{(n)} g^n,\qquad
f_a^b  =  g \sum_{n=0}^\infty \eta_a^{b(n)} g^n\\
e^\beta_\alpha & = \sum_{n=0}^\infty \xi^{\beta(n)}_\alpha g^n,\qquad
k_\alpha^{ab}=\sum_{n=0}^\infty \chi_\alpha^{ab(n)} g^n.
\end{split}
\eeq
The above expansion coefficients can be found by inserting these power
series into \refeqs{relation}, (\ref{relation_2}) and requiring the equations to be satisfied at each order of $g$.
It should be noted that the existence of a unique power series solution is a non-trivial matter: It depends on the theory
as well as on the choice of the set of independent parameters.

It should also be noted that in the case that there are no \textit{independent}
mass-dimension~1 parameters ($\hat h$) the reduction of these terms take
naturally the form
\[
\hat h_a = \sum_{b=1}^L f_a^b(g)M,
\]
where $M$ is a mass-dimension 1 parameter which could be a gaugino mass that
corresponds to the independent (gauge) coupling. Furthermore, if there
are no \textit{independent} mass-dimension 2
parameters ($\hat m^2$), the corresponding reduction takes the analogous form
\[
\hat m^2_a=\sum_{b=1}^M e_a^b(g) M^2.
\]

\section{Reduction of Couplings of Soft Breaking Terms in $N=1$ Suspersymmetric
 Theories}\label{sec:RCN=1}

The method of reducing the dimensionless couplings was
extended\cite{Kubo:1996js,Jack:1995gm}, as we have discussed in the introduction, to the soft
supersymmetry breaking (SSB) dimensionful parameters of $N = 1$
supersymmetric theories.  In addition it was found \cite{Kawamura:1997cw,Kobayashi:1997qx} that
RGI SSB scalar masses in Gauge-Yukawa unified models satisfy a
universal sum rule.

Consider the superpotential given by
\be
W= \frac{1}{2}\,\mu^{ij} \,\Phi_{i}\,\Phi_{j}+
\frac{1}{6}\,C^{ijk} \,\Phi_{i}\,\Phi_{j}\,\Phi_{k}~,
\label{supot}
\ee
along with the Lagrangian for SSB terms
\be
-{\cal L}_{\rm SSB} =
\frac{1}{6} \,h^{ijk}\,\phi_i \phi_j \phi_k
+
\frac{1}{2} \,b^{ij}\,\phi_i \phi_j
+
\frac{1}{2} \,(m^2)^{j}_{i}\,\phi^{*\,i} \phi_j+
\frac{1}{2} \,M\,\lambda \lambda+\mbox{H.c.},
\label{supot_l}
\ee
where the $\phi_i$ are the scalar parts of the chiral superfields $\Phi_i$, $\lambda$ are the gauginos
and $M$ their unified mass.

Let us recall (see Eqs.(\ref{betag}-\ref{gamay}))
that the one-loop $\beta$-function of the gauge coupling $g$ is given by
\cite{Parkes:1984dh,West:1984dg,Jones:1985ay,Jones:1984cx,Parkes:1985hh}
\bea
\beta^{(1)}_{g}=\frac{d g}{d t} =
  \frac{g^3}{16\pi^2}\,\left[\,\sum_{i}\,T(R_{i})-3\,C_{2}(G)\,\right]~,
\eea
the $\beta$-function of $C_{ijk}$ is given by
\be
\beta_C^{ijk} =
  \frac{d C_{ijk}}{d t}~=~C_{ijl}\,\gamma^{l}_{k}+
  C_{ikl}\,\gamma^{l}_{j}+
  C_{jkl}\,\gamma^{l}_{i}~,
\ee
and, at one-loop level, the anomalous dimension $\gamma^{(1)}\,^i_j$ of the chiral
superfield is
\be
\gamma^{(1)}\,^i_j=\frac{1}{32\pi^2}\,\left[\,
C^{ikl}\,C_{jkl}-2\,g^2\,C_{2}(R_{i})\delta^i_j\,\right].
\ee
Then, the $N = 1$ non-renormalization theorem \cite{Wess:1973kz,Iliopoulos:1974zv,Fujikawa:1974ay} ensures
there are no extra mass and cubic-interaction-term renormalizations,
implying that the $\beta$-functions of $C_{ijk}$ can be expressed as
linear combinations of the anomalous dimensions $\gamma^i_j$.

Here we assume that the reduction equations admit power series solutions of the form
\be
C^{ijk} = g\,\sum_{n=0}\,\rho^{ijk}_{(n)} g^{2n}~.
\label{Yg}
\ee
In order to obtain higher-loop results instead of knowledge of
explicit $\beta$-functions, which anyway are known only up to
two-loops, relations among $\beta$-functions are required.

Judicious use of  the spurion technique,
\cite{Fujikawa:1974ay,Delbourgo:1974jg,Salam:1974pp,Grisaru:1979wc,Girardello:1981wz} leads to
the following  all-loop relations among SSB $\beta$-functions (in an
obvious notation),
\cite{Yamada:1994id,Kazakov:1997nf,Jack:1997pa,Hisano:1997ua,Jack:1997eh,Avdeev:1997vx,Kazakov:1998uj}
\begin{align}
\beta_M &= 2{\cal O}\left(\frac{\beta_g}{g}\right)~,
\label{betaM}\\
\beta_h^{ijk}&=\gamma^i_l h^{ljk}+\gamma^j_l h^{ilk}
+\gamma^k_l h^{ijl}\non\\
&\,-2\left(\gamma_1\right)^i_l C^{ljk}
-2\left(\gamma_1\right)^j_l C^{ilk}-2\left(\gamma_1\right)^k_l C^{ijl}~,\label{betah}\\
(\beta_{m^2})^i_j &=\left[ \Delta
+ X \frac{\partial}{\partial g}\right]\gamma^i_j~,
\label{betam2}
\end{align}
where
\begin{align}
{\cal O} &=\left(Mg^2\frac{\partial}{\partial g^2}
-h^{lmn}\frac{\partial}{\partial C^{lmn}}\right)~,
\label{diffo}\\
\Delta &= 2{\cal O}{\cal O}^* +2|M|^2 g^2\frac{\partial}
{\partial g^2} +\tilde{C}_{lmn}
\frac{\partial}{\partial C_{lmn}} +
\tilde{C}^{lmn}\frac{\partial}{\partial C^{lmn}}~,\\
(\gamma_1)^i_j&={\cal O}\gamma^i_j,\\
\tilde{C}^{ijk}&=
(m^2)^i_l C^{ljk}+(m^2)^j_l C^{ilk}+(m^2)^k_l C^{ijl}~.
\label{tildeC}
\end{align}

The assumption, following \cite{Jack:1997pa}, that the relation among couplings
\be
h^{ijk} = -M (C^{ijk})'
\equiv -M \frac{d C^{ijk}(g)}{d \ln g}~,
\label{h2}
\ee
is RGI and furthermore, the use of the all-loop gauge $\beta$-function of Novikov  et al.
\cite{Novikov:1983ee,Novikov:1985rd,Shifman:1996iy}
given by
\be
\beta_g^{\rm NSVZ} =
\frac{g^3}{16\pi^2}
\left[ \frac{\sum_l T(R_l)(1-\gamma_l /2)
-3 C_2(G)}{ 1-g^2C_2(G)/8\pi^2}\right]~,
\label{bnsvz}
\ee
lead to the all-loop RGI sum rule \cite{Kobayashi:1998jq} (assuming $(m^2)^i_j=m^2_j\delta^i_j$),
\begin{equation}
\begin{split}
m^2_i+m^2_j+m^2_k &=
|M|^2 \left\{~
\frac{1}{1-g^2 C_2(G)/(8\pi^2)}\frac{d \ln C^{ijk}}{d \ln g}
+\frac{1}{2}\frac{d^2 \ln C^{ijk}}{d (\ln g)^2}~\right\}\\
& \qquad\qquad +\sum_l
\frac{m^2_l T(R_l)}{C_2(G)-8\pi^2/g^2}
\frac{d \ln C^{ijk}}{d \ln g}~.
\label{sum2}
\end{split}
\end{equation}
Surprisingly enough, the all-loop result of Eq.(\ref{sum2}) coincides with
the superstring result for the finite case in a certain class of orbifold models
\cite{Ibanez:1992hc,Brignole:1995fb,Kobayashi:1997qx}
if
\[
\frac{d \ln C^{ijk}}{d \ln g}=1~,
\]
as discussed in ref.~\cite{Mondragon:1993tw}.

Let us now see how the all-loop results on the SSB $\beta$-functions, Eqs.(\ref{betaM})-(\ref{tildeC}),
lead to all-loop RGI relations. We assume:\\
(a) the existence of a RGI surfaces on which $C = C(g)$, or equivalently that the expression
\be
\label{Cbeta}
\frac{dC^{ijk}}{dg} = \frac{\beta^{ijk}_C}{\beta_g}
\ee
holds,  i.e. reduction of couplings is possible, and\\
(b) the existence of a RGI surface on which
\be
\label{h2NEW}
h^{ijk} = - M \frac{dC(g)^{ijk}}{d\ln g}
\ee
holds too in all-orders.\\
Then one can prove \cite{Jack:1999aj,Kobayashi:1998iaa}, that the following relations are RGI to all-loops (note that in
both (a) and (b) assumptions above we do not rely on specific solutions of these equations)
\begin{align}
M &= M_0~\frac{\beta_g}{g} ,  \label{Mbeta} \\
h^{ijk}&=-M_0~\beta_C^{ijk},  \label{hbeta}  \\
b^{ij}&=-M_0~\beta_{\mu}^{ij},\label{bij}\\
(m^2)^i_j&= \frac{1}{2}~|M_0|^2~\mu\frac{d\gamma^i{}_j}{d\mu},
\label{scalmass}
\end{align}
where $M_0$ is an arbitrary reference mass scale to be specified shortly. The assumption that
\be
C_a\frac{\partial}{\partial C_a}
= C_a^*\frac{\partial}{\partial C_a^*} \label{dc/dc}
\ee
for a RGI surface $F(g,C^{ijk},C^{*ijk})$ leads to
\begin{equation}
\label{F}
\frac{d}{dg} = \left(\frac{\partial}{\partial g} + 2\frac{\partial}{\partial C}\,\frac{dC}{dg}\right)
= \left(\frac{\partial}{\partial g} + 2 \frac{\beta_C}{\beta_g}
\frac{\partial}{\partial C} \right)\, ,
\end{equation}
where Eq.(\ref{Cbeta}) has been used. Now let us consider the partial differential operator ${\cal O}$ in
Eq.(\ref{diffo}) which, assuming Eq.(\ref{h2}), becomes
\be
{\cal O} = \frac{1}{2}M\frac{d}{d\ln g}\, .
\ee
In turn, $\beta_M$ given in Eq.(\ref{betaM}), becomes
\be
\beta_M = M\frac{d}{d\ln g} \big( \frac{\beta_g}{g}\big) ~, \label{betaM2}
\ee
which by integration provides us \cite{Karch:1998qa,Jack:1999aj} with the
generalized, i.e. including Yukawa couplings, all-loop RGI Hisano - Shifman relation \cite{Hisano:1997ua}
\be
 M = \frac{\beta_g}{g} M_0~, \label{M-M0}
\ee
where $M_0$ is the integration constant and can be associated to the
unification scale $M_U$ in GUTs or to the gravitino mass $m_{3/2}$ in
a supergravity framework. Therefore, Eq.(\ref{M-M0}) becomes the
all-loop RGI Eq.(\ref{Mbeta}).  Note that $\beta_M$ using
Eqs.(\ref{betaM2}) and (\ref{M-M0}) can be written as
\be \beta_M =
M_0\frac{d}{dt} (\beta _g/g)~.
\ee
Similarly
\be (\gamma_1)^i_j =
{\cal O} \gamma^i_j = \frac{1}{2}~M_0~\frac{d
  \gamma^i_j}{dt}~. \label{gammaO}
\ee
Next, from Eq.(\ref{h2}) and Eq.(\ref{M-M0}) we obtain
\be
 h^{ijk} = - M_0 ~\beta_C^{ijk}~,  \label{hm32}
\ee
while $\beta^{ijk}_h$, given in Eq.(\ref{betah}) and using Eq.(\ref{gammaO}), becomes \cite{Jack:1999aj}
\be
  \beta_h^{ijk} = - M_0~\frac{d}{dt} \beta_C^{ijk},
\ee
which shows that Eq.(\ref{hm32}) is all-loop RGI. In a similar way
Eq.(\ref{bij}) can be shown to be all-loop RGI.

Finally we would like to emphasize that under the same assumptions (a)
and (b) the sum rule given in Eq.(\ref{sum2}) has been proven
\cite{Kobayashi:1998jq} to be all-loop RGI, which (using Eq.(\ref{M-M0})) gives us
a generalization of Eq.(\ref{scalmass}) to be applied in
considerations of non-universal soft scalar masses, which are
necessary in many cases including the MSSM.

Having obtained the Eqs.(\ref{Mbeta})-(\ref{scalmass}) from
Eqs.(\ref{betaM})-(\ref{tildeC}) with the assumptions (a) and (b), we
would like to conclude the present section with some remarks.  First
it is worth noting the difference, say in first order in $g$, among
the possibilities to consider specific solution of the reduction
equations or just assume the existence of a RGI surface, which is a
weaker assumption. So in the case we consider the reduction equation
(\ref{Cbeta}) without relying on a specific solution, the sum rule
of Eq.(\ref{sum2}) reads
\be m^2_i+m^2_j+m^2_k = |M|^2 \frac{d\ln C^{ijk}}{d\ln g},
\label{sumrulenow}
\ee
and we find that
\be
\frac{d\ln C^{ijk}}{d\ln g} =\frac {g}{C^{ijk}}\frac{dC^{ijk}}{dg} =
\frac{g}{C^{ijk}}\frac{ \beta^{ijk}_C}{\beta_g},
\ee
which is clearly
model dependent. However assuming a specific power series solution of
the reduction equation, as in Eq.(\ref{powerser}), which in first
order in $g$ is just a linear relation among $C^{ijk}$ and $g$, we
obtain that
\be
\label{dcdg}
\frac{d\ln C^{ijk}}{d\ln g} = 1
\ee
and therefore the sum rule of Eq.(\ref{sumrulenow}) becomes model
independent. We should also emphasize that in order to show
\cite{Jack:1997pa} that the relation
\be (m^2)^i_j =
\frac{1}{2}\frac{g^2}{\beta_g} |M|^2\frac{d\gamma^i_j}{dg},
\ee
which using Eq.(\ref{M-M0}) becomes Eq.(\ref{scalmass}), is RGI to
all-loops a specific solution of the reduction equations has to be
required. As it has already been pointed out above such a requirement
is not necessary in order to obtain the all-loop RG invariance of the
sum rule of Eq.(\ref{sum2}).

As it was emphasized in ref.~\cite{Jack:1999aj} the set of the all-loop
RGI relations (\ref{Mbeta})-(\ref{scalmass}) is the one obtained in
the \textit{Anomaly Mediated SB Scenario}
\cite{Randall:1998uk,Giudice:1998xp,Gherghetta:1999sw,Pomarol:1999ie,Chacko:1999am,Katz:1999uw},
by fixing
the $M_0$ to be $m_{3/2}$, which is the natural scale in the
supergravity framework.

A final remark concerns the resolution of the fatal problem of the
anomaly induced scenario in the supergravity framework, i.e.
the negative mass squared for sleptons, leading to  tachyonic sleptons.
Here, the problem is solved thanks to the sum rule of Eq.(\ref{sum2}), as it will become clear in
the next section.  Other solutions have been provided by introducing
Fayet-Iliopoulos terms \cite{Hodgson:2005en}.

%% file: PR2018_Chapter_3.tex
\chapter{Reduction of Couplings in the Standard Model and Predictions}
The first application of the idea of reduction of couplings in realistic models
was presented in the celebrated paper \cite{Kubo:1985up}.
We encourage the reader to study the
original article and here we limit ourselves to an introduction, comments and
updated remarks of the authors presented in the book
``Reduction of couplings and its application in particle physics, Finite theories, Higgs and top mass predictions''
\cite{Heinemeyer:2014vxa}.

Even today, more than twenty years after the first paper on reduction of couplings in
the Standard Model the original motivation for applying this
method to this Model has not become obsolete, neither by time nor by new insight.
The theoretical predictions originating from the Standard Model are in extremely
good agreement with experiment. Two decades of precision
measurement and precision calculation yielded essentially on all available observables
a truly astonishing coincidence\cite{Hollik:2010id}.
And, yet there is no convincing explanation why
the number of families is three; why the mass scales --the Planck mass and the electroweak
breaking scale-- differ so much in magnitude, why the Higgs mass is so small compared to
the Planck scale. And, quite generally, there is also no explanation for the mixing
of the families.

Reduction of couplings offers a way to understand at least to some degree masses and
mixings of charged leptons and quarks and the mass of the Higgs particle.  It extends
the well known case of closed renormalization orbits due to symmetry to other, more
general ones. Which structure these orbits have had to be learned, i.e. deduced from
the relevant renormalization group equation in the specific model. In particular, one had
to take into account the different behaviour of abelian versus non-abelian gauge groups
and of the Higgs self-coupling, say in the ultraviolet region. If asymptotic expansions
should make sense in the transition from a non-perturbative theory to a perturbative
version it should be possible to rely on common ultraviolet asymptotic freedom. One
also has to respect gross features coming from phenomenology. In mathematical terms this
is the problem of integrating partial differential equations by imposing suitable boundary
conditions (originating from physical requirements): partial reduction.

Perhaps the most important and not obvious result of the entire analysis is the fact that
reduction of couplings (even the version of ``partial reduction'') is extremely sensitive to
the model. If one accepts the integration ``paths'' as derived in the relevant papers,
the ordinary Standard Model can neither support a mass of the top quark nor of the Higgs
particle as large as they have been found experimentally.
There is an apparent mismatch among the the reduced Standard Model predictions and the experimental
findings of the top and Higgs masses. Renormalization group improvements of the original theoretical
predictions were concerning essentially the QCD sector, which was taken into account in the reduction.
Whereas the differences originating from the other couplings turned out to be negligibly small. Hence
it became clear that other model classes are to be studied and further constraining principles had to
be found. This will be the subject of Chapters~\ref{ch:fin} and \ref{ch:viable}.

In ref.~\cite{Kubo:1985up}
within the context of the Standard Model with one Higgs doublet and n families
the principle of reduction of couplings was applied. For simplicity mixing of the families
was assumed to be absent: the Yukawa couplings are diagonal and real. For the massless model
reduction solutions can be found to all orders of perturbation theory as power series in the
``primary'' coupling, thus superseding fixed point considerations based on one-loop approximations.
Due to the different asymptotic behaviour of the SU(3), SU(2) and U(1) couplings the space of
solutions is clearly structured and permits reduction in very distinct
ways only. Since reducing the gauge couplings relative to each other is either inconsistent
or phenomenologically not acceptable, $\alpha_s$ (the largest coupling) has been chosen as the expansion
parameter --the primary coupling-- and thus UV-asymptotic freedom as the relevant regime.
This allows to neglect in the lowest order approximation the other gauge
couplings and to take their effect into account as corrections.

In the matter sector (leptons, quarks, Higgs) discrete solutions emerge for the reduced
couplings which permit essentially only the Higgs self-coupling and the Yukawa coupling
to the top quark to be non-vanishing. Stability considerations (Liapunov's theory) show
how the power series solutions are embedded in the set of the general solutions.
Couplings
of the massless model were converted into masses in the tree approximation of the spontaneously
broken model. For three generations one finds $m_H = 61$~GeV, $m_t = 81$~GeV with an error
of about 10-15\%.

Reduction of couplings is based on the requirement that all reduced couplings vanish simultaneously
upon reduction of the primary coupling. This is clearly only possible if the couplings considered have
the same asymptotic behavior or have vanishing $\beta$-functions. Hence in the Standard model, based on
$SU(3)\times SU(2)\times U(1)$ straightforward reduction cannot be realized. Since however the strong coupling $\alpha_s$
is, say at the $W$-mass, considerably larger than the weak and electromagnetic coupling one may put those
equal to zero, reduce within the system of quantum chromodynamics including the Higgs and the Yukawa
couplings and subsequently take into account electroweak corrections as a kind of
perturbation. This is called ``partial reduction''. In \cite{Kubo:1988zu}
a new perturbation method was developed and then applied using the updated experimental values of the
strong coupling and the Weinberg angle at the time.

In asymptotically free theories the $\beta$-functions usually go to zero with some power of the couplings involved.
Thus, reduction equations are singular for vanishing coupling and require a case by case study at this
singular point. In particular this is true for the reduction equations of Yukawa and Higgs couplings
when reducing  to $\alpha_s$. It was shown in the paper that for the non-trivial reduction solution
(i.e. only the top Yukawa coupling and the Higgs coupling do not vanish) one can de-singularize the
system by a variable transformation and thereafter go over to a partial differential equation which
is easier to solve than the ordinary differential equations one started with. The reduction solutions
of the perturbed system are then in one-to-one correspondence with the unperturbed ones.

In terms of mass values the non-trivial reduction yields $m_t = 91.3$~GeV, $m_H = 64.3$~GeV. These mass
values are at the same time the upper bound for the trivial reduction, where the Higgs mass is a function
of the top mass. Here is used as definition for ``trivial'' that the ratios of top-Yukawa coupling
and Higgs coupling with respect to $\alpha_s$ go to zero for the weak coupling limit $\alpha_s$ going to zero.

Still there are corrections to the above values:\\
1. The above mass values depend on the SM parameters, in particular the strong coupling constant $\alpha_s$
and $\sin\theta_W$.  Since the values of $\alpha_s$ and $\sin\theta_W$ were updated, the above predictions
had to be updated, too.\\
2. Two-loop corrections could be important.\\
3. In ref.~\cite{Kubo:1985up}
the difference of the physical mass (pole mass) and the mass defined in the $\overline{\textrm{MS}}$ scheme has been ignored.
In ref.~\cite{Kubo:1991ui}
all these corrections are included. It was found that the correction coming from the $\overline{\textrm{MS}}$ to
the pole mass transition increases $m_t$ by about 4\%, while $m_H$ is increased by about 1\%.
The two-loop effect is non-negligible especially for $m_t: +2$\% for $m_t$ and 0.2\% for $m_H$.
Taking into account all these corrections it was found
\beq
   m_t =  98.6\pm 9.2\textrm{ GeV},    m_H = 64.5\pm 1.5\textrm{ GeV},
\eeq
where the 1991 values of $M_Z$, $\alpha_s(M_Z)$, $\sin^2\theta_W(M_Z)$ and $\alpha_{em}(M_Z)$
were used. Even with updated values it was found \cite{Heinemeyer:2014vxa}
that the change of the prediction is negligible. Obviously, this prediction
is inconsistent with the experimental observations.

An optimistic point of view is that the failure of the reduction of couplings programme in the
SM shows that it is not the final theory but only a very interesting part of it and therefore
we have to search further for the ultimate theory. 

%% file: PR2018_Chapter_4.tex
\chapter{Finiteness}
\label{ch:fin}
The principle of finiteness  requires perhaps some more motivation
to be considered and generally accepted these days than when it was
first envisaged.
It is however interesting to note that in the early days
of field theory the feeling was quite different.
Probably the well
known Dirac's phrase that ``...divergencies are hidden under the carpet'' is
representative of the views of that time.
In recent years we have a more relaxed attitude towards divergencies. Most
theorists believe that the divergencies are signals of the existence of a higher
scale, where new degrees of freedom are excited.
Even accepting this dogma, we are naturally led to the conclusion that
beyond the unification scale, i.e. when all interactions have been taken
into account in a unified scheme, the theory should be completely finite.
In fact, this is one of the main motivations and aims of string, non-commutative
geometry, and quantum group theories, which include also gravity in the unification
of the interactions.
In our work on reduction of couplings and finiteness we restricted ourselves to
unifying only the known gauge interactions, based on a lesson of the history of
Elementary Particle Physics (EPP) that if a nice idea works in physics, usually
it is realised in its simplest
form.

\section{The idea behind finiteness}
Finiteness is based on the fact that it is possible to find renormalization
group invariant (RGI) relations among couplings that keep finiteness in perturbation
theory, even to all orders.
Accepting finiteness as a guiding principle in constructing realistic theories of EPP,
the first thing that comes to mind is to look for an $N = 4$ supersymmetric unified
gauge theory,
since any ultraviolet (UV) divergencies are absent in these theories.
However nobody has managed so far to produce realistic models in the framework of
$N = 4$ SUSY. In the best of cases one could try to do a drastic truncation of the
theory like the orbifold projection of refs.~\cite{Kachru:1998ys,Chatzistavrakidis:2010xi},
but this is already a different theory than the original one.
The next possibility is to consider an $N = 2$ supersymmetric gauge theory,
whose beta-function receives corrections only at one-loop.
Then it is not hard to select a spectrum to make the theory all-loop
finite. However a serious obstacle in these theories is their mirror spectrum, which in
the absence of a mechanism to make it heavy, does not permit the construction of realistic
models. Therefore, we are naturally led to consider $N = 1$ supersymmetric gauge theories,
which can be chiral and in principle realistic.

Let us be clear at this point and state that in our approach (ultra violet, UV) finiteness
means the vanishing of all the $\beta$-functions, i.e. the non-renormalization
of the coupling constants, in contrast to a complete (UV) finiteness where even
field amplitude renormalization is absent.
Before our work the studies on $N = 1$ finite theories were following two directions:
(a)~construction of finite theories up to two-loops examining various possibilities to
make them phenomenologically viable,
(b)~construction of all-loop finite models without particular emphasis on the
phenomenological consequences. The success of our work was that we  constructed the
first realistic all-loop finite model, based on the theorem presented in the Sect.~\ref{sec:N=1_FIN},
realising in this way an old theoretical dream of field theorists. Equally
important was the correct prediction of the top quark mass one and half year before
the experimental discovery. It was the combination of these two facts that motivated
us to continue with the study of $N=1$ finite theories.  It is worth noting that nobody
expected at the time such a heavy mass for the top quark. Given that the analysis of
the experimental data changes over time, the comparison of our original prediction with
the updated analyses will be discussed later.

\section{Finiteness in N=1 Supersymmetric Gauge Theories}
\label{sec:N=1_FIN}
Let us, once more, consider a chiral, anomaly free, $N=1$ globally supersymmetric
gauge theory based on a group G with gauge coupling constant $g$.
The superpotential of the theory is given by (see Eq.(\ref{supot0}))
\beq
W= \frac{1}{2}\,m_{ij} \,\phi_{i}\,\phi_{j}+
\frac{1}{6}\,C_{ijk} \,\phi_{i}\,\phi_{j}\,\phi_{k}~.
\label{supot}
\eeq
The $N=1$ non-renormalization theorem, ensuring the absence of mass and
cubic-interaction-term infinities, leads to wave-function infinities.
The one-loop $\beta$-function is given by (see Eq.(\ref{betag})
\beq
\beta^{(1)}_{g}=\frac{d g}{d t} =
\frac{g^3}{16\pi^2}\left[\,\sum_{i}\,T(R_{i})-3\,C_{2}(G)\right]~,
\label{betagP}
\eeq
the $\beta$-function of $C_{ijk}$ by (see Eq. (\ref{betay}))
\beq
\beta_{ijk} =
 \frac{d C_{ijk}}{d t}~=~C_{ijl}\,\gamma^{l}_{k}+
 C_{ikl}\,\gamma^{l}_{j}+
 C_{jkl}\,\gamma^{l}_{i}~
\label{betayP}
\eeq
and the one-loop wave function anomalous dimensions by (see Eq. (\ref{gamay} ))
\beq
\gamma^{(1)}{}_{j}^{i}=\frac{1}{32\pi^2}\,[\,
C^{ikl}\,C_{jkl}-2\,g^2\,C_{2}(R)\delta_{j}^i\,]~. \label{gamayP}
\eeq

As one can see from Eqs.~(\ref{betagP}) and  (\ref{gamayP}),
all the  one-loop $\beta$-functions of the  theory vanish if
$\beta_g^{(1)}$ and  $\gamma^{(1)}{}_{j}^{i}$ vanish, i.e.
\begin{align}
\sum _i T(R_{i})& = 3 C_2(G) \,,
\label{1st}     \\
 C^{ikl} C_{jkl} &= 2\delta ^i_j g^2  C_2(R_i)\,,
\label{2nd}
\end{align}

The conditions for finiteness for $N=1$ field theories with $SU(N)$ gauge
symmetry are discussed in \cite{Rajpoot:1984zq}, and  the
analysis of the  anomaly-free and  no-charge renormalization
requirements for these theories can be found in \cite{Rajpoot:1985aq}.
A very interesting result is that the  conditions (\ref{1st},\ref{2nd})
are necessary and  sufficient for finiteness at the  two-loop level
\cite{Parkes:1984dh,West:1984dg,Jones:1985ay,Jones:1984cx,Parkes:1985hh}.

In case SUSY is broken by soft terms, the  requirement of
finiteness in the  one-loop soft breaking terms imposes further
constraints among themselves \cite{Jones:1984cu}.  In addition, the  same set
of conditions that are sufficient for one-loop finiteness of the  soft
breaking terms render the  soft sector of the  theory two-loop
finite\cite{Jack:1994kd}.

The one- and  two-loop finiteness conditions of Eqs. (\ref{1st},\ref{2nd}) restrict
considerably the  possible choices of the  irreducible representations
(irreps)
$R_i$ for a given
group $G$ as well as the  Yukawa couplings in the  superpotential
(\ref{supot}).  Note in particular that the  finiteness conditions cannot be
applied to the  minimal supersymmetric standard model (MSSM), since the  presence
of a $U(1)$ gauge group is incompatible with the  condition
(\ref{1st}), due to $C_2[U(1)]=0$.  This naturally leads to the
expectation that finiteness should be attained at the  grand unified
level only, the  MSSM being just the  corresponding, low-energy,
effective theory.

Another important consequence of one- and  two-loop finiteness is that
SUSY (most probably) can only be broken due to the  soft
breaking terms.  Indeed, due to the  unacceptability of gauge singlets,
F-type spontaneous symmetry breaking \cite{O'Raifeartaigh:1975pr}
terms are incompatible with finiteness, as well as D-type
\cite{Fayet:1974jb} spontaneous breaking which requires the  existence
of a $U(1)$ gauge group.

A natural question to ask is what happens at higher loop orders.  The
answer is contained in a theorem
\cite{Lucchesi:1987he,Lucchesi:1987ef} which states the  necessary and
sufficient conditions to achieve finiteness at all orders.  Before we
discuss the  theorem let us make some introductory remarks.  The
finiteness conditions impose relations between gauge and  Yukawa
couplings.  To require such relations which render the  couplings
mutually dependent at a given renormalization point is trivial.  What
is not trivial is to guarantee that relations leading to a reduction
of the  couplings hold at any renormalization point.  As we have seen
(see Eq. (\ref{Cbeta})),
the necessary and  also sufficient, condition for this to happen is to
require that such relations are solutions to the  REs
\beq \beta _g
\frac{d C_{ijk}}{dg} = \beta _{ijk}
\label{redeq2}
\eeq
and hold at all orders.   Remarkably, the  existence of
all-order power series solutions to (\ref{redeq2}) can be decided at
one-loop level, as already mentioned.

Let us now turn to the  all-order finiteness theorem
\cite{Lucchesi:1987he,Lucchesi:1987ef}, which states under which conditions an $N=1$
supersymmetric gauge theory can become finite to all orders in perturbation theory,
that is attain physical scale invariance.  It is based on (a) the  structure of the  supercurrent in
$N=1$ supersymmetric gauge theory
\cite{Ferrara:1974pz,Piguet:1981mu,Piguet:1981mw}, and  on (b) the
non-renormalization properties of $N=1$ chiral anomalies
\cite{Lucchesi:1987he,Lucchesi:1987ef,Piguet:1986td,Piguet:1986pk,Ensign:1987wy}.
Details of the  proof can be found in
refs.~\cite{Lucchesi:1987he,Lucchesi:1987ef} and  further discussion in
\citeres{Piguet:1986td,Piguet:1986pk,Ensign:1987wy,Lucchesi:1996ir,Piguet:1996mx}.
Here, following mostly \citere{Piguet:1996mx} we present a
comprehensible sketch of the  proof.

Consider an $N=1$ supersymmetric gauge theory, with simple Lie group
$G$.  The  content of this theory is given at the  classical level by
the matter supermultiplets $S_i$, which contain a scalar field
$\phi_i$ and  a Weyl spinor $\psi_{ia}$, and  the  vector supermultiplet
$V_a$, which contains a gauge vector field $A_{\mu}^a$ and  a gaugino
Weyl spinor $\lambda^a_{\alpha}$.

Let us first recall certain facts about the  theory:

\noindent (1)  A massless $N=1$ supersymmetric theory is invariant
under a $U(1)$ chiral transformation $R$ under which the  various fields
transform as follows
\beq
\begin{split}
A'_{\mu}&=A_{\mu},~~\lambda '_{\alpha}=\exp({-i\theta})\lambda_{\alpha}\\
\phi '&= \exp({-i\frac{2}{3}\theta})\phi,~~\psi_{\alpha}'= \exp({-i\frac{1}
    {3}\theta})\psi_{\alpha},~\cdots
\end{split}
\eeq
The corresponding axial Noether current $J^{\mu}_R(x)$ is
\beq
J^{\mu}_R(x)=\bar{\lambda}\gamma^{\mu}\gamma^5\lambda + \cdots
\label{noethcurr}
\eeq
is conserved classically, while in the  quantum case is violated by the
axial anomaly
\beq
\partial_{\mu} J^{\mu}_R =
r\left(\epsilon^{\mu\nu\sigma\rho}F_{\mu\nu}F_{\sigma\rho}+\cdots\right).
\label{anomaly}
\eeq

From its known topological origin in ordinary gauge theories
\cite{AlvarezGaume:1983cs,Bardeen:1984pm,Zumino:1983rz}, one would
expect the  axial vector current
$J^{\mu}_R$ to satisfy the  Adler-Bardeen theorem  and
receive corrections only at the  one-loop level.  Indeed it has been
shown that the  same non-renormalization theorem holds also in
supersymmetric theories \cite{Piguet:1986td,Piguet:1986pk,Ensign:1987wy}.  Therefore
\beq
r=\hbar \beta_g^{(1)}.
\label{r}
\eeq

\noindent (2)  The  massless theory we consider is scale invariant at
the classical level and, in general, there is a scale anomaly due to
radiative corrections.  The  scale anomaly appears in the  trace of the
energy momentum tensor $T_{\mu\nu}$, which is traceless classically.
It has the  form
\beq
T^{\mu}_{\mu} = \beta_g F^{\mu\nu}F_{\mu\nu} +\cdots
\label{Tmm}
\eeq

\noindent (3)  Massless, $N=1$ supersymmetric gauge theories are
classically invariant under the  supersymmetric extension of the
conformal group -- the  superconformal group.  Examining the
superconformal algebra, it can be seen that the  subset of
superconformal transformations consisting of translations,
SUSY transformations, and  axial $R$ transformations is closed
under SUSY, i.e. these transformations form a representation
of SUSY.  It follows that the  conserved currents
corresponding to these transformations make up a supermultiplet
represented by an axial vector superfield called the  supercurrent~$J$,
\beq
J \equiv \left\{ J'^{\mu}_R, ~Q^{\mu}_{\alpha}, ~T^{\mu}_{\nu} , ... \right\},
\label{J}
\eeq
where $J'^{\mu}_R$ is the  current associated to R invariance,
$Q^{\mu}_{\alpha}$ is the  one associated to SUSY invariance,
and $T^{\mu}_{\nu}$ the  one associated to translational invariance
(energy-momentum tensor).

The anomalies of the  R current $J'^{\mu}_R$, the  trace
anomalies of the
SUSY current, and  the  energy-momentum tensor, form also
a second supermultiplet, called the  supertrace anomaly
\[
S =\left\{ Re~ S, ~Im~ S,~S_{\alpha}\right\}=
\left\{T^{\mu}_{\mu},~\partial _{\mu} J'^{\mu}_R,~\sigma^{\mu}_{\alpha
  \dot{\beta}} \bar{Q}^{\dot\beta}_{\mu}~+~\cdots \right\}
\]
where $T^{\mu}_{\mu}$ is given in Eq.(\ref{Tmm}) and
\begin{align}
\partial _{\mu} J'^{\mu}_R &~=~\beta_g\epsilon^{\mu\nu\sigma\rho}
F_{\mu\nu}F_{\sigma \rho}+\cdots\\
\sigma^{\mu}_{\alpha \dot{\beta}} \bar{Q}^{\dot\beta}_{\mu}&~=~\beta_g
\lambda^{\beta}\sigma ^{\mu\nu}_{\alpha\beta}F_{\mu\nu}+\cdots
\end{align}

\noindent (4) It is very important to note that
the Noether current defined in (\ref{noethcurr}) is not the  same as the
current associated to R invariance that appears in the
supercurrent
$J$ in (\ref{J}), but they coincide in the  tree approximation.
So starting from a unique classical Noether current
$J^{\mu}_{R(class)}$,  the  Noether
current $J^{\mu}_R$ is defined as the  quantum extension of
$J^{\mu}_{R(class)}$ which allows for the
validity of the  non-renormalization theorem.  On the  other hand
$J'^{\mu}_R$, is defined to belong to the  supercurrent $J$,
together with the  energy-momentum tensor.  The  two requirements
cannot be fulfilled by a single current operator at the  same time.

Although the  Noether current $J^{\mu}_R$ which obeys (\ref{anomaly})
and the  current $J'^{\mu}_R$ belonging to the  supercurrent multiplet
$J$ are not the  same, there is a relation
\cite{Lucchesi:1987he,Lucchesi:1987ef} between quantities associated
with them
\beq
r=\beta_g(1+x_g)+\beta_{ijk}x^{ijk}-\gamma_Ar^A
\label{rbeta}
\eeq
where $r$ was given in Eq.~(\ref{r}).  The  $r^A$ are the
non-renormalized coefficients of
the anomalies of the  Noether currents associated to the  chiral
invariances of the  superpotential, and  --like $r$-- are strictly
one-loop quantities. The  $\gamma_A$'s are linear
combinations of the  anomalous dimensions of the  matter fields, and
$x_g$, and  $x^{ijk}$ are radiative correction quantities.
The structure of Eq. (\ref{rbeta}) is independent of the
renormalization scheme.

One-loop finiteness, i.e. vanishing of the  $\beta$-functions at one-loop,
implies that the  Yukawa couplings $\lambda_{ijk}$ must be functions of
the gauge coupling $g$. To find a similar condition to all orders it
is necessary and  sufficient for the  Yukawa couplings to be a formal
power series in $g$, which is solution of the  REs (\ref{redeq2}).

We can now state the  theorem for all-order vanishing
$\beta$-functions~\cite{Lucchesi:1987he}.
\bigskip

{\bf Theorem:}

Consider an $N=1$ supersymmetric Yang-Mills theory, with simple gauge
group. If the  following conditions are satisfied
\begin{enumerate}
\item There is no gauge anomaly.
\item The  gauge $\beta$-function vanishes at one-loop
  \beq
  \beta^{(1)}_g = 0 =\sum_i T(R_{i})-3\,C_{2}(G).
  \eeq
\item There exist solutions of the  form
  \beq
  C_{ijk}=\rho_{ijk}g,~\qquad \rho_{ijk}\in\complex
  \label{soltheo}
  \eeq
to the   conditions of vanishing one-loop matter fields anomalous
dimensions
\beq
  \gamma^{(1)}{}_{j}^{i}~=~0
  =\frac{1}{32\pi^2}~[ ~
  C^{ikl}\,C_{jkl}-2~g^2~C_{2}(R)\delta_j^i ].
\eeq
\item These solutions are isolated and  non-degenerate when considered
  as solutions of vanishing one-loop Yukawa $\beta$-functions:
   \beq
   \beta_{ijk}=0.
   \eeq
\end{enumerate}
Then, each of the  solutions (\ref{soltheo}) can be uniquely extended
to a formal power series in $g$, and  the  associated super Yang-Mills
models depend on the  single coupling constant $g$ with a $\beta$-function
which vanishes at all-orders.

\bigskip

It is important to note a few things:
The requirement of isolated and  non-degenerate
solutions guarantees the
existence of a unique formal power series solution to the  reduction
equations.
The vanishing of the  gauge $\beta$~function at one-loop,
$\beta_g^{(1)}$, is equivalent to the
vanishing of the  R current anomaly (\ref{anomaly}).  The  vanishing of
the anomalous
dimensions at one-loop implies the  vanishing of the  Yukawa couplings
$\beta$~functions at that order.  It also implies the  vanishing of the
chiral anomaly coefficients $r^A$.  This last property is a necessary
condition for having $\beta$ functions vanishing at all orders.\footnote{There is an alternative way to find finite theories
\cite{Ermushev:1986cu,Kazakov:1987vg,Jones:1986vp,Leigh:1995ep}.}

\bigskip

{\bf Proof:}

Insert $\beta_{ijk}$ as given by the  REs into the relationship (\ref{rbeta}).
Since these chiral anomalies vanish, we get
for $\beta_g$ an homogeneous equation of the  form
\beq
0=\beta_g(1+O(\hbar)).
\label{prooftheo}
\eeq
The solution of this equation in the  sense of a formal power series in
$\hbar$ is $\beta_g=0$, order by order.  Therefore, due to the
REs (\ref{redeq2}), $\beta_{ijk}=0$ too.

Thus we see that finiteness and  reduction of couplings are intimately
related. Since an equation like eq.~(\ref{rbeta}) is lacking in
non-supersymmetric theories, one cannot extend the  validity of a
similar theorem in such theories.

A very interesting development was done in ref~\cite{Kazakov:1997nf}.
Based on the all-loop relations among the beta-functions of the soft supersymmetry breaking terms
and those of the rigid supersymmetric theory with the help of the differential operators,
discussed in Sect. \ref{sec:RCN=1}, it was shown that certain RGI surfaces can be chosen, so as to
reach all-loop finiteness of the full theory. More specifically it was shown that on certain RGI
surfaces the partial differential operators appearing in Eqs.~(\ref{betaM},\ref{betah}) acting on the
beta- and gamma-functions of the rigid theory can be transformed to total derivatives.
Then the all-loop finiteness of the beta and gamma functions of the rigid theory can be
transferred to the beta functions of the soft supersymmetry breaking terms. Therefore a
totally all-loop finite $N=1$ SUSY gauge theory can be constructed,
including the soft supersymmetry breaking terms. 

%% file: PR2018_Chapter_5.tex
\chapter{Reduction of Couplings in Phenomenologically Viable Models}
\label{ch:viable}

In this chapter we apply the idea of reduction of couplings
to phenomenologically viable supersymmetric
models. These models make clear predictions for the top and bottom quark
masses. Confronting the models with the experimental values allows to
restrict the parameter space and to single out the viable models. The
full set of experimental constraints for a subset of these models will
then be discussed in Chapter~\ref{chap:FUT_LOW}.


\section{Finite Unified Models}
\label{sec:fut}
From the  classification of theories with vanishing one-loop gauge
$\beta$-function~\cite{Hamidi:1984ft},
one can easily see that there
exist only two candidate possibilities to construct $SU(5)$ GUTs with
three generations. These possibilities require that the  theory should
contain as matter fields the  chiral supermultiplets ${\bf
  5},~\overline{\bf 5},~{\bf 10}, ~\overline{\bf 10},~{\bf 24}$ with
the multiplicities $(6,9,4,1,0)$ or $(4,7,3,0,1)$, respectively. Only
the second one contains a ${\bf 24}$-plet which can be used to provide
the spontaneous symmetry breaking (SB) of $SU(5)$ down to $SU(3)\times
SU(2)\times U(1)$. For the  first model one has to incorporate another
way, such as the  Wilson flux breaking mechanism in higher dimensional
theories, to achieve the  desired
SB of $SU(5)$ \cite{Kapetanakis:1992vx,Mondragon:1993tw}. Therefore,
for a self-consistent field theory discussion we would like to
concentrate only on the  second possibility.

The particle content of the  models we will study consists of the
following supermultiplets: three ($\overline{\bf 5} + {\bf 10}$),
needed for each of the  three generations of quarks and  leptons, four
($\overline{\bf 5} + {\bf 5}$) and  one ${\bf 24}$ considered as Higgs
supermultiplets.
When the  gauge group of the  finite GUT is broken the  theory is no
longer finite, and  we will assume that we are left with the  MSSM.

Therefore, a predictive Gauge-Yukawa unified $SU(5)$
model which is finite to all orders, in addition to the  requirements
mentioned already, should also have the  following properties:

\begin{enumerate}

\item
The one-loop anomalous dimensions are diagonal,
i.e.,  $\gamma^{(1)}{}_{j}^{i} \propto \delta^{j}_{i} $.
\item The  three fermion generations, in the  irreducible representations
  $\overline{\bf 5}_{i},{\bf 10}_i$ $(i=1,2,3)$,  should
  not couple to the  adjoint ${\bf 24}$.
\item The  two Higgs doublets of the  MSSM should mostly be made out of a
pair of Higgs quintet and  anti-quintet, which couple to the  third
generation.
\end{enumerate}

In the  following we discuss two versions of the  all-order finite
model:  the  model of \citere{Kapetanakis:1992vx,Mondragon:1993tw},
which will be labeled ${\bf A}$, and  a slight variation of this model
(labeled ${\bf B}$), which can also be obtained from the  class of the
models suggested in \citere{Avdeev:1997vx,Kazakov:1998uj} with a
modification to suppress non-diagonal anomalous dimensions
\cite{Kobayashi:1997qx}.

The superpotential which describes the  two models, before the  reduction
of couplings takes place, is of the  form
\cite{Kapetanakis:1992vx,Mondragon:1993tw,Kobayashi:1997qx,Jones:1984qd,Leon:1985jm}
\bea
W &=&
\sum_{i=1}^{3}\,[~\frac{1}{2}g_{i}^{u} \,{\bf 10}_i{\bf 10}_i H_{i}+
g_{i}^{d}\,{\bf 10}_i \overline{\bf 5}_{i}\, \overline{H}_{i}~] +
g_{23}^{u}\,{\bf 10}_2{\bf 10}_3 H_{4} \label{zoup-super1}\\
& &+g_{23}^{d}\,{\bf 10}_2 \overline{\bf 5}_{3}\, \overline{H}_{4}+
g_{32}^{d}\,{\bf 10}_3 \overline{\bf 5}_{2}\, \overline{H}_{4}+
\sum_{a=1}^{4}g_{a}^{f}\,H_{a}\, {\bf 24}\,\overline{H}_{a}+
\frac{g^{\lambda}}{3}\,({\bf 24})^3~,\nonumber
\eea
where
$H_{a}$ and  $\overline{H}_{a}~~(a=1,\dots,4)$
stand for the  Higgs quintets and  anti-quintets.

The main difference between model ${\bf A}$ and  model
${\bf B}$ is that two pairs of Higgs quintets and  anti-quintets couple
to the  ${\bf 24}$ in ${\bf B}$, so that it is not necessary to mix
them with $H_{4}$ and  $\overline{H}_{4}$ in order to achieve the
triplet-doublet splitting after the  symmetry breaking of $SU(5)$
\cite{Kobayashi:1997qx}.  Thus, although the  particle content is the
same, the  solutions to Eqs.~(\ref{betayP},\ref{gamayP}) and  the  sum rules
are different, which will be reflected in the  phenomenology, discussed
in \refse{sec:num-fut}.


\subsection{FUT A}
\label{sec:futA}

\begin{table}
\begin{center}
\renewcommand{\arraystretch}{1.3}
\begin{tabular}{|l|l|l|l|l|l|l|l|l|l|l|l|l|l|l|l|}
\hline
& $\overline{{\bf 5}}_{1} $ & $\overline{{\bf 5}}_{2} $& $\overline{{\bf
    5}}_{3}$ & ${\bf 10}_{1} $ &  ${\bf 10}_{2}$ & ${\bf
  10}_{3} $ & $H_{1} $ & $H_{2} $ & $H_{3} $ &$H_{4 }$&  $\overline H_{1} $ &
$\overline H_{2} $ & $\overline H_{3} $ &$\overline H_{4 }$& ${\bf 24} $\\ \hline
$Z_7$ & 4 & 1 & 2 & 1 & 2 & 4 & 5 & 3 & 6 & -5 & -3 & -6 &0& 0 & 0 \\\hline
$Z_3$ & 0 & 0 & 0 & 1 & 2 & 0 & 1 & 2 & 0 & -1 & -2 & 0 & 0 & 0&0  \\\hline
$Z_2$ & 1 & 1 & 1 & 1 & 1 & 1 & 0 & 0 & 0 & 0 & 0 & 0 &  0 & 0 &0 \\\hline
\end{tabular}
  \caption{Charges of the  $Z_7\times Z_3\times Z_2$ symmetry for Model
    {\bf FUT A}. }
\renewcommand{\arraystretch}{1.0}
\label{tableA}
\end{center}
\end{table}

This model was introduced and examined first in refs~\cite{Kapetanakis:1992vx,Mondragon:1993tw}.
After the  reduction of couplings
the symmetry of the  superpotential $W$ (Eq. (\ref{zoup-super1})), is enhanced.
For  model ${\bf A}$ one finds that
the superpotential has the
$Z_7\times Z_3\times Z_2$ discrete symmetry with the  charge assignment
shown in Tab.~\ref{tableA}, and  with the  following superpotential
\beq
W_A = \sum_{i=1}^{3}\,[~\frac{1}{2}g_{i}^{u}
\,{\bf 10}_i{\bf 10}_i H_{i}+
g_{i}^{d}\,{\bf 10}_i \overline{\bf 5}_{i}\,
\overline{H}_{i}~] +
g_{4}^{f}\,H_{4}\,
{\bf 24}\,\overline{H}_{4}+
\frac{g^{\lambda}}{3}\,({\bf 24})^3~.
\label{w-futa}
\eeq

The non-degenerate and  isolated solutions to $\gamma^{(1)}_{i}=0$ for
model {\bf FUT A} , which are the  boundary conditions for the  Yukawa
couplings at the  GUT scale, are:
\begin{eqnarray}
&& (g_{1}^{u})^2
=\frac{8}{5}~g^2~, ~(g_{1}^{d})^2
=\frac{6}{5}~g^2~,~
(g_{2}^{u})^2=(g_{3}^{u})^2=\frac{8}{5}~g^2~,\label{zoup-SOL5}\\
&& (g^{\lambda})^2 =\frac{15}{7}g^2~,~
 (g_{2}^{d})^2 = (g_{3}^{d})^2=\frac{6}{5}~g^2~,~(g_{4}^{f})^2= g^2\nonumber\\
&&(g_{23}^{u})^2 =
(g_{23}^{d})^2=(g_{32}^{d})^2=
 (g_{2}^{f})^2=
(g_{3}^{f})^2=(g_{1}^{f})^2=0~.~\nonumber
\end{eqnarray}
In the  dimensionful sector, the  sum rule gives us the  following
boundary conditions at the  GUT scale for this model
\cite{Kobayashi:1997qx}:
\begin{equation}
m^{2}_{H_u}+
2  m^{2}_{{\bf 10}} =
m^{2}_{H_d}+ m^{2}_{\overline{{\bf 5}}}+
m^{2}_{{\bf 10}}=M^2 ~~,
\end{equation}
and thus we are left with only three free parameters, namely
$m_{\overline{{\bf 5}}}\equiv m_{\overline{{\bf 5}}_3}$,
$m_{{\bf 10}}\equiv m_{{\bf 10}_3}$
and $M$.


\subsection{FUT B}
\label{sec:futB}
This model was introduced and was presented its first study in ref~\cite{Kobayashi:1997qx}.
Also in the  case of {\bf FUT B}  the  symmetry is enhanced after the  reduction
of couplings.  The  superpotential has now a
  $Z_4\times Z_4\times Z_4$ symmetry with charges shown in Tab.~\ref{tableB} and   with the
following superpotential
\begin{multline}
W_B = \sum_{i=1}^{3}\,[~\frac{1}{2}g_{i}^{u}
\,{\bf 10}_i{\bf 10}_i H_{i}+
g_{i}^{d}\,{\bf 10}_i \overline{\bf 5}_{i}\,
\overline{H}_{i}~] +
g_{23}^{u}\,{\bf 10}_2{\bf 10}_3 H_{4} \\
  +g_{23}^{d}\,{\bf 10}_2 \overline{\bf 5}_{3}\,
\overline{H}_{4}+
g_{32}^{d}\,{\bf 10}_3 \overline{\bf 5}_{2}\,
\overline{H}_{4}+
g_{2}^{f}\,H_{2}\,
{\bf 24}\,\overline{H}_{2}+ g_{3}^{f}\,H_{3}\,
{\bf 24}\,\overline{H}_{3}+
\frac{g^{\lambda}}{3}\,({\bf 24})^3~,
\label{w-futb}
\end{multline}
For this model the  non-degenerate and  isolated solutions to
$\gamma^{(1)}_{i}=0$ give us:
\begin{eqnarray}
&& (g_{1}^{u})^2
=\frac{8}{5}~ g^2~, ~(g_{1}^{d})^2
=\frac{6}{5}~g^2~,~
(g_{2}^{u})^2=(g_{3}^{u})^2=(g_{23}^{u})^2 =\frac{4}{5}~g^2~,\label{zoup-SOL52}
\nonumber\\
&& (g_{2}^{d})^2 = (g_{3}^{d})^2=
(g_{23}^{d})^2=(g_{32}^{d})^2=\frac{3}{5}~g^2~,
\\
&& (g^{\lambda})^2 =\frac{15}{7}g^2~,~ (g_{2}^{f})^2
=(g_{3}^{f})^2=\frac{1}{2}~g^2~,~ (g_{1}^{f})^2=
(g_{4}^{f})^2=0~,\nonumber
\end{eqnarray}
and from the  sum rule we obtain \cite{Kobayashi:1997qx}:
\begin{equation}
m^{2}_{H_u}+
2  m^{2}_{{\bf 10}} =M^2~,\quad
m^{2}_{H_d}-2m^{2}_{{\bf 10}}=-\frac{M^2}{3}~,\quad
m^{2}_{\overline{{\bf 5}}}+
3m^{2}_{{\bf 10}}=\frac{4M^2}{3}~,
\end{equation}
i.e., in this case we have only two free parameters
$m_{{\bf 10}}\equiv m_{{\bf 10}_3}$  and  $M$ for the  dimensionful sector.

\begin{table}
\begin{center}
\renewcommand{\arraystretch}{1.3}
\begin{tabular}{|l|l|l|l|l|l|l|l|l|l|l|l|l|l|l|l|}
\hline
& $\overline{{\bf 5}}_{1} $ & $\overline{{\bf 5}}_{2} $& $\overline{{\bf
    5}}_{3}$ & ${\bf 10}_{1} $ &  ${\bf 10}_{2}$ &  ${\bf
  10}_{3} $ & $ H_{1} $ & $H_{2} $ & $ H_{3}
$ &$H_{4}$&   $\overline H_{1} $ &
$\overline H_{2} $ & $\overline H_{3} $ &$\overline H_{4 }$&${\bf 24} $\\\hline
$Z_4$ & 1 & 0 & 0 & 1 & 0 & 0 & 2 & 0 & 0 & 0 & -2 & 0 & 0 & 0 &0  \\\hline
$Z_4$ & 0 & 1 & 0 & 0 & 1 & 0 & 0 & 2 & 0 & 3 & 0 & -2 & 0 & -3& 0  \\\hline
$Z_4$ & 0 & 0 & 1 & 0 & 0 & 1 & 0 & 0 & 2 & 3 & 0 & 0 & -2& -3 & 0 \\\hline
\end{tabular}
  \caption{Charges of the  $Z_4\times Z_4\times Z_4$ symmetry for Model
    {\bf FUT B}.}
\label{tableB}
\renewcommand{\arraystretch}{1.0}
\end{center}

\end{table}

As already mentioned, after the  $SU(5)$ gauge symmetry breaking we
assume we have the  MSSM, i.e. only two Higgs doublets.  This can be
achieved by introducing appropriate mass terms that allow to perform a
rotation of the  Higgs sector \cite{Leon:1985jm,Kapetanakis:1992vx,Mondragon:1993tw,Hamidi:1984gd,Jones:1984qd},
in such a way that only one pair of Higgs doublets,
coupled mostly to the  third family, remains light and  acquire vacuum
expectation values.  To avoid fast proton decay the  usual fine tuning
to achieve doublet-triplet splitting is performed.  Notice that,
although similar, the  mechanism is not identical to minimal $SU(5)$,
since we have an extended Higgs sector.

Thus, after the  gauge symmetry of the  GUT theory is broken we are left
with the  MSSM, with the  boundary conditions for the  third family given
by the  finiteness conditions, while the  other two families are basically
decoupled.


\subsection{Predictions for Quark Masses}
\label{sec:mq}
We will now examine the prediction of such all-loop Finite Unified
theories with $SU(5)$ gauge group for the third generation quark masses
(for the reasons expressed above). An extension to three
families, and  the  generation of quark mixing angles and  masses in
Finite Unified Theories has been addressed in~\cite{Babu:2002in},
where several examples are given. These extensions are not discussed
here.

Since the  gauge symmetry is spontaneously broken below $M_{\rm GUT}$,
the finiteness conditions do not restrict the  renormalization
properties at low energies, and  all it remains are boundary conditions
on the  gauge and  Yukawa couplings (\ref{zoup-SOL5}) or (\ref{zoup-SOL52}),
the relation
\[
h^{ijk}=-MC^{ijk}+...=-M\rho_{(0)}^{ijk}g+O(g^5)~,
\]
which follow from Eq.(\ref{h2NEW}) and the power series solution Eq.(\ref{Yg})
and the soft scalar-mass sum rule \cite{Kobayashi:1999pn,Kobayashi:1997qx,Mondragon:2003bp}
\beq
(~m_{i}^{2}+m_{j}^{2}+m_{k}^{2}~)/
M M^{\dag} =
1+\frac{g^2}{16 \pi^2}\,\Delta^{(2)}+O(g^4)~,
\label{sumr}
\eeq
where the $g^2$ term is given by
\[
\Delta^{(2)}=-2\sum_l\left[\frac{m^2_l}{MM^\dagger}-\frac 13\right]T(R_l)~,
\]
all taken at $M_{\rm GUT}$, as applied in
the two models.  Thus we examine the  evolution of these parameters
according to their RGEs up to two-loops for dimensionless parameters
and at one-loop for dimensionful ones with the  relevant boundary
conditions.  Below $M_{\rm GUT}$ their evolution is assumed to be
governed by the  MSSM.  We further assume a unique SUSY
breaking scale $M_{\rm SUSY}$ (which we define as the  geometrical average
of the  stop masses) and  therefore below that scale the  effective
theory is just the  SM.  This allows to evaluate observables at or
below the  electroweak scale.

In the  following, we review the derivation of the  prediction for the
third generation of quark masses that allows
for a direct comparison with experimental data and  to determine the
models that are in good agreement with the  observed quark mass
values~\cite{Heinemeyer:2007tz,Heinemeyer:2012yj,Heinemeyer:2013fga}.

\begin{figure}[htb!]
\begin{center}
\includegraphics[width=0.67\textwidth]{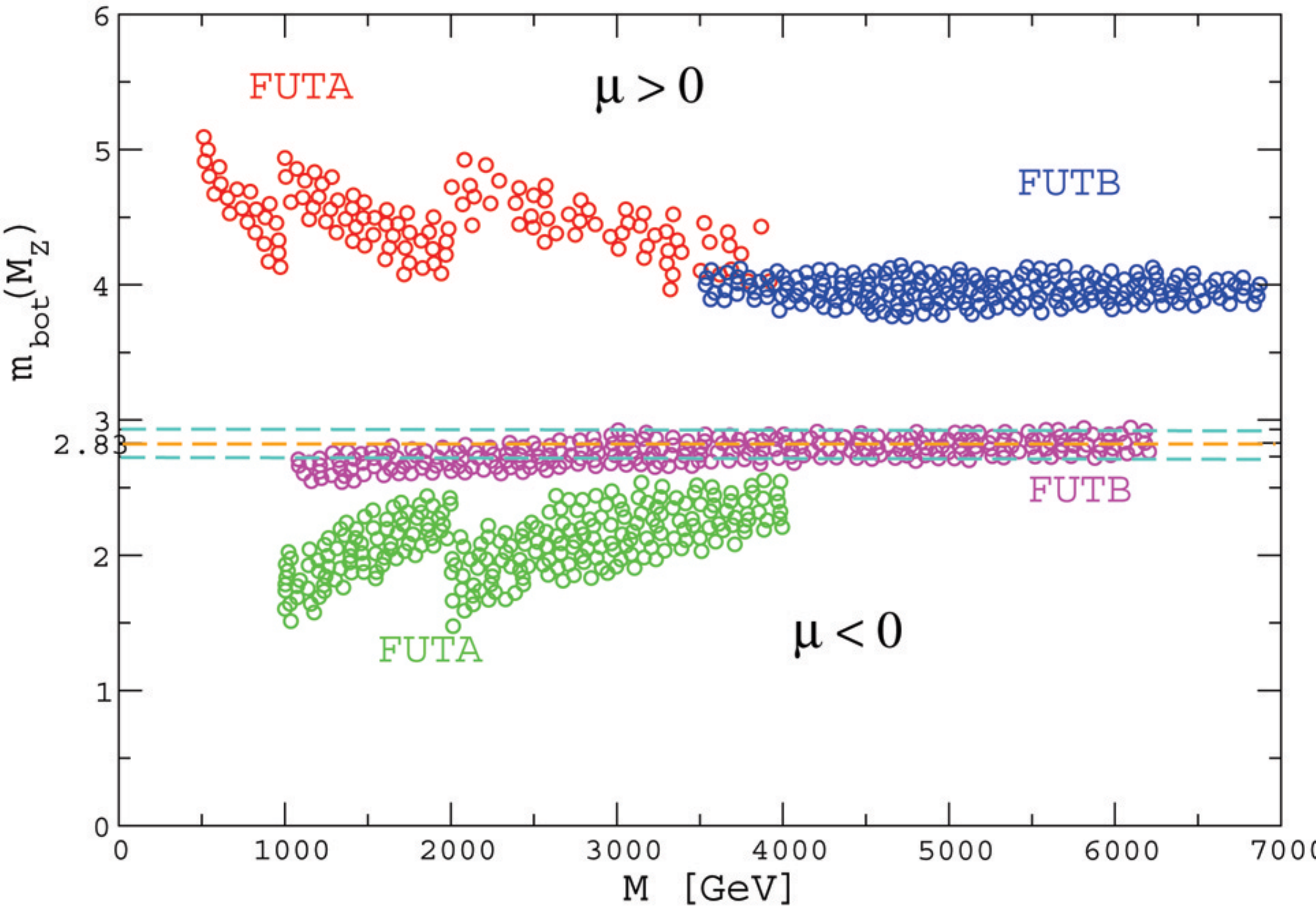}\\[2em]
\includegraphics[width=0.68\textwidth]{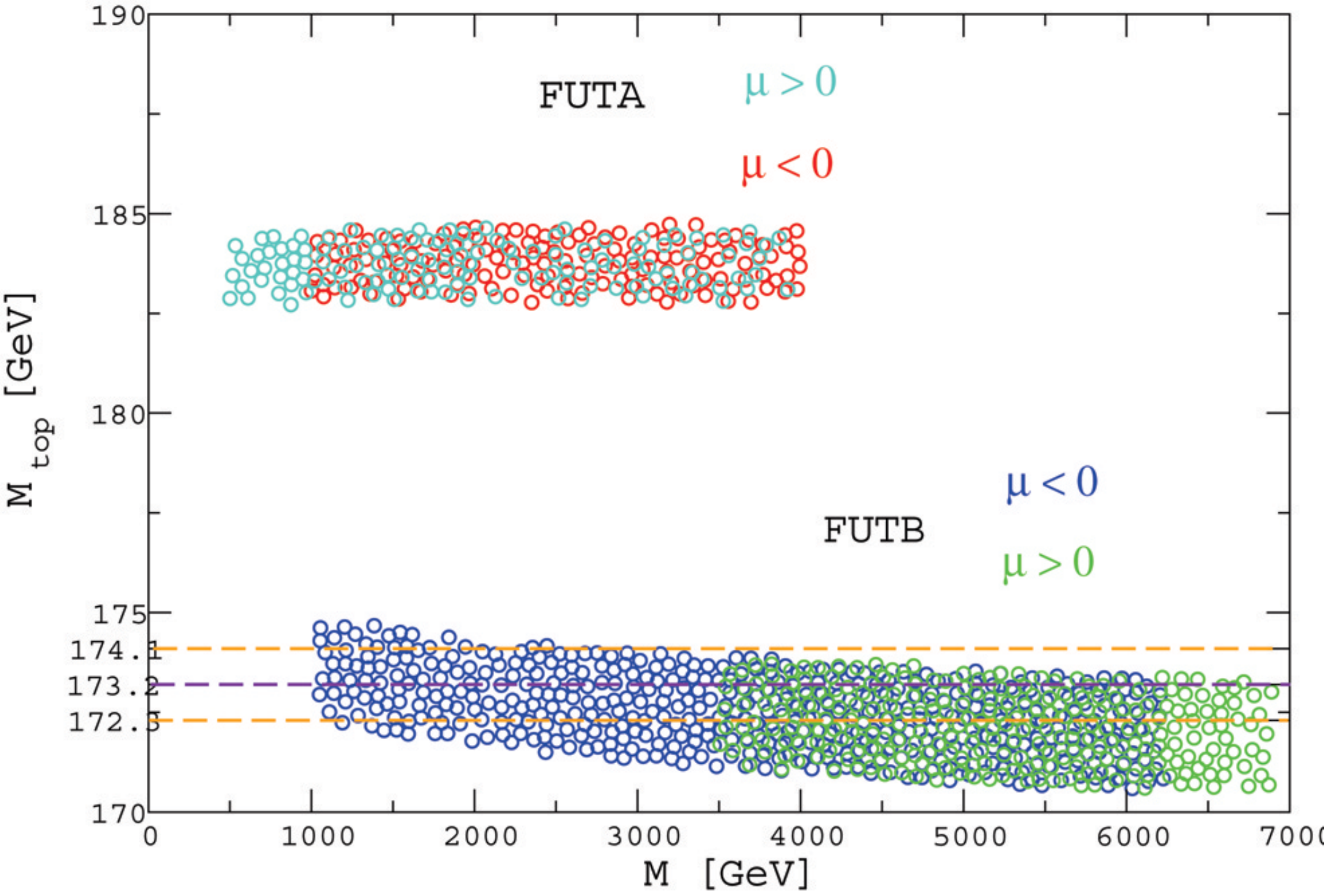}
\caption{The bottom quark mass at the  $Z$~boson scale (upper)
and top quark pole mass (lower plot) are shown
as function of $M$ for both models and  both signs of $\mu$.}
\label{fig:MtopbotvsM5}
\end{center}
\vspace{-1em}
\end{figure}

In \reffi{fig:MtopbotvsM5} we show the  {\bf FUT A} and  {\bf FUT B}
predictions for the  top pole mass, $M_{\rm top}$, and  the  running bottom
mass at the  scale $\MZ$, $m_{\rm bot}(M_Z)$,
as a function of the   unified gaugino mass $M$, for the  two cases
$\mu <0$ and  $\mu >0$.
The running bottom mass is used to avoid the  large
QCD uncertainties inherent to the  pole mass.
In the  evaluation of the  bottom mass $m_{\rm bot}$,
we have included the  corrections coming from bottom
squark-gluino loops and  top squark-chargino loops~\cite{Carena:1999py}.
The prediction is compared to the experimental values~\cite{pdg}%
\footnote{
These values correspond to the experimental measurements at the time of
the original evaluation. However, the small change to the current values
would not change the phenomenological analysis in a relevant way.}
\beq
\mb(M_Z) = 2.83 \pm 0.10 \gev ~.
\label{mbexp}
\eeq
and
\beq
\mt^{\rm exp} = (173.2 \pm 0.9) \gev~.
\label{mtexp}
\eeq
One can see that the  value of $\mb$ depends strongly on the  sign of
$\mu$ due to the  above mentioned radiative corrections involving SUSY
particles.  For both models ${\bf A}$ and  ${\bf B}$ the  values for
$\mu >0$ are above the  central experimental value, with
$\mb(M_Z) \sim 4.0 - 5.0$~GeV.  For $\mu < 0$, on the  other hand,
model~${\bf B}$ shows overlap with the  experimentally measured values,
$\mb(M_Z) \sim 2.5-2.8$~GeV. For model~${\bf A}$ we find
$\mb (M_Z) \sim 1.5 - 2.6$~GeV, and there is only a small region of
allowed parameter space at large $M$ where we find agreement with the
experimental value at the two~$\sigma $ level.
Therefore, the  experimental determination of
$\mb(\MZ)$ clearly selects the negative sign of $\mu$.

Now we turn to the  top quark mass. The  predictions for the  top quark
mass $M_t$ are $\sim 183$~GeV and  $\sim 172$~GeV in the  models
${\bf A}$ and  ${\bf B}$ respectively, as shown in the  lower plot of
Fig.~\ref{fig:MtopbotvsM5}. (Here it should be kept in mind that
theoretical values for $M_t$ may suffer from a correction of
$\sim 4 \%$ \cite{Kubo:1997fi,Kobayashi:2001me,Mondragon:2003bp}).
One can see clearly that model~${\bf B}$ is singled out. In addition the
value of $\tan \beta$ is found to be $\tan \beta \sim 54$ and
$\sim 48$ for models~${\bf A}$ and~${\bf B}$, respectively. Thus from the
comparison of the  predictions of the  two models with experimental data
only {\bf FUT B} with $\mu < 0$ survives.


\section{Reduction of Couplings in the Minimal Supersymmetric $SU(5)$ GUT}
\label{sec:su5}

In this section we consider the partial reduction of couplings in the minimal $N = 1$ supersymmetric
gauge model based on the group SU(5) according to refs~\cite{Kubo:1994bj,Kubo:1996js}
The three
generations of quarks and leptons   are accommodated by
three chiral superfields in
$\Psi^{I}({\bf 10})$ and $\Phi^{I}(\overline{\bf 5})$,
where $I$ runs over the three generations.
A $\Sigma({\bf 24})$ is used to break $SU(5)$ down to $SU(3)_{\rm C}
\times SU(2)_{\rm L} \times U(1)_{\rm Y}$,  and
$H({\bf 5})$ and $\overline{H}({\overline{\bf 5}})$
 to describe the
two Higgs superfields appropriate for
electroweak symmetry breaking  \cite{dimop,sakai}.
Note that the Finite Unified Models discussed in Sect.~\ref{sec:fut} contain
four $({\bf 5} + {\bf \bar{5}})$ to describe the Higgs superfields appropriate for electroweak symmetry breaking
instead of one set of $({\bf 5} + {\bf \bar{5}})$ used here in the minimal $N = 1$ supersymmetric $SU(5)$ version.
This minimality makes the present version asymptotically free (negative $\beta_g$) instead of finite at one loop, which was
the case of the models in Sect.~\ref{sec:fut}.
The superpotential of the model is  \cite{dimop,sakai}
\be
\begin{split}
W &= \frac{g_{t}}{4}\,
\epsilon^{\alpha\beta\gamma\delta\tau}\,
\Psi^{(3)}_{\alpha\beta}\Psi^{(3)}_{\gamma\delta}H_{\tau}+
\sqrt{2}g_b\,\Phi^{(3) \alpha}
\Psi^{(3)}_{\alpha\beta}\overline{H}^{\beta}+
\frac{g_{\lambda}}{3}\,\Sigma_{\alpha}^{\beta}
\Sigma_{\beta}^{\gamma}\Sigma_{\gamma}^{\alpha}+
g_{f}\,\overline{H}^{\alpha}\Sigma_{\alpha}^{\beta} H_{\beta}\\
&+ \frac{\mu_{\Sigma}}{2}\,
\Sigma_{\alpha}^{\gamma}\Sigma_{\gamma}^{\alpha}+
+\mu_{H}\,\overline{H}^{\alpha} H_{\alpha}~,
\end{split}
\ee
where $\alpha,\beta,\ldots$ are the $SU(5)$
indices, and we have suppressed the Yukawa couplings of the
first two generations.
The Lagrangian containing the SSB terms
is
\be
\begin{split}
-{\cal L}_{\rm soft} &=
m_{H_u}^{2}{\hat H}^{* \alpha}{\hat H}_{\alpha}
+m_{H_d}^{2}
\hat{\overline {H}}^{*}_{\alpha}\hat{\overline {H}}^{\alpha}
+m_{\Sigma}^{2}{\hat \Sigma}^{\dag~\alpha}_{\beta}
{\hat \Sigma}_{\alpha}^{\beta}
+\sum_{I=1,2,3}\,[\,
m_{\Phi^I}^{2}{\hat \Phi}^{* ~(I)}_{\alpha}{\hat \Phi}^{(I)\alpha}\\
& +\,m_{\Psi^I}^{2}{\hat \Psi}^{\dag~(I)\alpha\beta}
{\hat \Psi}^{(I)}_{\beta\alpha}\,]
+\{ \,
 \frac{1}{2}M\lambda \lambda+
B_H\hat{\overline {H}}^{\alpha}{\hat H}_{\alpha}
+B_{\Sigma}{\hat \Sigma}^{\alpha}_{\beta}
{\hat \Sigma}_{\alpha}^{\beta}
+h_{f}\,\hat{\overline{H}}^{\alpha}
{\hat \Sigma}_{\alpha}^{\beta} {\hat H}_{\beta}\\
& +\frac{h_{\lambda}}{3}\,{\hat \Sigma}_{\alpha}^{\beta}
{\hat \Sigma}_{\beta}^{\gamma}{\hat \Sigma}_{\gamma}^{\alpha}+
\frac{h_{t}}{4}\,
\epsilon^{\alpha\beta\gamma\delta\tau}\,
{\hat \Psi}^{(3)}_{\alpha\beta}
{\hat \Psi}^{(3)}_{\gamma\delta}{\hat H}_{\tau}+
\sqrt{2}h_{b}\,{\hat \Phi}^{(3) \alpha}
{\hat \Psi^{(3)}}_{\alpha\beta}\hat{\overline{H}}^{\beta}
+\mbox{h.c.}\, \}~,
\end{split}
\ee
where a hat is used to denote the scalar
component of each chiral superfield.

The RG functions of this model may be found
in refs.~\cite{mondragon1,polonsky1,Kazakov:1995cy}, and
we employ the usual
normalization  of the RG functions,
$d {\rm A}/d \ln \mu ~=~
[\beta^{(1)}(A) ~~\mbox{or}~~   \gamma^{(1)}(A)]/16
\pi^2+\dots$,
where $\dots$ are higher orders, and $\mu$ is the renormalization scale:
\begin{align}
\beta^{(1)}(g) &= -3 g^3~,~
\beta^{(1)}(g_t) =
[\,-\frac{96}{5}\,g^2+9\,g_{t}^{2}+\frac{24}{5}\,g_{f}^{2}+
4\,g_{b}^{2}\,]\,g_{t}~, \nn \\
\beta^{(1)}(g_b) &=
[\,-\frac{84}{5}\,g^2+3\,g_{t}^{2}+\frac{24}{5}\,g_{f}^{2}+
10\,g_{b}^{2}\,]\,g_{b}~, \nn \\
\beta^{(1)}(g_\lambda) &=
[\,-30\,g^2+\frac{63}{5}\,g_{\lambda}^2+3\,g_{f}^{2}
\,]\,g_{\lambda}~, \nn \\
\beta^{(1)}(g_f) &=
[\,-\frac{98}{5}\,g^2+3\,g_{t}^{2}
+4\,g_{b}^{2}
+\frac{53}{5}\,g_{f}^{2}+\frac{21}{5}\,g_{\lambda}^{2}
\,]\,g_{f}~,~
\gamma^{(1)}(M) = -6g^2 \,M~, \nn \\
\gamma^{(1)}(\mu_{\Sigma}) &= [\, -20g^2 +2g_{f}^{2}
+\frac{42}{5} g_{\lambda}^{2}\,]\,\mu_{\Sigma} ~,~
\gamma^{(1)}(\mu_H) =
[\, -\frac{48}{5}g^2 +\frac{48}{5}g_{f}^{2}
+4 g_{b}^{2}+3g_{t}^{2}\, ]\,\mu_H ~, \nn \\
\gamma^{(1)}(B_H) &=
[\, -\frac{48}{5}g^2 +\frac{48}{5}g_{f}^{2}
+4 g_{b}^{2}+3g_{t}^{2} \,]\, B_H \nn \\
 &+
[\,\frac{96}{5}g^2 M+\frac{96}{5}h_{f}g_{f}
+8 g_b h_{b}+6 g_t h_{t}]\, \mu_H ~, \nn \\
\gamma^{(1)}(B_{\Sigma}) &=
[\, -20 g^2 +2g_{f}^{2}
+\frac{42}{5} g_{\lambda}^{2} \,]\, B_{\Sigma}+
[\,40 g^2 M+4 h_{f}g_{f}
+\frac{84}{5} g_{\lambda} h_{\lambda}]\, \mu_{\Sigma} ~, \nn \\
\gamma^{(1)}(h_t) &=
[\,-\frac{96}{5}\,g^2+9\,g_{t}^{2}+\frac{24}{5}\,g_{f}^{2}+
4\,g_{b}^{2}\,]\,h_t \nn \\
& +[\, \frac{192}{5} M g^2+
18 h_t g_t+8 h_b g_b +\frac{48}{5}h_f g_f\,] \, g_t~, \nn \\
\gamma^{(1)}(h_b) &=
[\,-\frac{84}{5}\,g^2+3\,g_{t}^{2}+\frac{24}{5}\,g_{f}^{2}+
10\,g_{b}^{2}\,]\, h_b \nn \\
& +[\, \frac{168}{5} M g^2+
6 h_t g_t+20 h_b g_b +\frac{48}{5}h_f g_f\,]\, g_b~~, \nn \\
\gamma^{(1)}(h_{\lambda}) &=
[\,-30\,g^2+\frac{63}{5}\,g_{\lambda}^2+3\,g_{f}^{2}
\,]\,h_{\lambda}
+[\, 60 M g^2+
\frac{126}{5}h_{\lambda} g_{\lambda}
 +6h_f g_f\,] \,g_{\lambda}~, \nn \\
\gamma^{(1)}(h_f) &=
[\,-\frac{98}{5}\,g^2+3\,g_{t}^{2}
+4\,g_{b}^{2}
+\frac{53}{5}\,g_{f}^{2}+\frac{21}{5}\,g_{\lambda}^{2}
\,]\, h_f \nn \\
& ~+[\, \frac{196}{5} M g^2+6 h_t g_t+8 h_b g_b+
\frac{42}{5}h_{\lambda} g_{\lambda}
 +\frac{106}{5}h_f g_f\,] \, g_{f} ~, \nn \\
\gamma^{(1)}(m_{H_d}^{2}) &=
 -\frac{96}{5} g^2 M^{2}+\frac{48}{5}g_{f}^{2}(m_{H_u}^{2}+
m_{H_d}^{2}+m_{\Sigma}^{2}) \nn
\\
& +
8 g_{b}^{2}(
m_{H_d}^{2}+m_{\Psi^3}^{2}+m_{\Phi^3}^{2})+
\frac{48}{5} h_{f}^{2}+8 h_{b}^{2}~, \nn \\
\gamma^{(1)}(m_{H_u}^{2}) &=
 -\frac{96}{5} g^2 M^{2}+\frac{48}{5}g_{f}^{2}(m_{H_u}^{2}+
m_{H_d}^{2}+m_{\Sigma^3}^{2})+
6 g_{t}^{2}(
m_{H_u}^{2}+2m_{\Psi^3}^{2})+
\frac{48}{5} h_{f}^{2}+6 h_{t}^{2}~, \nn \\
\gamma^{(1)}(m_{\Sigma}^{2}) &=
 -40 g^2 M^{2}+2g_{f}^{2}(m_{H_u}^{2}+
m_{H_d}^{2}+m_{\Sigma}^{2})
+
\frac{126}{5} g_{\lambda}^{2}m_{\Sigma}^{2}+
2 h_{f}^{2}+\frac{42}{5} h_{\lambda}^{2}~, \nn \\
\gamma^{(1)}(m_{\Phi^3}^{2}) &=
 -\frac{96}{5} g^2 M^{2}+
8 g_{b}^{2}(
m_{H_d}^{2}+m_{\Psi^3}^{2}+m_{\Phi^3}^{2})+
8 h_{b}^{2} ~, \nn \\
\gamma^{(1)}(m_{\Psi^3}^{2}) &=
 -\frac{144}{5} g^2 M^{2}+6g_{t}^{2}(
m_{H_u}^{2}+2m_{\Psi^3}^{2})+
4 g_{b}^{2}(
m_{H_d}^{2}+m_{\Psi^3}^{2}+m_{\Phi^3}^{2})+
6h_{t}^{2}+4  h_{b}^{2}~, \nn \\
\gamma^{(1)}(m_{\Phi^{1,2}}^{2}) &=
-\frac{96}{5} g^2 M^{2} ~,~
\gamma^{(1)}(m_{\Psi^{1,2}}^{2}) =
-\frac{144}{5} g^2 M^{2} ~,
\end{align}
where $g$ stands for the gauge coupling.

\vspace{0.3cm}

\noindent
The reduction solution is found as follows.
We require that the reduced theory
should contain the minimal number of the SSB parameters that are
consistent with   perturbative renormalizability. We will find
that the set of the  perturbatively unified SSB parameters
significantly differ from the so-called universal SSB parameters.

Without loss of generality,
one can assume that the gauge coupling $g$ is the primary coupling.
Note that the reduction solutions in the dimension-zero sector
is independent
of the dimensionful sector (under the assumption of a
mass independent renormalization scheme).
It has been found \cite{mondragon1} that there exist
two asymptotically free (AF) solutions that make
a Gauge-Yukawa Unification possible in the present model:
\be
\label{two_sol}
\begin{split}
a & : g_t=\sqrt{\frac{2533}{2605}} g + \order{g^3}~,~
g_b=\sqrt{\frac{1491}{2605}} g + \order{g^3}~,~
g_{\lambda}=0~,~
g_f=\sqrt{\frac{560}{521}} g + \order{g^3}~,\\
b & : g_t=\sqrt{\frac{89}{65}} g + \order{g^3}~,~
g_b=\sqrt{\frac{63}{65}} g + \order{g^3}~,~
g_{\lambda}=0~,~g_f=0~,
\end{split}
\ee
where the higher order terms denote uniquely computable
power series in $g$.
It has been also found that the two solutions in (\ref{two_sol})
describe the boundaries of an
asymptotically free RG-invariant surface in the space
of the couplings, on which $g_{\lambda}$ and $g_f$
can be different from zero. This observation
has enabled us to obtain a partial reduction of couplings
for which the $g_{\lambda}$ and $g_f$ can be treated
as (non-vanishing) independent parameters without loosing
AF.
Later we have found \cite{Kubo:1995cg} that the region on the
AF surface consistent with the
proton decay constraint
has to be very close to the solution $a$.
Therefore, we assume in the following discussion
that we are exactly at the boundary defined by the solution $a$
\footnote{
Note that
$ g_{\lambda}=0 $ is inconsistent, but $g_{\lambda} < \sim 0.005$
has to be fulfilled to satisfy the proton
decay constraint \cite{Kubo:1995cg}.
We expect that the inclusion of a small $g_{\lambda} $ will
not affect the prediction of the perturbative unification of
the SSB parameters.}.

In the dimensionful sector, we seek the reduction
of the parameters in the form of Eqs. (\ref{reduction}).
First, one can realize that the
supersymmetric mass parameters, $\mu_{\Sigma}$ and
$\mu_H$, and the gaugino mass parameter
$M$ cannot be reduced; that is, there
is no solution in the desired form. Therefore,  they
should be treated as independent parameters.
We find the following lowest order
reduction solution:
\begin{equation}\label{red_sol_1}
B_H = \frac{1029}{521}\,\mu_H M~,~
B_{\Sigma}=-\frac{3100}{521}\,\mu_{\Sigma} M~,
\end{equation}
\be
\begin{split}
\label{red_sol}
h_t &=-g_t\,M~,~h_b =-g_b\,M~,
~h_f =-g_f\,M~,~h_{\lambda}=0~,\\
m_{H_u}^{2} &=-\frac{569}{521} M^{2}~,~
m_{H_d}^{2} =-\frac{460}{521} M^{2}~,
~m_{\Sigma}^{2} = \frac{1550}{521} M^{2}~,\\
m_{\Phi^3}^{2} & = \frac{436}{521} M^{2}~,~
m_{\Phi^{1,2}}^{2} =\frac{8}{5} M^{2}~,~
m_{\Psi^3}^{2} =\frac{545}{521} M^{2}~,~
m_{\Psi^{1,2}}^{2} =\frac{12}{5} M^{2}~.
\end{split}
\ee
So, the gaugino mass parameter $M$ plays
a similar role as the gravitino mass $m_{2/3}$ in
supergravity coupled GUTs and characterizes
the scale of the supersymmetry--breaking.

In addition to the $\mu_{\Sigma}$,
$\mu_H$ and $M$, it is possible to include  also
$B_H$ and $B_{\Sigma}$ as independent parameters
without changing the one-loop
reduction solution (\ref{red_sol}).

The prediction of the minimal supersymmetric $SU(5)$ GUT, following the
Gauge-Yukawa Unification of the solution a in Eqs (\ref{two_sol}) is:
\[
\begin{split}
M_{t} &\simeq 1.8 \times 10^2 ~\mbox{GeV}~,~
M_{b} \simeq 5.4 ~\mbox{GeV}~,~\alpha_3 (M_Z)\simeq 0.12~,\\
M_{\rm GUT} &\simeq 1.7\times 10^{16}~\mbox{GeV}~,
\alpha_{\rm GUT} \simeq 0.040~,
~\tan\beta (M_{\rm SUSY})\simeq 48~,
\end{split}
\]
where $M_{t}$ and $M_b$ are the physical top and bottom quark masses.
These values suffer from corrections coming from different sources
such as threshold effects, which are partly taken into account and
estimated in \citere{Kubo:1995cg}. In \citere{Kubo:1996js}, Tab.~1,
also the prediction of the specific model for several SSB parameters can be
found. Just for completeness we mention that the input parameters for the
above prediction were:
\[
\alpha_1(M_Z) =0.0169~,~
\alpha_2(M_Z) =0.0337~,~
\alpha_{\tau}(M_Z) =8.005\times 10^{-6}~
\]
while the SUSY scale was fixed at $500$ GeV.

The present model has very good chances to survive the recent experimental
constraints.
A more detailed examination is in order to determine its viability.


\section{Gauge-Yukawa Unification in other Supersymmetric GUTs by Reducing
the Couplings}

\subsection{Asymptotically Non-Free Pati-Salam Model}

In this section a model is discussed, where a partial reduction of couplings is achieved,
which however is not based on a single gauge group, but on a product of simple groups.
In order for the RGI method for the gauge coupling
unification to  work,
the gauge couplings should
have the same asymptotic behavior.
Recall that this common behavior is absent
in the standard model with three families.
A way to achieve a common asymptotic behavior of all the
different gauge couplings is to embed
$SU(3)_{C}\times SU(2)_{L}\times U(1)_{Y}$ to some
non-abelian gauge group, as it was done in the previous sections.
However, in this case still a major r\^ole in the
GYU is due to the group theoretical aspects of the covering GUT. Here
we would like  to examine the power of RGI method by considering
theories without covering GUTs.
We found \cite{Kubo:1994xa}
that the minimal phenomenologically viable model is based on the gauge
group of Pati and Salam \cite{pati1}-- ${\cal G}_{\rm PS}\equiv
SU(4)\times SU(2)_{R}\times SU(2)_{L}$.
We recall that $N=1$ supersymmetric  models based on this
gauge group have been studied with renewed interest because they could
in principle be derived from superstrings \cite{anton1,leont}.

In our supersymmetric, Gauge-Yukawa unified model
based on $ {\cal G}_{\rm PS}$ \cite{Kubo:1994xa}, three generations of
quarks and leptons  are accommodated by six chiral supermultiplets, three
in $({\bf 4},{\bf 2},{\bf 1})$ and three  $({\bf \overline{4}},
{\bf 1},{\bf 2})$, which we denote by $\Psi^{(I)\mu~ i_R}$ and
$\overline{\Psi}_{\mu}^{(I) i_L}$, ($I$ runs over the three generations,
and $\mu,\nu~(=1,2,3,4)$ are the $SU(4)$ indices while $i_R~,~i_L~(=1,2)$
stand for the $SU(2)_{L,R}$ indices.)
The Higgs supermultiplets
in $({\bf 4},{\bf 2},{\bf 1})$,
$({\bf \overline{4}},{\bf 2},{\bf 1})$
and  $({\bf 15},{\bf 1},{\bf 1})$ are denoted by
$ H^{\mu ~i_R}~,~
\overline{H}_{\mu ~i_R} $ and $\Sigma^{\mu}_{\nu}$, respectively. They
 are responsible for the spontaneous
symmetry breaking (SSB) of $SU(4)\times SU(2)_{R}$ down
to $SU(3)_{C}\times U(1)_{Y}$.
The SSB of $U(1)_{Y}\times
SU(2)_{L}$ is then achieved by the nonzero VEV of
$h_{i_R i_L}$ which is in $({\bf 1},{\bf 2},{\bf 2})$. In addition to
these Higgs supermultiplets, we introduce $G^{\mu}_{\nu~i_R i_L}~
({\bf 15},{\bf 2},{\bf 2})$, $\phi~({\bf 1},{\bf 1},{\bf 1})$ and
$\Sigma^{' \mu}_{\nu}~({\bf 15},{\bf 1},{\bf 1})$.
The $G^{\mu}_{\nu~i_R i_L}$ is introduced to realize
the $SU(4)\times SU(2)_{R}\times
SU(2)_{L}$ version of the Georgi-Jarlskog type
ansatz \cite{georgi4} for
the mass matrix of leptons and quarks while $\phi$
is supposed to mix with the right-handed neutrino
supermultiplets at a high energy scale.
With these things in mind, we write down
the superpotential of the model
$W$, which is the sum of the following superpotentials:
\[
\begin{split}
W_{Y} &=\sum_{I,J=1}^{3}g_{IJ}\,\overline{\Psi}^{(I) i_R}_{\mu}
\,\Psi^{(J)\mu~ i_L}~h_{i_R i_L}~,~\\
W_{GJ} &= g_{GJ}\,
\overline{\Psi}^{(2)i_R}_{\mu}\,
G^{\mu}_{\nu~i_R j_L}\,\Psi^{(2)\nu~ j_L}~,\\
W_{NM} &=
\sum_{I=1,2,3}\,g_{I\phi}~\epsilon_{i_R j_R}\,\overline{\Psi}^{(I)
i_R}_{\mu} ~H^{\mu ~j_R}\,\phi~,\\
W_{SB} &=
g_{H}\,\overline{H}_{\mu~ i_R}\,
\Sigma^{\mu}_{\nu}\,H^{\nu ~i_R}+\frac{g_{\Sigma}}{3}\,
\mbox{Tr}~[~\Sigma^3~]+
\frac{g_{\Sigma '}}{2}\,\mbox{Tr}~[~(\Sigma ')^2\,\Sigma~]~,\\
W_{TDS} &=
\frac{g_{G}}{2}\,\epsilon^{i_R j_R}\epsilon^{i_L j_L}\,\mbox{Tr}~
[~G_{i_R i_L}\,
\Sigma\,G_{j_R j_L}~]~,\\
W_{M}&= m_{h}\,h^2+m_{G}\,G^2+m_{\phi}\,
\phi^2+m_{H}\,\overline{H}\,H+
m_{\Sigma}\,\Sigma^2+
m_{\Sigma '}\,(\Sigma ')^2~.\label{12}
\end{split}
\]
Although $W$ has the parity, $\phi\to -\phi$
and $\Sigma ' \to -\Sigma '$,
it is not the most general potential, but,
as we already mentioned,
this is not a problem in N = 1 SUSY theories.

We denote the gauge couplings of $SU(4)\times SU(2)_{R}\times SU(2)_{L}$
by $\alpha_{4}~,~\alpha_{2R}$ and $\alpha_{2L}$,
respectively. The gauge coupling for $U(1)_{Y}$, $\alpha_1$, normalized
in the usual GUT inspired manner, is given by
$1/\alpha_{1} ~=~2/5\alpha_{4}+
3/5 \alpha_{2R}~$.
In principle, the primary coupling can be any one of the couplings.
But it is more convenient to choose a gauge coupling as the primary
one because the one-loop $\beta$ functions for a gauge coupling
depends only on its own gauge coupling. For the present model,
we use $\alpha_{2L}$ as the primary one.
Since the gauge sector for the one-loop $\beta$ functions is closed,
the solutions of the fixed point equations (\ref{fixpt}) 
are
independent on the Yukawa and Higgs couplings. One easily obtains
$
\rho_{4}^{(1)} =8/9~,~\rho_{2R}^{(1)}~=~4/5$,
so that the
RGI relations (\ref{algeq})
at the one-loop level become
\bea
\tilde{\alpha}_{4} &=&\frac{\alpha_4}{\alpha_{2L}}~=~
\frac{8}{9}~,~\tilde{\alpha}_{1} ~=~\frac{\alpha_1}{\alpha_{2L}}~=~
\frac{5}{6}~.
\label{13}
\eea

The solutions in the Yukawa-Higgs sector strongly
depend on the result of the gauge sector. After slightly involved
algebraic computations, one finds that
most predictive solutions contain at least
three vanishing $\rho_{i}^{(1)}$'s.
Out of these solutions, there are two that
 exhibit the most predictive
power and moreover they satisfy
the neutrino mass relation
$m_{\nu_{\tau}}>m_{\nu_{\mu}}~,~
m_{\nu_{e}}$.
For the first solution we have $\rho_{1\phi}^{(1)}=
\rho_{2\phi}^{(1)}=
\rho_{\Sigma}^{(1)}=0$, while for the second solution,
$  \rho_{1\phi}^{(1)}=
\rho_{2\phi}^{(1)}=
\rho_{G}^{(1)}=0 $,
 and one finds that for the cases above the power series solutions
(\ref{algeq}) take the form
\be
\begin{split}
\tilde{\alpha}_{GJ} &\simeq
\left\{
\begin{array}{l} 1.67 - 0.05 \tilde{\alpha}_{1\phi}
+
0.004 \tilde{\alpha}_{2\phi}
 - 0.90\tilde{\alpha}_{\Sigma}+\cdots \\
 2.20 - 0.08 \tilde{\alpha}_{2\phi}
 - 0.05\tilde{\alpha}_{G}+\cdots
\end{array} \right. ~~,\\
\tilde{\alpha}_{33} &\simeq \left\{
\begin{array}{l}  3.33 + 0.05 \tilde{\alpha}_{1\phi}
+
0.21 \tilde{\alpha}_{2\phi}-0.02 \tilde{\alpha}_{\Sigma}+ \cdots\\
3.40 + 0.05 \tilde{\alpha}_{1\phi}
-1.63 \tilde{\alpha}_{2\phi}- 0.001 \tilde{\alpha}_{G}+
\cdots \end{array} \right. ~~,\\
\tilde{\alpha}_{3\phi} &\simeq
\left\{
\begin{array}{l}  1.43 -0.58 \tilde{\alpha}_{1\phi}
-
1.43 \tilde{\alpha}_{2\phi}-0.03 \tilde{\alpha}_{\Sigma}+
\cdots\\
 0.88 -0.48 \tilde{\alpha}_{1\phi}
+8.83 \tilde{\alpha}_{2\phi}+ 0.01 \tilde{\alpha}_{G}+
\cdots\end{array} \right. ~~,\\
\tilde{\alpha}_{H} &\simeq \left\{
\begin{array}{l}
 1.08 -0.03 \tilde{\alpha}_{1\phi}
+0.10 \tilde{\alpha}_{2\phi}- 0.07 \tilde{\alpha}_{\Sigma}+
\cdots\\
2.51 -0.04 \tilde{\alpha}_{1\phi}
-1.68 \tilde{\alpha}_{2\phi}- 0.12 \tilde{\alpha}_{G}+
\cdots\end{array} \right. ~~,~~\\
\tilde{\alpha}_{\Sigma} &\simeq \left\{
\begin{array}{l}
---\\
0.40 +0.01 \tilde{\alpha}_{1\phi}
-0.45 \tilde{\alpha}_{2\phi}-0.10 \tilde{\alpha}_{G}+
\cdots \end{array} \right. ~,\\
\tilde{\alpha}_{\Sigma '} &\simeq \left\{
\begin{array}{ll}
4.91 - 0.001 \tilde{\alpha}_{1\phi}
-0.03 \tilde{\alpha}_{2\phi}- 0.46 \tilde{\alpha}_{\Sigma}+
\cdots\\
8.30 + 0.01 \tilde{\alpha}_{1\phi}
+1.72 \tilde{\alpha}_{2\phi}- 0.36 \tilde{\alpha}_{G}+
\cdots \end{array} \right. ~~,  \\
\tilde{\alpha}_{G} &\simeq \left\{
\begin{array}{ll}
5.59 + 0.02 \tilde{\alpha}_{1\phi}
-0.04 \tilde{\alpha}_{2\phi}- 1.33 \tilde{\alpha}_{\Sigma}+
\cdots
\\---
\end{array} \right.   ~ ~.
\label{14}
\end{split}
\ee
We have assumed that the Yukawa couplings $g_{IJ}$ except for
$g_{33}$ vanish. They can be included into RGI relations
as small
perturbations,
but their numerical effects
will be rather small.

\begin{figure}[htb!]
\begin{center}
\includegraphics[scale=0.4,angle=-90]{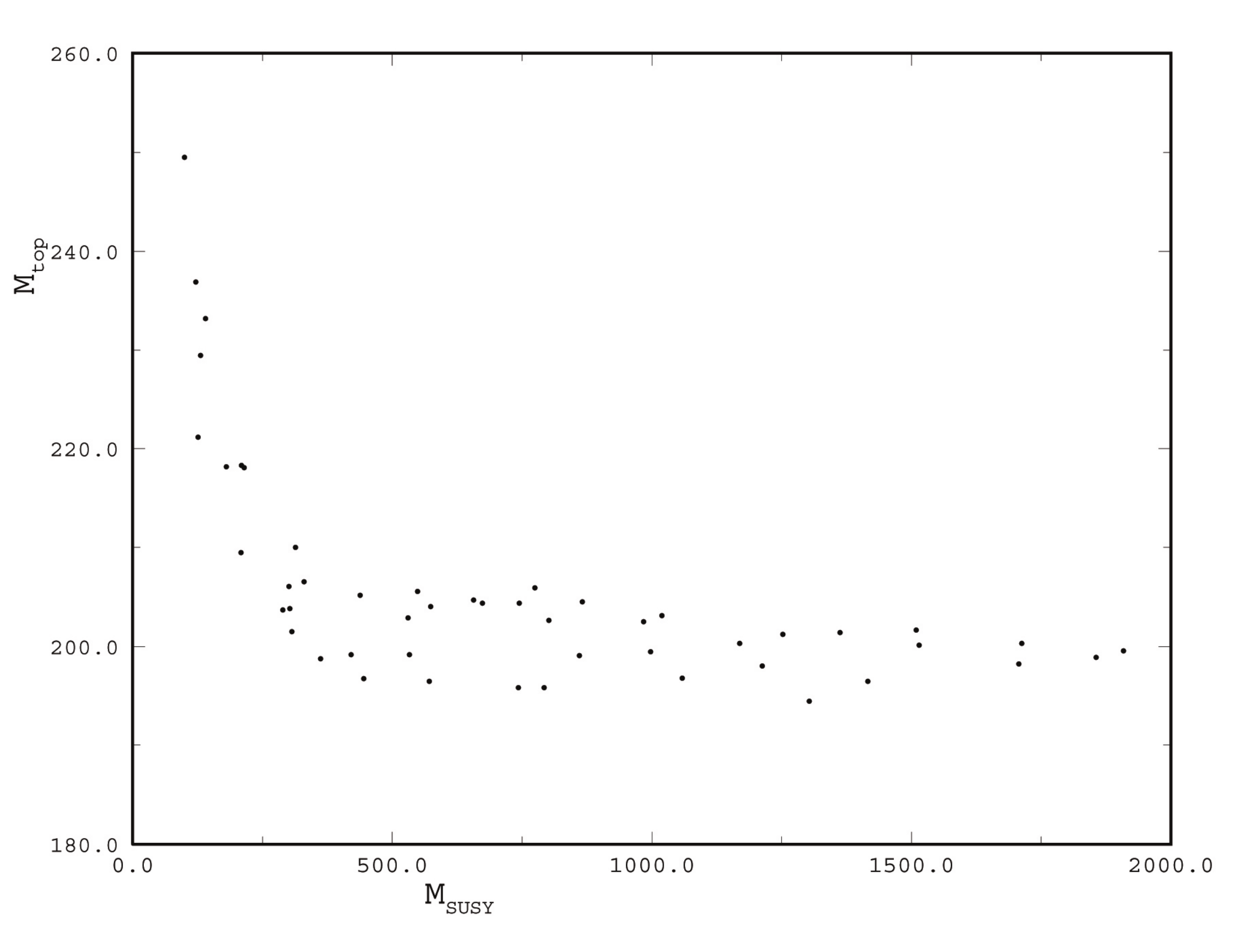}
\caption[\ ]{The values for $M_t$ predicted by the Pati-Salam model
    for different $M_{\rm SUSY}$ scales.
}
\label{fig:2}
\end{center}
\end{figure}

The number $ N_{H}$ of the Higgses lighter
than $M_{\rm SUSY}$ could vary from one to four while the number of
those to be taken into account above $M_{\rm SUSY}$ is fixed at four.
We have assumed here that $N_{H}=1$. The dependence of the top mass on
$M_{\rm SUSY}$ in this model is shown in Fig.~\ref{fig:2}.
One can see that for any reasonable supersymmetry breaking scale in the TeV
region the experimentally found top quark mass cannot be reproduced within this model.

\subsection{Asymptotically Non-Free $SO(10)$ Model}

In this section a model based on SO(10) is discussed, which also admits a partial reduction
of couplings \cite{Kubo:1995zg}.
We denote the hermitean
$SO(10)$-gamma matrices  by
$\Gamma_{\alpha}~,~\alpha=1,\cdots,10$.
The charge conjugation matrix $C$ satisfies
$C = C^{-1}~,~C^{-1}\,\Gamma_{\alpha}^{T}\,C ~=~
-\Gamma_{\alpha}$, and
the $\Gamma_{11}$ is defined as
$\Gamma_{11} \equiv (-i)^5 \,\Pi_{\alpha =1}^{10}
\Gamma_{\alpha} ~~ \mbox{with} ~~(\Gamma_{11})^2 ~=~1$.
The chiral projection operators are given by
${\cal P}_{\pm} = \frac{1}{2}(\,1\pm \Gamma_{11})$.

In $SO(10)$ GUTs \cite{fritzsch1,georgi2,mohapatra1},
three generations of quarks and leptons are
 accommodated by
three chiral supermultiplets in
${\bf 16}$ which we
denote by
\be
\Psi^{I}({\bf 16})~~\mbox{with}~~{\cal P}_{+}\,
\Psi^{I}~=~\Psi^{I}~,
\ee
where $I$ runs over the three generations
and the spinor index is suppressed.
To break $SO(10)$ down to $SU(3)_{\rm C}
\times SU(2)_{\rm L} \times U(1)_{\rm Y}$, we use
the following set of chiral superfields:
\be
S_{\{\alpha\beta\}}({\bf 54})~,~
A_{[\alpha\beta]}({\bf 45})~,~
\phi({\bf 16})~,~\overline{\phi}({\overline{\bf 16}})~.
\ee
The two $SU(2)_{\rm L}$ doublets which are responsible for
the spontaneous symmetry breaking (SSB) of
$SU(2)_{\rm L} \times U(1)_{\rm Y}$ down to $U(1)_{\rm EM}$
are contained in
$ H_{\alpha}({\bf 10})$.
We further introduce a singlet $\varphi$ which after
the  SSB of $SO(10)$ will
mix with the right-handed neutrinos
so that they will become superheavy.

The superpotential of  the model is
given by
\bea
W &=& W_{Y}+W_{SB}+ W_{HS}+ W_{NM}+W_{M}~,
\eea
where
\begin{align}
W_{Y} &=\frac{1}{2}\sum_{I,J=1}^{3}g_{IJ}\,\Psi^{I}\,C\Gamma_{\alpha}
\,\Psi^{J}~H_{\alpha}~, \nn \\
W_{SB} &=\frac{g_{\phi}}{2}
\,\overline{\phi}\,
\Gamma_{[\alpha\beta]}\,\phi~A_{[\alpha\beta]}
+\frac{g_{S}}{3!}\,
\mbox{Tr}~S^3+
\frac{g_{A}}{2}\,\mbox{Tr}~A^2\,S~, \nn \\
W_{HS} &=
\frac{g_{HS}}{2}\,\
H_{\alpha}\,S_{ \{\alpha\beta \} }\,H_{\beta}~,~
W_{NM}^{I} ~=~ \sum_{I=1}^{3}\,g_{INM}\,\Psi^I\,
\overline{\phi}\,\varphi~, \nn \\
W_{M}&=\frac{m_{H}}{2}\,H^2+m_{\varphi}\,\varphi^2
+m_{\phi}\,\overline{\phi} \phi+\frac{m_{S}}{2}\,S^2+
\frac{m_{A}}{2}\,A^2~ ,
\end{align}
and $\Gamma_{[\alpha\beta]}=i
 (\Gamma_{\alpha}\Gamma_{\beta} - \Gamma_{\beta}\Gamma_{\alpha})/2$.
As in the case of the $SU(5)$ minimal model, the superpotential is not
the most general one, but
this does not contradict the philosophy of
the coupling unification by the reduction
method.
$W_{SB}$  is responsible for
the SSB of $SO(10)$ down to
$SU(3)_{C}\times SU(2)_{W}\times U(1)_{Y}$,
and this can be achieved without breaking supersymmetry,
while $W_{HS}$ is responsible for
the triplet-doublet splitting of $H$.
The right-handed neutrinos
obtain a superheavy mass through $W_{NM}$ after the
SSB, and the Yukawa couplings
for the leptons and quarks are contained in $W_{Y}$.
We assume  that
 there exists a choice of soft supersymmetry breaking terms so that
all the vacuum expectation values necessary for the desired SSB
corresponds to the minimum of
the potential.

Given the supermultiplet content and the superpotential $W$,
we  can  compute the $\beta$ functions of the model.
The gauge coupling of $SO(10)$ is denoted by $g$, and
our  normalization of the $\beta$ functions
is as usual, i.e.,
$d g_{i}/d \ln \mu ~=~
\beta^{(1)}_{i}/16 \pi^2+O(g^5)$,
where $\mu$ is the renormalization
scale. We find:
\be
\begin{split}
  \beta^{(1)}_{g} &= 7\,g^3~,\\
\beta^{\left(1\right)}_{g_T} &= g_T\,\left(\,14  g_T ^2+\frac{27}{5} g_{HS} ^2+
 g_{3NM} ^2-\frac{63}{2}g^2\,\right)~,\\
\beta^{\left(1\right)}_{g_{\phi}} &= g_{\phi}\left(\,53  g_{\phi ^2}+
\frac{48}{5} g_{A} ^2+\frac{1}{2} g_{1NM} ^2+\frac{1}{2} g_{2NM} ^2+
\frac{1}{2} g_{3NM} ^2-\frac{77}{2}g^2\,\right),\\
\beta^{\left(1\right)}_{S} &= g_{S}\left(\,\frac{84}{5} g_{S} ^2
+12 g_{A} ^2+\frac{3}{2} g_{HS} ^2-60 g^2\,\right)~,\\
\beta^{\left(1\right)}_{A} &=
g_{A}\left(\,16 g_{\phi} ^2+
\frac{28}{5} g_{S} ^2+\frac{116}{5} g_{A} ^2+\frac{1}{2} g_{HS} ^2 -52
g^2\,\right)~,\\
\beta^{\left(1\right)}_{HS} &= g_{HS}\left(\,8 g_{T} ^2+
\frac{28}{5} g_{S} ^2+4 g_{A} ^2+\frac{113}{10} g_{HS} ^2
-38 g^2\,\right) ~,\\
\beta^{\left(1\right)}_{1NM} &= g_{1NM}\left(\,\frac{45}{2} g_{\phi} ^2+
9  g_{1NM} ^2+\frac{17}{2} g_{2NM} ^2+
\frac{17}{2} g_{3NM} ^2-\frac{45}{2}g^2\,\right)~,\\
\beta^{\left(1\right)}_{2NM} &= g_{2NM}\left(\,\frac{45}{2} g_{\phi} ^2+
\frac{17}{2}  g_{1NM} ^2+9  g_{2NM} ^2+
\frac{17}{2} g_{3NM} ^2-\frac{45}{2}g^2\,\right)~,\\
\beta^{\left(1\right)}_{3NM} &= g_{3NM}\left(\,5  g_T ^2+\frac{45}{2} g_{\phi} ^2+
\frac{17}{2}  g_{1NM} ^2+\frac{17}{2} g_{2NM} ^2
+9 g_{3NM} ^2-\frac{45}{2}g^2\,\right)~.
\end{split}
\ee
We have assumed that the Yukawa couplings $g_{IJ}$ except for
$g_T \equiv g_{33}$ vanish. They can be included as small
perturbations. Needless to say that the
soft susy breaking terms do not alter the $\beta$ functions
above.

We find that
there exist two independent
solutions,
$A$ and $B$, that have the most predictive
power, where we have chosen the $SO(10)$ gauge coupling as
the primary coupling:
\begin{align}
\rho_{T} &= \left\{
\begin{array}{ll} 163/60 &\simeq 2.717 \\
0 &  \end{array} \right. ~~,~~
\rho_{\phi} ~= ~\left\{
\begin{array}{ll} 5351/9180 &\simeq 0.583 \\
1589/2727 &\simeq 0.583 \end{array} \right. ~,\nn \\
\rho_{S} &= \left\{
\begin{array}{ll} 152335/51408 &\simeq 2.963 \\
850135/305424 &\simeq 2.783 \end{array} \right. ,
\rho_{A}=\left\{
\begin{array}{ll} 31373/22032 &\simeq 1.424 \\
186415/130896 &\simeq 1.424 \end{array} \right., \nn \\
\rho_{HS}&= \left\{
\begin{array}{ll} 7/81 & \simeq 0.086 \\
170/81 &\simeq 2.099 \end{array} \right. ~,~
\rho_{1NM} = \rho_{2NM} =\left\{
\begin{array}{ll} 191/204 &\simeq 0.936 \\
191/303 &\simeq 0.630 \end{array} \right. ~~, \nn \\
\rho_{3NM}&= \left\{
\begin{array}{ll} 0 &  \\
191/303 &\simeq 0.630 \end{array} \right. ~
~~\mbox{for}~~\left\{\begin{array}{l} A   \\
B\end{array} \right.~.
\end{align}
Clearly, the solution B has less predictive power because
$\rho_T =0$. So, we consider below only the solution A,
in which the coupling $\alpha_{3NM}$ should be
regarded as a small perturbation because $\rho_{3NM}=0$.

Given this solution it is possible to show, as in the case of
$SU(5)$, that the $\rho$'s can be uniquely computed in any finite order
in perturbation theory.

The corrections to the reduced couplings coming from
the small perturbations up to and including terms of
$O(\tilde{\alpha}_{3NM}^2$) are:
\be
\begin{split}
\label{alfas-so10}
\tilde{\alpha}_T &=
(\,163/60 - 0.108\cdots  \tilde{\alpha}_{3NM} +
0.482 \cdots  \tilde{\alpha}_{3NM}^2+\cdots\,)
+\cdots~,\\
\tilde{\alpha}_{\phi} &=
(\,5351/9180 + 0.316\cdots  \tilde{\alpha}_{3NM} +
0.857\cdots  \tilde{\alpha}_{3NM}^2+\cdots\,)
+\cdots~,\\
\tilde{\alpha}_{S} &=
(\,152335/51408 + 0.573\cdots  \tilde{\alpha}_{3NM}
+ 5.7504\cdots  \tilde{\alpha}_{3NM}^2+\cdots\,)
+\cdots~,\\
\tilde{\alpha}_{A} &=
(\,31373/22032 - 0.591\cdots  \tilde{\alpha}_{3NM}
- 4.832\cdots  \tilde{\alpha}_{3NM}^2+\cdots\,)
+\cdots~,\\
\tilde{\alpha}_{HS} &=
(7/81 - 0.00017\cdots  \tilde{\alpha}_{3NM}
+ 0.056\cdots  \tilde{\alpha}_{3NM}^2+\cdots\,)
+\cdots~,\\
\tilde{\alpha}_{1NM}&= \tilde{\alpha}_{2NM}=
(\,191/204 - 4.473\cdots  \tilde{\alpha}_{3NM}
+ 2.831\cdots  \tilde{\alpha}_{3NM}^2+\cdots\,)
+\cdots,
\end{split}
\ee
where $\cdots$ indicates higher order terms which
can be uniquely computed.
In the partially reduced theory defined above,
we have two independent couplings,
$\alpha$ and $\alpha_{3NM}$ (along with the
Yukawa couplings $\alpha_{IJ}~,~ I,J\neq T$).

At the one-loop level,
Eq.~(\ref{alfas-so10}) defines a line parametrized by $\tilde{\alpha}_{3NM}$
in the $7$ dimensional space of couplings.
A numerical analysis shows  that this line blows up
in the direction of $\tilde{\alpha}_{S}$ at
a finite value of $\tilde{\alpha}_{3NM}$ \cite{Kubo:1995zg}.
So if we require $\tilde{\alpha}_{S}$ to remain within
the perturbative regime (i.e., $g_S \leq 2$,
which means $ \tilde{\alpha}_{S} \leq 8$ because
$\alpha_{\rm GUT} \sim 0.04$),
the $\tilde{\alpha}_{3NM}$ should be restricted to be below
$\sim 0.067$. As a consequence, the value of
$\tilde{\alpha}_{T}$ is also bounded
\be
2.714 \leq \tilde{\alpha}_{T} \le 2.736~.
\ee
This defines
GYU boundary conditions holding
at the unification scale $M_{\rm GUT}$ in addition to the
group theoretical one,
$\alpha_{T}=\alpha_{t} ~=~\alpha_{b} ~=~\alpha_{\tau}$.
The value of $ \tilde{\alpha}_{T}$ is practically fixed so that
we may assume that
$\tilde{\alpha}_{T}=163/60 \simeq 2.72$,
which is the unperturbed value.

\begin{figure}[htb!]
\begin{center}
\includegraphics[scale=0.7,angle=0]{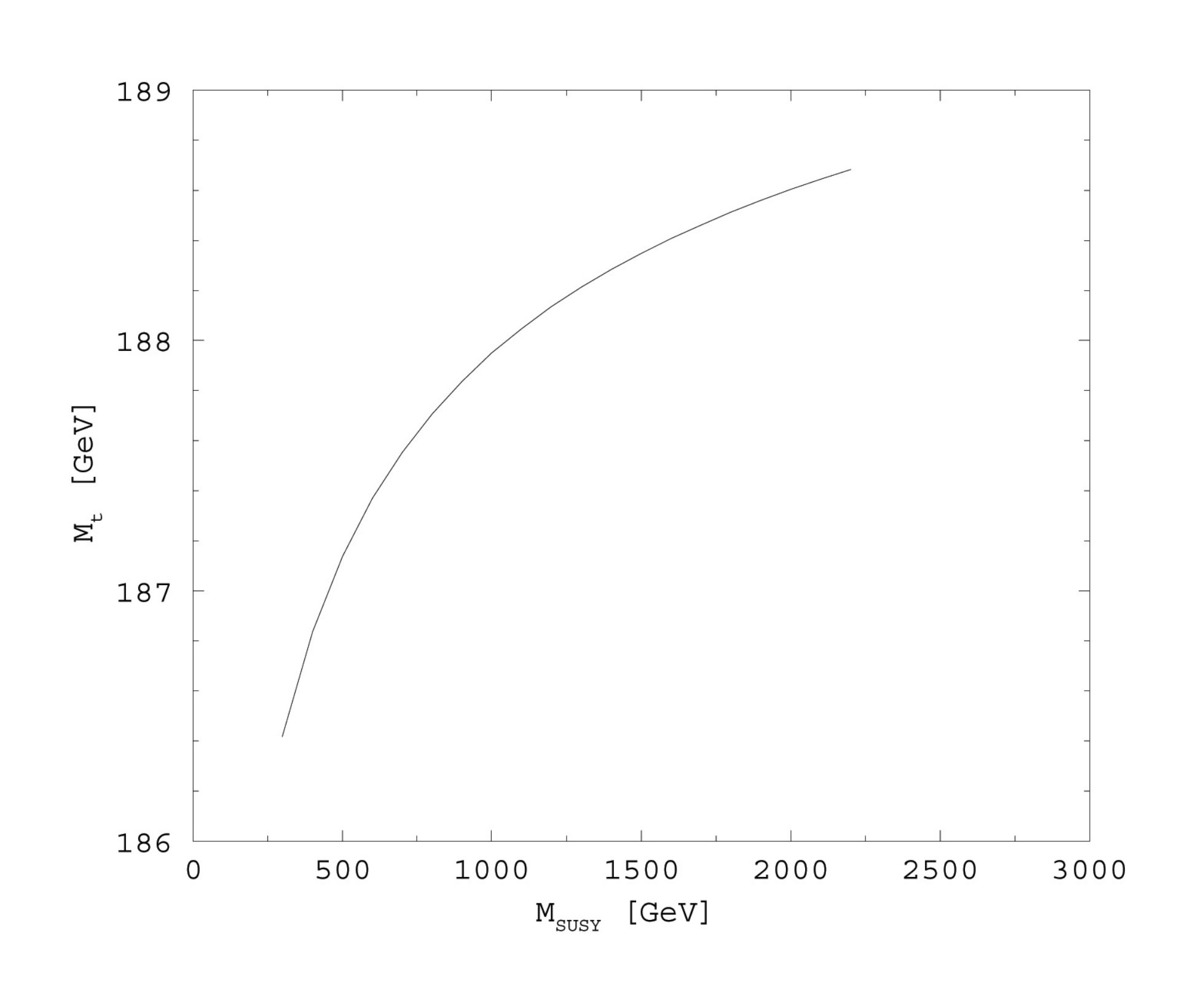}
\caption{Top quark mass prediction versus
$M_{\rm SUSY}$ for $\tilde{\alpha}_{T}=2.717$.}
\label{fig:sotms}
\end{center}
\end{figure}

Fig.~\ref{fig:sotms} shows the prediction for the top quark mass
in this model
for different values of the supersymmetry
breaking scale $M_{\rm SUSY}$.
While the value for the top quark mass predicted is below its infrared
value ($\sim 189 \gev$)~\cite{Kubo:1995zg}, it is above the experimental
value~\cite{PDG18}. Consequently, also this particular model has difficulties
to meet the experimental data on the
top quark mass, despite the theoretical uncertainties involved.


\newpage

\section{Finite $SU(N)^3$ Unification}
\label{sec:sun3}

We continue examining the possibility of constructing realistic FUTs
based on product gauge groups. Consider an $N=1$ supersymmetric theory, with
gauge group $SU(N)_1 \times SU(N)_2 \times \cdots \times SU(N)_k$,
with $n_f$ copies (number of families) 
of the
supersymmetric multiplets $(N,N^*,1,\dots,1) + (1,N,N^*,\dots,1) +
\cdots + (N^*,1,1,\dots,N)$.  The one-loop $\beta$-function
coefficient in the renormalization-group equation of each $SU(N)$
gauge coupling is simply given by
\begin{equation}
b = \left( -\frac{11}{3} + \frac{2}{3} \right) N + n_f \left( \frac{2}{3}
 + \frac{1}{3} \right) \left( \frac{1}{2} \right) 2 N = -3 N + n_f
N\,.
\label{3gen}
\end{equation}
This means that $n_f = 3$ is the only solution of Eq.(\ref{3gen}) that
yields $b = 0$.  Since $b=0$ is a
necessary condition for a finite field theory, the existence of three
families of quarks and leptons is natural in such models, provided the
matter content is exactly as given above.

The model of this type with best phenomenology is the $SU(3)^3$ model
discussed in \citere{Ma:2004mi}, where the details of the model are
given. It corresponds to the well-known example of $SU(3)_C
\times SU(3)_L \times
SU(3)_R$~\cite{Derujula:1984gu,Lazarides:1993sn,Lazarides:1993uw,Ma:1986we},
 with quarks transforming as
\begin{equation}
  q = \begin{pmatrix} d & u & h \\ d & u & h \\ d & u & h \end{pmatrix}
\sim (3,3^*,1), ~~~
    q^c = \begin{pmatrix} d^c & d^c & d^c \\ u^c & u^c & u^c \\ h^c & h^c & h^c
\end{pmatrix}
    \sim (3^*,1,3),
\label{2quarks}
\end{equation}
and leptons transforming as
\begin{equation}
\lambda = \begin{pmatrix} N & E^c & \nu \\ E & N^c & e \\ \nu^c & e^c & S
\end{pmatrix}
\sim (1,3,3^*).
\label{3leptons}
\end{equation}
Switching the first and third rows of $q^c$ together with the first and
third columns of $\lambda$, we obtain the alternative left-right model first
proposed in \citere{Ma:1986we} in the context of superstring-inspired $E_6$.

In order for all the gauge couplings to be equal at an energy scale, $M_{\rm GUT}$,
the cyclic symmetry $Z_3$ must be imposed, i.e.
\begin{equation}
q \to \lambda \to q^c \to q,
\label{15}
\end{equation}
where $q$ and $q^c$ are given in eq.~(\ref{2quarks}) and $\lambda$ in
eq.~(\ref{3leptons}).  Then,
the first of the finiteness conditions (\ref{1st}) for one-loop
finiteness, namely the vanishing of the gauge $\beta$-function is
satisfied.

Next let us consider the second condition, i.e. the vanishing of the
anomalous dimensions of all superfields, eq.~(\ref{2nd}).  To do that
first we have to write down the superpotential.  If there is just one
family, then there are only two trilinear invariants, which can be
constructed respecting the symmetries of the theory, and therefore can
be used in the superpotential as follows
\begin{equation}
f ~Tr (\lambda q^c q) + \frac{1}{6} f' ~\epsilon_{ijk} \epsilon_{abc}
(\lambda_{ia} \lambda_{jb} \lambda_{kc} + q^c_{ia} q^c_{jb} q^c_{kc} +
q_{ia} q_{jb} q_{kc}),
\label{16}
\end{equation}
where $f$ and $f'$ are the Yukawa couplings associated to each invariant.
%
Quark and leptons obtain masses when the scalar parts of the
superfields $(\tilde N,\tilde N^c)$ obtain vacuum expectation values (vevs),
\begin{equation}
m_d = f \langle \tilde N \rangle, ~~ m_u = f \langle \tilde N^c \rangle, ~~
m_e = f' \langle \tilde N \rangle, ~~ m_\nu = f' \langle \tilde N^c \rangle.
\label{18}
\end{equation}

With three families, the most general superpotential contains 11 $f$
couplings, and 10 $f'$ couplings, subject to 9 conditions, due to the
vanishing of the anomalous dimensions of each superfield.  The
conditions are the following
\begin{equation}
\sum_{j,k} f_{ijk} (f_{ljk})^* + \frac{2}{3} \sum_{j,k} f'_{ijk}
(f'_{ljk})^* = \frac{16}{9} g^2 \delta_{il}\,,
\label{19}
\end{equation}
where
\begin{eqnarray}
&& f_{ijk} = f_{jki} = f_{kij}, \label{20}\\
&& f'_{ijk} = f'_{jki} = f'_{kij} = f'_{ikj} = f'_{kji} = f'_{jik}.
\label{21}
\end{eqnarray}
Quarks and leptons receive  masses when  the scalar part of the
superfields $\tilde N_{1,2,3}$ and $\tilde N^c_{1,2,3}$ obtain vevs as follows
\begin{eqnarray}
&& ({\cal M}_d)_{ij} = \sum_k f_{kij} \langle \tilde N_k \rangle, ~~~
   ({\cal M}_u)_{ij} = \sum_k f_{kij} \langle \tilde N^c_k \rangle, \label{22} \\
&& ({\cal M}_e)_{ij} = \sum_k f'_{kij} \langle \tilde N_k \rangle, ~~~
   ({\cal M}_\nu)_{ij} = \sum_k f'_{kij} \langle \tilde N^c_k \rangle.
\label{23}
\end{eqnarray}

We will assume that the below $M_{\rm GUT}$ we have the usual MSSM
\footnote{For details of how the spontaneous breaking of $SU(3)^3$ to MSSM can be achieved see
refs \cite{Irges:2011de,Irges:2012ze} and refs therein.},
with the two Higgs doublets coupled maximally to the third generation.
Therefore we have to choose
the linear combinations $\tilde N^c = \sum_i a_i \tilde N^c_i$ and
$\tilde N = \sum_i b_i \tilde N_i$ to play the role of the two Higgs
doublets, which will be responsible for the electroweak symmetry
breaking.  This can be done by choosing appropriately the masses in
the superpotential \cite{Leon:1985jm}, since they are not
constrained by the finiteness conditions.  We choose that
the two Higgs doublets are predominately coupled to the third
generation. Then these two Higgs doublets couple to the three
families differently, thus providing the freedom to understand
their different masses and mixings.
The remnants of the $SU(3)^3$ FUT are the boundary conditions on the
gauge and Yukawa couplings, i.e. Eq.(\ref{19}), the $h=-MC$
relation, and the soft scalar-mass sum rule eq.~(\ref{sumr})
at $M_{\rm GUT}$, which, when applied to the present model, takes the form
\begin{eqnarray}
m^2_{H_u} + m^2_{\tilde t^c} + m^2_{\tilde q} = M^2 =
m^2_{H_d} + m^2_{\tilde b^c} + m^2_{\tilde q}~,
\end{eqnarray}
where   ${\tilde t^c}, ~{\tilde b^c}$, and ${\tilde q}$ are the
scalar parts of the corresponding
superfields.

Concerning the solution to Eq.(\ref{19}) we consider two versions of the model:\\
I) An all-loop finite model with a unique and isolated solution, in
which $f'$ vanishes, which leads to the following relations
\begin{equation}
f^2 = f^2_{111} = f^2_{222} = f^2_{333} = \frac{16}{9} g^2\,.
\label{isosol}
\end{equation}
 As for
the lepton masses, since all $f'$ couplings have been fixed to be
zero at this order, in principle they would be expected to appear
radiatively induced by the scalar lepton masses appearing in the SSB
sector of the theory.  However, due to the finiteness
conditions they cannot appear radiatively and remain as a
problem for further study.  \\
II) A two-loop finite solution, in which we keep $f'$ non-vanishing
and we use it to introduce the lepton masses. The model in turn
becomes finite only up to two-loops since the corresponding solution
of Eq.(\ref{19}) is not an isolated one any more, i.e. it is a parametric
one.  In this case we have the following boundary conditions for the
Yukawa couplings
\begin{eqnarray}
f^2 = r \left(\frac{16}{9}\right) g^2\,,\quad
f'^2 = (1-r) \left(\frac{8}{3}\right) g^2\,,
\label{fprime}
\end{eqnarray}
where $r$ is a free parameter which parametrizes the different
solutions to the finiteness conditions. 
As for the boundary conditions of the soft scalars, we have the
universal case.

Below $M_{\rm GUT}$ all couplings and masses of the theory run according
to the RGEs of the MSSM.  Thus we examine the evolution of these
parameters according to their RGEs up to two-loops for dimensionless
parameters and at one-loop for dimensionful ones imposing the
corresponding boundary conditions.  We further assume a unique
SUSY breaking scale $M_{\rm SUSY}$ and below that scale the
effective theory is just the SM.

We compare our predictions with the experimental value of $m_t^{\rm exp}$
\footnote{
As before, these values correspond to the experimental measurements at
the time of the original evaluation. Again, the small change to the
current values
would not change the phenomenological analysis in a relevant way.}
and recall that the theoretical values for
$m_t$ suffer from a correction of
$\sim 4 \%$~\cite{Kubo:1997fi,Kobayashi:2001me,Mondragon:2003bp}.
In the case of the bottom quark, we take again the value evaluated at
$M_Z$, see \refeq{mbexp}. In the case of
model I, the predictions for the top quark mass (in this case $m_b$ is
an input) $m_t$ are $\sim 183\,{\rm ~GeV}$ for $\mu < 0 $, which is
above the experimental value, and there are no solutions for $\mu>0$.

\begin{figure}[htb!]
\begin{center}
\includegraphics[width=7cm,angle=0]{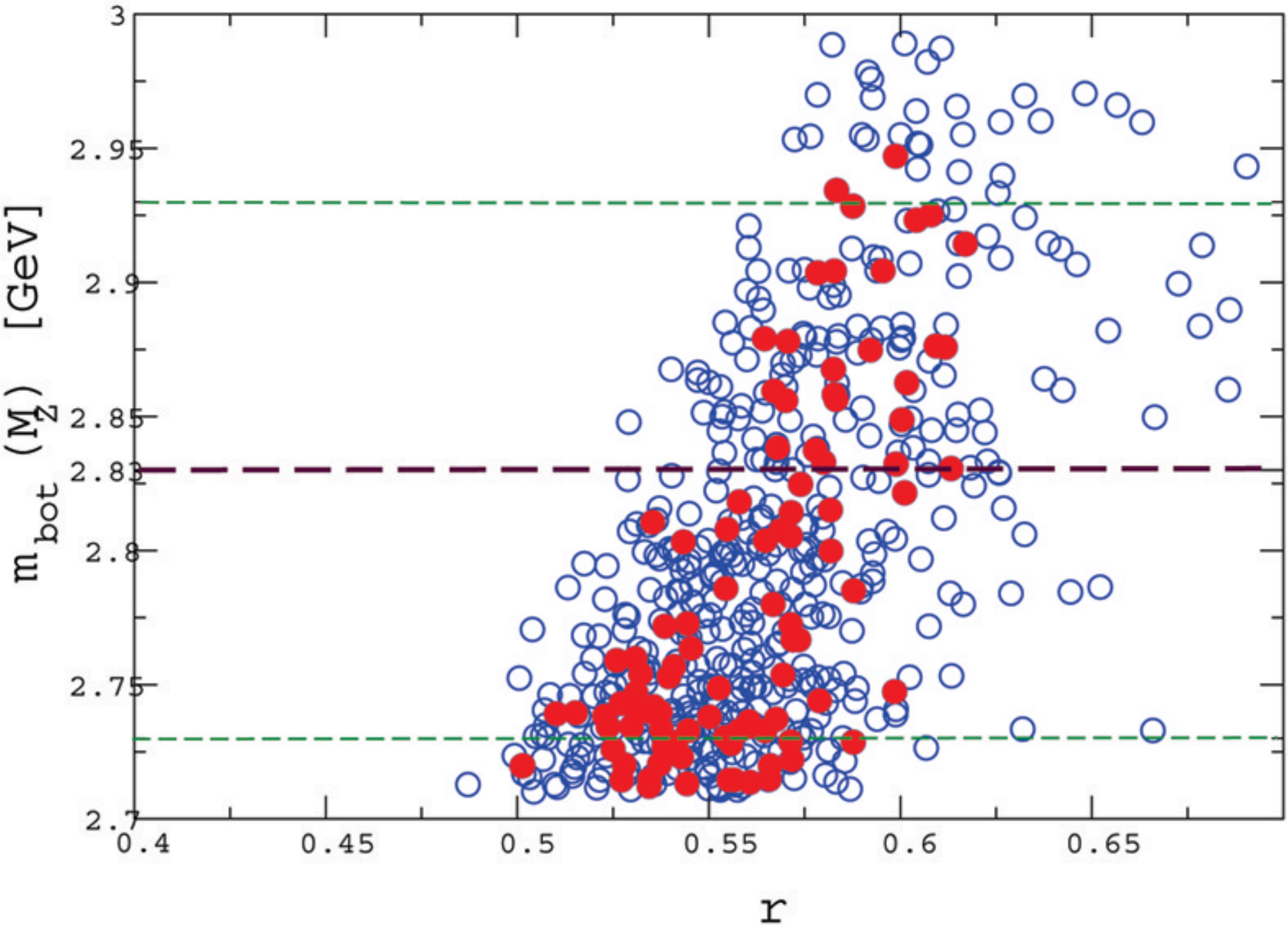} 
\includegraphics[width=7cm,angle=0]{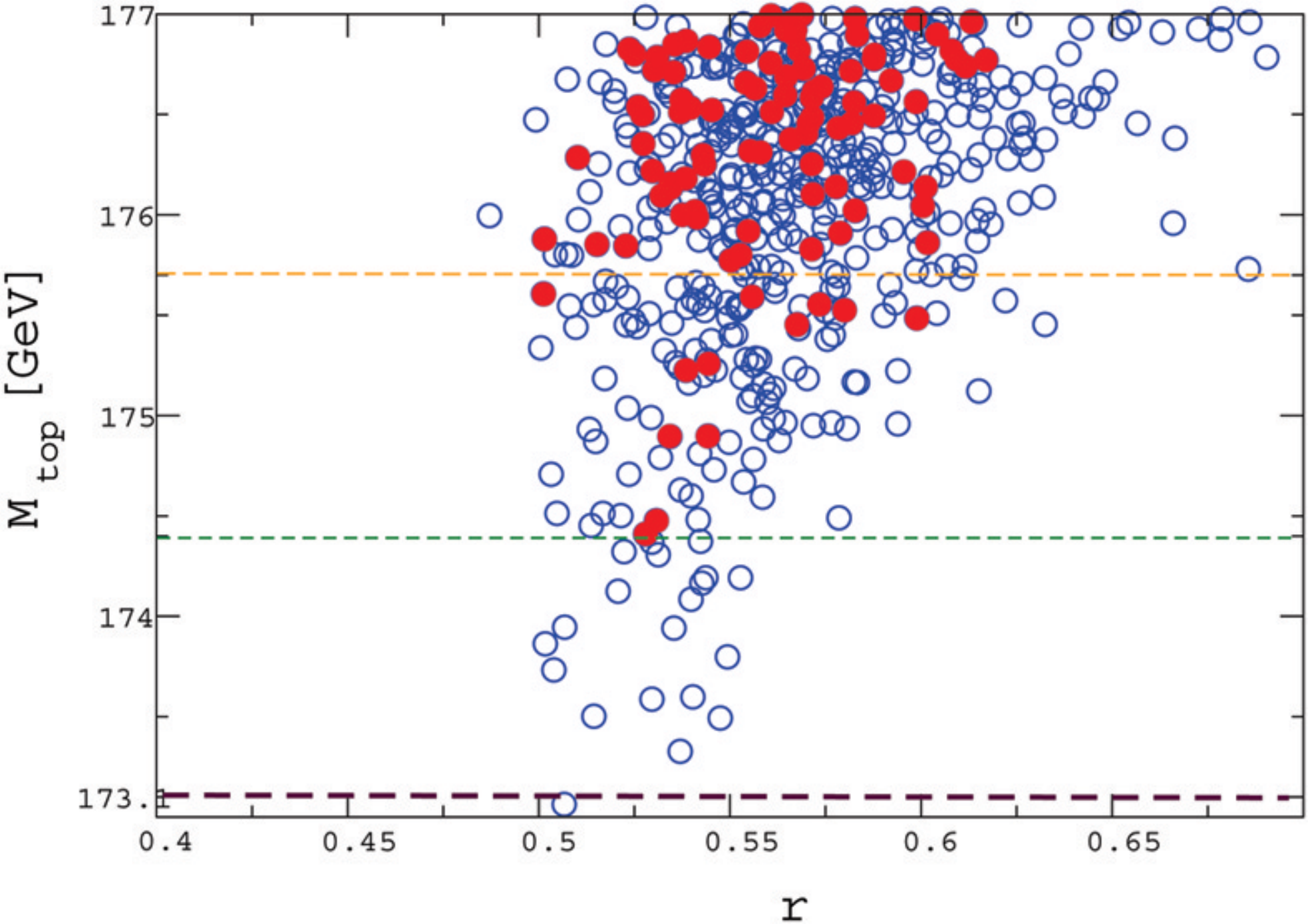}
\caption{The figures show the values for the top and bottom quark
  masses for the FUT model $SU(3)^3$, with $\mu <0 $, vs the parameter
  $r$. The thicker horizontal line is the experimental central value,
  and the lighter green and orange ones are the one and two sigma
  limits respectively.  The red points are the ones that satisfy the
  $B$-physics constraints, as discussed in Chapter~\ref{chap:FUT_LOW}.}
\label{fig:Higgs-m5}
\end{center}
\end{figure}

For the two-loop model {II}, we look for the values of the parameter
$r$ which comply with the experimental limits given above for top and
bottom quarks masses. In the case of $\mu >0$, for the bottom quark,
the values of $r$ lie in the range $0.15 \lesssim r \lesssim 0.32$.
For the top mass, the range of values for r is $0.35 \lesssim r
\lesssim 0.6$. From these values we can see that there is a very small
region where both top and bottom quark masses are in the experimental
range for the same value of $r$.  In the case of $\mu<0$ the situation
is similar, although slightly better, with the range of values $0.62
\lesssim r \lesssim 0.77$ for the bottom mass, and $0.4 \lesssim r
\lesssim 0.62$ for the top quark mass.  In the above mentioned analysis, the
masses of the new particles $h$'s and $E$'s of all families were taken
to be at the $M_{\rm GUT}$ scale.

Taking into account new thresholds for
these exotic particles below $M_{\rm GUT}$ we  find a wider
phenomenologically viable parameter space~\cite{Mondragon:2011zzb}.
This can be seen in Fig.~\ref{fig:Higgs-m5}, where we took only one down-like
exotic particle decoupling at $10^{14}$ GeV, below than the usual MSSM.

In this case, for $r\sim 0.5 \sim 0.62$ we have reasonable agreement with
experimental data for both top and bottom quark masses, where the red points
in the figure are the ones that satisfy the $B$-physics constraints
(at the time of the analysis)~\cite{Amsler:2008zzb}.
The above analysis shows that it is worth returning with a fresh examination
of this model taking into account all new experimental constraints.


\section{Reduction of Couplings in the MSSM}
\label{sec:mssm}

In this section we are working in the framework of MSSM, assuming though
the existence of a covering GUT.

The analysis of the partial reduction of couplings in this framework was first done in
refs~\cite{Mondragon:2013aea,Mondragon:2017hki}.

The superpotential of the MSSM (where again we restrict ourselves to the
  third generation of fermions) is defined by
\be
\label{supot2}
W = Y_tH_2Qt^c+Y_bH_1Qb^c+Y_\tau H_1L\tau^c+ \mu H_1H_2\, ,
\ee
where $Q,L,t,b,\tau, H_1,H_2$ are the usual superfields of MSSM,
while the SSB Lagrangian is given by
\be
\label{SSB_L}
\begin{split}
-\mathcal{L}_{\rm SSB} &= \sum_\phi m^2_\phi\hat{\phi^*}\hat{\phi}+
\left[m^2_3\hat{H_1}\hat{H_2}+\sum_{i=1}^3 \frac 12 M_i\lambda_i\lambda_i +\textrm{h.c}\right]\\
&+\left[h_t\hat{H_2}\hat{Q}\hat{t^c}+h_b\hat{H_1}\hat{Q}\hat{b^c}+h_\tau \hat{H_1}\hat{L}\hat{\tau^c}+\textrm{h.c.}\right] ,
\end{split}
\ee
where $\hat{\phi}$ represents the scalar component of all superfields, $\lambda$ refers to the gaugino fields
while all hatted fields refer to the scalar components of the corresponding superfield.
The Yukawa $Y_{t,b,\tau}$ and the trilinear $h_{t,b,\tau}$ couplings  refer to the third generator only,
neglecting the first two generations.

Let us start with the dimensionless couplings, i.e. gauge and Yukawa. As a first step we consider
only the strong coupling and the top and bottom Yukawa couplings, while the other two gauge couplings and the tau
Yukawa will be treated as corrections.
Following the above line, we reduce the Yukawa couplings in favour of
the strong coupling~$\al_3$
\[
\frac{Y^2_i}{4\pi}\equiv \al_i=G_i^2\al_3,\qquad i=t,b,
\]
and using the RGE for the Yukawa, we get
\[
G_i^2=\frac 13 ,\qquad i=t,b.
\]
This system of the top and bottom Yukawa couplings reduced with the strong one
is dictated by (i) the different running behaviour of the
$SU(2)$ and $U(1)$ coupling compared to the strong one
\cite{Kubo:1985up}
and (ii) the incompatibility of applying the above reduction
for the tau Yukawa since the corresponding $G^2$ turns negative
\cite{{MTZ:14}}.
Adding now the two other gauge couplings and the tau Yukawa in the RGE as
corrections, we obtain
\be
\label{Gt2_Gb2}
G_t^2=\frac 13+\frac{71}{525}\rho_1+\frac 37 \rho_2 +\frac 1{35}\rho_\tau,\qquad
G_b^2=\frac 13+\frac{29}{525}\rho_1+\frac 37 \rho_2 -\frac 6{35}\rho_\tau
\ee
where
\be
\label{r1_r2_rtau}
\rho_{1,2}=\frac{g_{1,2}^2}{g_3^2}=\frac{\al_{1,2}}{\al_3},\qquad
\rho_\tau=\frac{g_\tau^2}{g_3^2}=\frac{\displaystyle{\frac{Y^2_\tau}{4\pi}}}{\al_3}
\ee

Note that the corrections in Eq.(\ref{Gt2_Gb2}) are taken at the GUT scale and under the assumption that
\[
\frac{d}{dg_3}\left(\frac {Y_{t,b}^2}{g_3^2}\right)=0.
\]

Let us comment on our assumption above, which led to the Eq.(\ref{Gt2_Gb2}).
In practice we assume that even including the corrections from the rest of the gauge as well as the tau Yukawa couplings,
at the GUT scale the ratio of the top and bottom couplings $\al_{t,b}$ over the strong coupling are still constant,
i.e. their scale dependence is negligible.
Or, rephrasing it, our assumption can be understood as a requirement that in the ultraviolet (close to the GUT scale)
the ratios of the top and bottom Yukawa couplings over the strong coupling become least sensitive against the change
of the renormalization scale. This requirement sets the boundary condition at the GUT scale, given in Eq.(\ref{Gt2_Gb2}).
Alternatively one could follow the systematic method to include the corrections to a non-trivially reduced system developed
in ref.~\cite{Kubo:1988zu}, but considering two reduced systems: the first one consisting of the ``top, bottom'' couplings and the second of the
``strong, bottom'' ones.

In the next order the corrections are assumed to be in the form
\[
\al_i=G_i^2\al_3+J_i^2 \al_3^2,\qquad i=t,b.
\]
Then, the coefficients $J_i$ are given by
\[
J_i^2=\frac 1{4\pi}\,\frac{17}{24},\qquad i=t,b
\]
for the case where only the strong gauge and the top and bottom Yukawa couplings are active,
while for the case where the other two gauge and the tau Yukawa couplings are
added as corrections we obtain
\[
J_t^2=\frac 1{4\pi}\frac{N_t}{D},\quad
J_b^2=\frac 1{4\pi}\frac{N_b}{5D},
\]
where
\[
\begin{split}
D=&
257250 (196000 + 44500 \rho_1 + 2059 \rho_1^2 + 200250 \rho_2 + 22500 \rho_1 \rho_2 +
   50625 \rho_2^2 - \\
&33375 \rho_\tau - 5955 \rho_1 \rho_\tau - 16875 \rho_2 \rho_\tau -
   1350 \rho_\tau^2),
\end{split}
\]

\[
\begin{split}
N_t=&
-(-35714875000 - 10349167500 \rho_1 + 21077903700 \rho_1^2 +9057172327 \rho_1^3 +\\
&   481651575 \rho_1^4 - 55566000000 \rho_2 +2857680000 \rho_1 \rho_2 + 34588894725 \rho_1^2 \rho_2 +\\
&   5202716130 \rho_1^3 \rho_2 +3913875000 \rho_2^2 + 8104595625 \rho_1 \rho_2^2 + 11497621500 \rho_1^2 \rho_2^2 +\\
&   27047671875 \rho_2^3 + 1977918750 \rho_1 \rho_2^3 + 7802578125 \rho_2^4 +3678675000 \rho_\tau +\\
&   1269418500 \rho_1 \rho_\tau - 2827765710 \rho_1^2 \rho_\tau -1420498671 \rho_1^3 \rho_\tau +7557637500 \rho_2 \rho_\tau -\\
&   2378187000 \rho_1 \rho_2 \rho_\tau - 4066909425 \rho_1^2 \rho_2 \rho_\tau -1284018750 \rho_2^2 \rho_\tau - 1035973125 \rho_1 \rho_2^2 \rho_\tau -\\
&   2464171875 \rho_2^3 \rho_\tau + 1230757500 \rho_\tau^2 + 442136100 \rho_1 \rho_\tau^2 -186425070 \rho_1^2 \rho_\tau^2 +\\
&   1727460000 \rho_2 \rho_\tau^2 +794232000 \rho_1 \rho_2 \rho_\tau^2 + 973518750 \rho_2^2 \rho_\tau^2 -\\
&   325804500 \rho_\tau^3 - 126334800 \rho_1 \rho_\tau^3 - 412695000 \rho_2 \rho_\tau^3 -
   32724000 \rho_\tau^4),
\end{split}
\]

\[
\begin{split}
N_b=&
-(-178574375000 - 71734162500 \rho_1 + 36055498500 \rho_1^2 +13029194465 \rho_1^3 +\\
&   977219931 \rho_1^4 - 277830000000 \rho_2 -69523650000 \rho_1 \rho_2 + 72621383625 \rho_1^2 \rho_2 +\\
&   10648126350 \rho_1^3 \rho_2 +19569375000 \rho_2^2 + 13062459375 \rho_1 \rho_2^2 + 25279672500 \rho_1^2 \rho_2^2 +\\
&   135238359375 \rho_2^3 + 16587281250 \rho_1 \rho_2^3 + 39012890625 \rho_2^4 +58460062500 \rho_\tau +\\
&   35924411250 \rho_1 \rho_\tau - 13544261325 \rho_1^2 \rho_\tau -2152509435 \rho_1^3 \rho_\tau - 13050843750 \rho_2 \rho_\tau +\\
&   45805646250 \rho_1 \rho_2 \rho_\tau - 75889125 \rho_1^2 \rho_2 \rho_\tau -24218578125 \rho_2^2 \rho_\tau + 17493046875 \rho_1 \rho_2^2 \rho_\tau -\\
&   1158046875 \rho_2^3 \rho_\tau - 36356775000 \rho_\tau^2 -26724138000 \rho_1 \rho_\tau^2 - 4004587050 \rho_1^2 \rho_\tau^2 -\\
&   97864200000 \rho_2 \rho_\tau^2 - 22359847500 \rho_1 \rho_2 \rho_\tau^2 -39783656250 \rho_2^2 \rho_\tau^2 + 25721797500 \rho_\tau^3 +\\
&   3651097500 \rho_1 \rho_\tau^3 + 11282287500 \rho_2 \rho_\tau^3 + 927855000 \rho_\tau^4).
\end{split}
\]

We move now to the dimension-1 parameters of the SSB Lagrangian, namely
the trilinear couplings $h_{t,b,\tau}$ of the SSB Lagrangian,
\refeq{SSB_L}. Again,
following the pattern in the Yukawa reduction,
in the first stage we reduce $h_{t,b}$, while $h_\tau$ will be
treated as a correction.
\[
h_i=c_i Y_i M_3 = c_i G_i M_3 g_3,\qquad i=t,b,
\]
where $M_3$ is the gluino mass.
Using the RGE for the two $h$ we get
\[
c_t=c_b=-1,
\]
where we have also used the 1-loop relation between the gaugino mass and the gauge coupling RGE
\[
2M_i\frac {dg_i}{dt}=g_i\frac {dM_i}{dt},\qquad i=1,2,3.
\]
Adding the other two gauge couplings as well as the tau Yukawa $h_\tau$
  as correction we get
\[
c_t=-\frac{A_A A_{bb} + A_{tb} B_B}{A_{bt} A_{tb} - A_{bb} A_{tt}},\qquad
c_b=-\frac{A_A A_{bt} + A_{tt} B_B}{A_{bt} A_{tb} - A_{bb} A_{tt}},
\]
where
\be
\label{rhtau}
\begin{split}
A_{tt}& = G_b^2 - \frac{16}3 - 3\rho_2 - \frac{13}{15}\rho_1,\quad
A_A = \frac{16}3 + 3\rho_2^2 + \frac{13}{15}\rho_1^2\\
A_{bb} &= G_t^2 + \rho_\tau - \frac{16}3 - 3\rho_2 - \frac 7{15}\rho_1,\quad
B_B = \frac{16}3 + 3\rho_2^2 + \frac7{15}\rho_1^2 + \rho_{h_\tau}\rho_\tau^{1/2}\\
A_{tb} &= G_b^2,\quad A_{bt} = G_t^2,\quad \rho_{h_\tau}=\frac{h_\tau}{g_3 M_3}.
\end{split}
\ee
Finally we consider the soft squared masses $m^2_\phi$ of the SSB
Lagrangian. Their reduction, according to the discussion in Sect.~\ref{sec:dimful}, takes
the form
\be\label{mM_rel}
m_i^2=c_i M_3^2,\quad i=Q,u,d,H_u,H_d.
\ee
The 1-loop RGE for the scalar masses reduce to the following algebraic system (where we have added
the corrections from the two gauge couplings, the tau Yukawa and $h_\tau$)
\[
\begin{split}
-12c_Q&=X_t+X_b-\frac{32}3-6\rho_2^3-\frac 2{15}\rho_1^3+\frac 15\rho_1 S,\\
-12c_u&=2X_t-\frac{32}3-\frac{32}{15}\rho_1^3-\frac 45\rho_1 S,\\
-12c_d&=2X_b-\frac{32}3-\frac 8{15}\rho_1^3+\frac 25\rho_1 S,\\
-12c_{H_u}&=3X_t-6\rho_2^3-\frac 65\rho_1^3+\frac 35\rho_1 S,\\
-12c_{H_d}&=3X_b+X_\tau-6\rho_2^3-\frac 65\rho_1^3-\frac 35\rho_1 S,
\end{split}
\]
where
\[
\begin{split}
X_t&=2G_t^2\left(c_{H_u}+c_Q+c_u\right)+2c_t^2G_t^2,\\
X_b&=2G_b^2\left(c_{H_d}+c_Q+c_d\right)+2c_b^2G_b^2,\\
X_\tau&=2\rho_\tau c_{H_d}+2\rho_{h_\tau}^2,\\
S&=c_{H_u}-c_{H_d}+c_Q-2c_u+c_d.
\end{split}
\]
Solving the above system for the coefficients $c_{Q,u,d,H_u,H_d}$ we get
\[
\begin{split}
c_Q=&-\frac{c_{Q{\rm Num}}}{D_m},\quad
c_u=-\frac 13\frac{c_{u{\rm Num}}}{D_m},\quad
c_d=-\frac{c_{d{\rm Num}}}{D_m},\\
c_{H_u}=&-\frac 23\frac{c_{Hu{\rm Num}}}{D_m},\quad
c_{H_d}=-\frac{c_{Hd{\rm Num}}}{D_m},
\end{split}
\]
where
\[
\begin{split}
D_m=&
4 (6480 + 6480 G_b^2 + 6480 G_t^2 + 6300 G_b^2 G_t^2 +
\rho_1(1836  + 1836 G_b^2  + 1836 G_t^2  +1785 G_b^2 G_t^2 )+ \\
&     \rho_\tau \left[1080  + 540 G_b^2  + 1080 G_t^2  + 510 G_b^2 G_t^2  +
 252 \rho_1  + 99 G_b^2 \rho_1  +252 G_t^2 \rho_1  + 92 G_b^2 G_t^2 \rho_1 \right]),
\end{split}
\]

\[
\begin{split}
c_{Q{\rm Num}}=&
2160 F_Q +
G_b^2(- 360 F_d - 360 F_{H_d}  + 1800 F_Q) +
G_t^2(- 360 F_{H_u} + 1800 F_Q  - 360 F_u)+\\
&G_b^2 G_t^2(- 300 F_d - 300 F_{H_d} - 300 F_{H_u} + 1500 F_Q  - 300 F_u ) +\\
&\rho_1(- 36 F_d +   36 F_{H_d}  - 36 F_{H_u}  + 576 F_Q  + 72 F_u  )+\\
&G_b^2 \rho_1(- 138 F_d  -66 F_{H_d} - 36 F_{H_u}  + 474 F_Q  + 72 F_u ) +\\
&G_t^2 \rho_1(- 36 F_d + 36 F_{H_d}  - 138 F_{H_u}  + 474 F_Q  - 30 F_u) +\\
&G_b^2 G_t^2 \rho_1(- 120 F_d - 50 F_{H_d}  -  120 F_{H_u}  + 390 F_Q  - 15 F_u ) +\\
&\rho_\tau\left[
360 F_Q  +G_b^2 (- 60 F_d + 120 F_Q )+
G_t^2 (- 60 F_{H_u} + 300 F_Q  - 60 F_u )+
\right.
\\
&G_b^2 G_t^2 (- 50 F_d - 20 F_{H_u}  + 100 F_Q  - 20 F_u )+
\rho_1 (-6 F_d - 6 F_{H_u}  + 78 F_Q  + 12 F_u )+\\
&G_b^2 \rho_1 ( - 11 F_d  + 22 F_Q ) +
G_t^2 \rho_1 ( - 6 F_d  -20 F_{H_u}  + 64 F_Q  - 2 F_u ) +\\
&\left.
G_b^2 G_t^2 \rho_1 ( -9 F_d  - 4 F_{H_u}  +18 F_Q  - 3 F_u )
\right],
\end{split}
\]

\[
\begin{split}
c_{u{\rm Num}}=&
6480 F_u + 6480 F_u G_b^2 +
G_t^2(- 2160 F_{H_u}  - 2160 F_Q  + 4320 F_u ) +\\
& G_b^2 G_t^2(  360 F_d  + 360 F_{H_d}  - 2160 F_{H_u}  -1800 F_Q  + 4140 F_u )+\\
&    \rho_1( 432 F_d  - 432 F_{H_d} +432 F_{H_u}  + 432 F_Q  + 972 F_u )+\\
&    G_b^2 \rho_1( 432 F_d  -432 F_{H_d}  + 432 F_{H_u}  + 432 F_Q  + 972 F_u )+\\
&   G_t^2 \rho_1(432 F_d  - 432 F_{H_d}  - 180 F_{H_u}  - 180 F_Q  +360 F_u )+\\
&    G_b^2 G_t^2 \rho_1( 522 F_d  - 318 F_{H_d}  - 192 F_{H_u}  - 90 F_Q  + 333 F_u ) +\\
&   \rho_\tau \left[
     1080 F_u  +
     540 G_b^2 F_u + G_t^2(  - 360 F_{H_u}   - 360 F_Q   + 720 F_u )+ \right.\\
&      G_b^2 G_t^2 (60 F_d   - 180 F_{H_u}   - 120 F_Q   + 330 F_u )  +
  \rho_1( 72 F_d   + 72 F_{H_u}   + 72 F_Q   + 108 F_u )  +\\
&   G_b^2 \rho_1( 36 F_{H_u}   + 27 F_u ) +
72 G_t^2 \rho_1 (F_d   - 12 F_{H_u}   - 12 F_Q   + 24 F_u )  +\\
&   \left.
G_b^2 G_t^2 \rho_1 (9 F_d   + 4 F_{H_u}   - 18 F_Q   + 3 F_u )
    \right],
\end{split}
\]

\[
\begin{split}
c_{d{\rm Num}}=&
2160 F_d + G_b^2(1440 F_d  - 720 F_{H_d}  - 720 F_Q ) + 2160 F_d G_t^2 +\\
&   G_b^2 G_t^2 (1380 F_d  - 720 F_{H_d}  + 120 F_{H_u}  -  600 F_Q  + 120 F_u ) +\\
&   \rho_1( 540 F_d  + 72 F_{H_d}  - 72 F_{H_u}  - 72 F_Q  + 144 F_u )+\\
&    G_b^2 \rho_1( 336 F_d  -132 F_{H_d}  - 72 F_{H_u}  - 276 F_Q  + 144 F_u ) +\\
&   G_t^2 \rho_1( 540 F_d  + 72 F_{H_d}  - 72 F_{H_u}  - 72 F_Q  + 144 F_u )+\\
&   G_b^2 G_t^2 \rho_1( 321 F_d  - 134 F_{H_d}  -36 F_{H_u}  - 240 F_Q  + 174 F_u ) +\\
&   \rho_\tau\left[
     360 F_d  + G_b^2( 60 F_d   - 120 F_Q )  + 360 F_d G_t^2  +
   G_b^2 G_t^2(50 F_d   + 20 F_{H_u}   - 100 F_Q   +20 F_u )+ \right.\\
&     \rho_1(72 F_d   - 12 F_{H_u}  -12 F_Q   + 24 F_u)   +  G_b^2 \rho_1 ( 11 F_d -22 F_Q)    +\\
&   \left.  G_t^2 \rho_1( 72 F_d   - 12 F_{H_u}   -12 F_Q   + 24 F_u )+
     G_b^2 G_t^2 \rho_1(9 F_d   + 4 F_{H_u}  - 18 F_Q   + 3 F_u )
    \right],
\end{split}
\]

\[
\begin{split}
c_{Hu{\rm Num}}=&
3240 F_{H_u} + 3240 F_{H_u} G_b^2 + G_t^2( 1620 F_{H_u}  - 1620 F_Q  - 1620 F_u )+\\
&    G_b^2 G_t^2( 270 F_d  + 270 F_{H_d}  + 1530 F_{H_u}  - 1350 F_Q  - 1620 F_u )+ \\
&    \rho_1(- 162 F_d  + 162 F_{H_d}  + 756 F_{H_u}  - 162 F_Q  + 324 F_u )+\\
&    G_b^2 \rho_1(- 162 F_d  + 162 F_{H_d}  + 756 F_{H_u}  - 162 F_Q  + 324 F_u )+\\
&   G_t^2 \rho_1(-162 F_d  + 162 F_{H_d}  + 297 F_{H_u}  - 621 F_Q  - 135 F_u )+\\
&   G_b^2 G_t^2 \rho_1 (- 81 F_d  + 234 F_{H_d}  + 276 F_{H_u}  - 540 F_Q  - 144 F_u ) +\\
&  \rho_\tau \left[
  540 F_{H_u}  + 270 F_{H_u} G_b^2   + G_t^2(270 F_{H_u}   - 270 F_Q    - 270 F_u )+ \right.\\
&       G_b^2 G_t^2( 45 F_d    + 120 F_{H_u}    - 90 F_Q    - 135 F_u )+
   \rho_1(-27 F_d    + 99 F_{H_u}    - 27 F_Q    + 54 F_u )   +\\
&   G_b^2 \rho_1( 36 F_{H_u}    + 27 F_u    - 27 F_d )   + G_t^2 \rho_1( 36 F_{H_u}    - 90 F_Q    - 9 F_u )   +\\
&  \left.  G_b^2 G_t^2 \rho_1( 9 F_d    + 4 F_{H_u}    - 18 F_Q    + 3 F_u )
\right],
\end{split}
\]

\[
\begin{split}
c_{Hd{\rm Num}}=&
2160 F_{H_d}+ G_b^2 (- 1080 F_d + 1080 F_{H_d}  - 1080 F_Q ) + 2160 F_{H_d} G_t^2+\qquad\qquad\qquad\qquad\\
&    G_b^2 G_t^2(- 1080 F_d  + 1020 F_{H_d}  +180 F_{H_u}  - 900 F_Q  + 180 F_u )+\\
&    \rho_1( 108 F_d  +504 F_{H_d} + 108 F_{H_u}  + 108 F_Q  - 216 F_u )+\\
&    G_b^2 \rho_1(- 198 F_d  +198 F_{H_d}  + 108 F_{H_u}  - 198 F_Q  - 216 F_u ) +\\
&   G_t^2 \rho_(108 F_d 1 + 504 F_{H_d}  + 108 F_{H_u}  + 108 F_Q  -216 F_u )+\\
&    G_b^2 G_t^2 \rho_1(- 201 F_d  + 184 F_{H_d}  +156 F_{H_u}  - 150 F_Q  - 159 F_u )
\end{split}
\]
and
\[
\begin{split}
F_Q &= 2 c_t^2 G_t^2 + 2 c_b^2 G_b^2 - \frac{32}{3} - 6 \rho_2^3 - \frac{2}{15} \rho_1^3,\\
F_u &= 4 c_t^2 G_t^2 - \frac{32}{3} - \frac{32}{15} \rho_1^3,\\
F_d &= 4 c_b^2 G_b^2 - \frac{32}{3} - \frac{8}{15} \rho_1^3,\\
F_{H_u}& = 6 c_t^2 G_t^2 - 6 \rho_2^3 - \frac{6}{5} \rho_1^3,\\
F_{H_d}& = 6 c_b^2 G_b^2 + 2 \rho_{h_\tau}^2 - 6 \rho_2^3 - \frac{6}{5} \rho_1^3,
\end{split}
\]
while $G_{t,b}^2$, $\rho_{1,2,\tau}$ and $\rho_{h_\tau}$ has been defined in Eqs.(\ref{Gt2_Gb2},\ref{r1_r2_rtau},\ref{rhtau})
respectively.
For our completely reduced system, i.e. $g_3,Y_t,Y_b,h_t,h_b$, the coefficients of the soft masses
become
\[
c_Q=c_u=c_d=\frac 23,\quad c_{H_u}=c_{H_d}=-1/3,
\]
obeying the celebrated sum rules
\[
\frac{m_Q^2+m_u^2+m_{H_u}^2}{M_3^2}=c_Q+c_u+c_{H_u}=1,\qquad
\frac{m_Q^2+m_d^2+m_{H_d}^2}{M_3^2}=c_Q+c_d+c_{H_d}=1.
\]

\smallskip

The selection of free parameters in this model, which tightly connected
to the prediction of the fermion masses, will be discussed in
Sect.~\ref{sec:para-red}. Subsequently, the corresponding phenomenological
implications of the quark mass predictions are analyzed in
\ref{sec:mf-red}.


\section{Comments on the Gauge-Yukawa unification}
\textit{Dimensionless sector}.
As has been already noted a natural extension of the GUT idea is to find a way to relate the gauge and Yukawa
sectors of a theory, that is to achieve GYU. A symmetry which naturally relates the two sectors is
supersymmetry, in particular $N = 2$ supersymmetry \cite{Fayet:1978ig}. However,
$N = 2$ supersymmetric theories have serious phenomenological problems due to light mirror fermions.
Also in superstring theories and in composite models there exist relations among the gauge and Yukawa
couplings, but both kind of theories have phenomenological problems, which we are not going to address here.

There have been other attempts in the past to relate the gauge and Yukawa sectors in the perturbative
renormalizable framework of the SM and MSSM which we recall and update for completeness here.
One was proposed by Decker, Pestieau \cite{Decker:1979nk} and Veltman \cite{Veltman:1980mj}.
By requiring the absence of quadratic divergencies in the SM,
they found a relationship among the squared masses appearing in the Yukawa and in the gauge sectors of the theory.
A very similar relation is obtained  by applying naively in the SM the general formula derived from demanding
spontaneous supersymmetry breaking via F-terms. In both cases a prediction for the top quark was possible only
when it was permitted experimentally to assume the $M_H  \ll M_{W,Z}$ with the result $M_t \simeq 69$ GeV
\cite{Veltman:1980mj}.
Otherwise there is only a quadratic relation among $M_t$ and $M_H$. Taking this relationship in the former case
and a version of naturalness into account, i.e. that the quadratic corrections to the Higgs mass be at most equal
to the physical mass, the Higgs mass is found to be
$\sim 260$ GeV, for a top quark mass of around 176 GeV, in complete disagreement with the recent findings at
LHC \cite{Aad:2012tfa,ATLAS:2013mma,Chatrchyan:2012ufa,Chatrchyan:2013lba}.

 Another well known relation among gauge and Yukawa couplings is the
Pendleton-Ross (P-R) infrared fixed point \cite{Pendleton:1980as}. The P-R proposal, involving the Yukawa coupling
of the top quark $g_t$ and the strong gauge coupling $\alpha_3$, was that the ratio
$\alpha_t/\alpha_3$, where $\alpha_t = g_t^2/4\pi$, has an
infrared fixed point.
This assumption predicted $M_t \sim 100$ GeV. In addition, it has been shown \cite{Zimmermann:1992eg}
that the P-R
conjecture is not justified at two-loops, since the ratio $\alpha_t/\alpha_3$ diverges in the infrared.
Another interesting conjecture, made by Hill \cite{Hill:1980sq,Bardeen:1989ds}, is that $\alpha_t$ itself
develops a quasi-infrared fixed point, leading to the prediction
$M_t \sim 280$ GeV.
 The P-R and Hill conjectures have been done in the framework of the SM.
The same conjectures within the MSSM lead to the following relations
(see also ref.~\cite{Bardeen:1993rv}):
\[
   M_t \approx 140 \gev \sin\beta\quad \textrm{(P-R)},\quad M_t \approx 200 \gev \sin\beta\quad \textrm{(Hill)},
\]
where $\tan\beta = v_u/v_d$ is the ratio of the two vacuum expectation values (vev's) of the Higgs fields of the MSSM. From theoretical considerations one can expect
\[
    1 < \tan\beta < 50  \Leftrightarrow 1/\sqrt{2} < \sin\beta < 1.
\]
This corresponds to
\[
 100 ~\gev < M_t < 140 \gev ~ \textrm{ (P-R)}, \quad 140 \gev < M_t < 200 \gev \textrm{ (Hill)}.
\]

Thus, the MSSM P-R conjecture is ruled out, while within the MSSM,
the Hill conjecture predicts a well defined range for $M_t$,
since the value of $\sin\beta$ is not fixed by other considerations.
The Hill model can accommodate the correct value of $M_t \sim 173$ GeV for $\sin\beta \approx 0.865$
corresponding to $\tan\beta \approx 1.7$. Such small values, however, are strongly
challenged if the newly discovered Higgs particle is identified with the lightest MSSM Higgs boson
\cite{Heinemeyer:2011aa,Capdevila:2017bsm,bsgth,HFAG}.
Only a very heavy scalar top spectrum with large mixing could accommodate such a small $\tan\beta$ value.

 \begin{figure}[t!]
  \begin{center}
\includegraphics[scale=0.45,angle=-90]{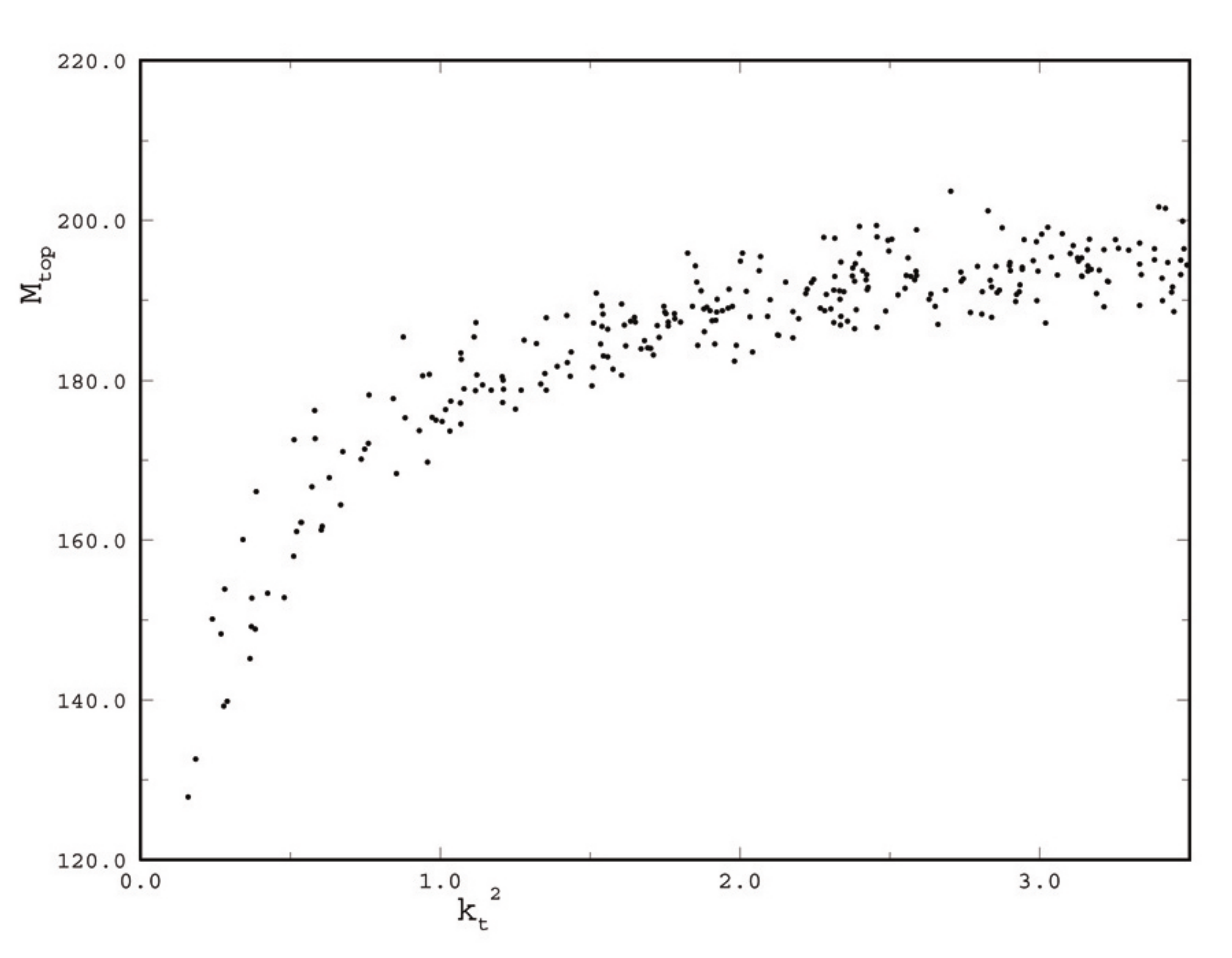}
  \caption[\ ]{The dependence of the top mass $M_t$ with $\kappa^2_t$, at
    fixed $M_{\rm SUSY}=500$ GeV.
    As we can see, after $\kappa^2_t\sim 2.0$ the top mass goes to its infrared
    fixed point value.
    Taken from~\cite{Kubo:1997fi}.}
  \label{fig:m_tVSk_t_2}
\end{center}
\end{figure}

In the GUT models examined in this chapter following the philosophy of reduction of couplings
a general consequence concerning GYU is that in the lowest order in perturbation theory the
gauge and Yukawa couplings at $M_{GUT}$ are related in the form
\begin{equation}
	g_i = \kappa_i g_{GUT}, \hspace{3mm}i = 1, 2, 3, e, ..., \tau, b, t,
\label{eq:fut-cond}
\end{equation}
where $g_i$ (i = 1,...,t) stand for the gauge and Yukawa couplings, $g_{GUT}$ is
the unified coupling and we have neglected the Cabbibo-Kobayashi-Maskawa
mixing of the quarks.
Thus,  Eq.(\ref{eq:fut-cond})
corresponds to a set of boundary conditions on the renormalization group evolution for the
theory below $M_{GUT}$, which in all cases is the MSSM.
As we have seen in the previous sections it is possible to obtain GYU in the third generation
that can predict the bottom and top quark masses in accordance with the experimental data
in certain cases.
This means that the top-bottom hierarchy could be explained in the successful models,
in a similar way as the hierarchy of the gauge couplings of the SM can be explained if
one assumes the existence of a unifying gauge symmetry at $M_{GUT}$.
It is clear that the GYU scenario based on the reduction of couplings in the
dimensionaless sector of the theory is the most predictive scheme as far as the
mass of the top quark is concerned.
It may be worth recalling the
predictions for $M_t$ of ordinary GUTs, in particular of supersymmetric
$SU(5)$ and $SO(10)$. The MSSM with $SU(5)$ Yukawa boundary unification allows
$M_t$ to be anywhere in the interval between 100-200 GeV for varying $\tan\beta$,
which is  a free parameter. Similarly, the MSSM with $SO(10)$ Yukawa
boundary conditions, i.e. $t-b -\tau$  Yukawa Unification, gives $M_t$
in the interval 160-200 GeV.
In addition in Ref.~\cite{Kubo:1995cg} we have analyzed the infrared quasi-fixed-point
behaviour of the $M_t$ prediction  in various models in some detail.
In particular it was found that the infrared value
exhibits a stronger dependance on $\tan\beta$ with increasing $\tan\beta$, and
its lowest value is $\sim 188$ GeV.

This is demonstrated in Fig.~\ref{fig:m_tVSk_t_2}, where the top quark mass prediction is
shown as a function of $\kappa_t^2$,  see Eq.~\ref{eq:fut-cond}.
Comparing the infra-red fixed point value, reached for large $\kappa_t^2$, with the experimental value
$m_t=(173.2\pm 0.9)$ GeV \cite{Lancaster:2011wr}  one can conclude
that the present data on $M_t$
cannot be explained from the infrared quasi-fixed-point behaviour alone
(see Fig.~\ref{fig:m_tVSk_t_2}) .
An estimate of the theoretical uncertainties involved in GYU has been done in ref.~\cite{Kubo:1995cg}.
Although a fresh look  is in order in the case of the minimal $N = $1 supersymmetric $SU(5)$, we can conclude
that the studies on the GYU of the asymptotically non-free supersymmetric Pati-Salam \cite{Kubo:1994xa} and
asymptotically non-free $SO(10)$ \cite{Kubo:1995zg} models have ruled them out on the basis of the top quark mass prediction.

It should be emphasized once more that only one of the Finite Unified models
(discussed in Sect.~\ref{sec:futB} and which will be further discussed  in Sect.~\ref{sec:num-fut})
not only predicted correctly the top and bottom quark masses but in addition predicted the Higgs mass in
striking agreement with the recent findings
at LHC \cite{Aad:2012tfa,ATLAS:2013mma,Chatrchyan:2012ufa,Chatrchyan:2013lba}.

\noindent
\textit{Dimensionful sector}.
As we have seen in Chapter~\ref{ch:theory} in the dimensionful sector of a reduced
$N = 1$ supersymmetric theory and in the lowest in perturbation theory the dimensionless
and dimensionful parameters, defined in Eqs.~(\ref{supot}) and (\ref{supot_l}) are related as follows
\begin{align}
h^{ijk} = - M C^{ijk},\label{h_C_rel}\\
\left(m^2_i + m^2_j + m^2_k \right)/M M^\dag = 1,\label{sum_2}
\end{align}
resulting from Eqs.~(\ref{h2}) and (\ref{sum2}) respectively.
We also recall that the sum rule was introduced  in order to overcome the problems introduced
by the universal relation and the scalar and gaugino masses in finite models. The sum rule obviously
enlarge the parameter space to overcome the problems, but in the successful $SU(5)$ FUT adds only one free parameter.

We would like to note though that in other models with reduced couplings, such as in the minimal
supersymmetric $SU(5)$ discussed in Sect.~\ref{sec:su5} and in the MSSM with reduced couplings discussed in Sect.~\ref{sec:mssm},
although the sum rule of Eq.~(\ref{sum_2}) is satisfied, there exist in addition exact relations among the scalar and gaugino masses;
see Eqs.~({\ref{red_sol_1},\ref{red_sol},\ref{mM_rel}).
This is not the case in finite theories \cite{Kawamura:1997cw}.

Therefore in ordinary (non-finite) theories the
reduction of couplings leads to exact relations among couplings as in the dimensionless sector. 

%% file: PR2018_Chapter_6.tex
\chapter{Low Energy Phenomenology of the Finite Unified Model and the Reduced MSSM}
\label{chap:FUT_LOW}

In this chapter we confront the Finite Unified model and the reduced
MSSM with current phenomenological constraints. We review how the
experimentally favoured parameter space can be tested with current and
future accelerator experiments.


\section{Phenomenological Constraints}
\label{sec:pheno-constraints}

Here we outline the various constraints that are taken into account in
our phenomenological analysis. We first consider four types of flavour
constraints, in which SUSY is known to have significant impact%
\footnote{
Over the past years several ``flavor anomalies'' appeared. The most
significant ones are given by the measurements of
$R(K^{(*)} = \br(B \to K^{(*)}\mu^+\mu^-)/\br(B \to K^{(*)}e^+e^-))$ and
$R(D^{(*)} = \br(B \to D^{(*)}\tau\nu)/\br(B \to D^{(*)}\mu\nu))$ as well
as the measurement of $P'_5$ capturing the momentum dependance of the
$B \to K^*\mu^+\mu^-$ decay~\cite{PDG18}.
While (a combination of) these anomalies may turn out to be
significant (see, e.g., \citere{Capdevila:2017bsm}),
our models do not provide any solution to them. Consequently, they do
not present an additional constraint on our preferred parameter space.
}%
. Specifically, we consider the
flavour observables $\br(b \to s \ga)$, $\br(B_s \to \mu^+ \mu^-)$, $\br(B_u
\to \tau \nu)$ and $\Delta M_{B_s}$.
It should be noted that for this review we have not used the latest
experimental and theoretical values. However, this has a minor impact on
the presented results. The uncertainties below are the linear
combination of the experimental~error and twice the theoretical
uncertainty in the~MSSM. The constraints are:
\begin{itemize}
\item
For the branching ratio $\br(b \to s \gamma)$ we take a value
from the Heavy Flavor Averaging Group (HFAG)~\cite{bsgth,HFAG}:
\beq
\frac{\br(b \to s \gamma )^{\rm exp}}{\br(b \to s \gamma )^{\rm SM}} = 1.089 \pm 0.27~.
\label{bsgaexp}
\eeq

\item
For the branching ratio $\br(B_s \to \mu^+ \mu^-)$ we use a combination of CMS and LHCb data~\cite{Bobeth:2013uxa,RmmMFV,Aaij:2012nna,CMSBsmm,BsmmComb}:
\beq
\br(B_s \to \mu^+ \mu^-) = (2.9\pm1.4) \times 10^{-9}~.
\eeq

\item
For the $B_u$ decay to $\tau\nu$ we use the limit~\cite{SuFla,HFAG,PDG14}:
\beq
\frac{\br(B_u\to\tau\nu)^{\rm exp}}{\br(B_u\to\tau\nu)^{\rm SM}}=1.39\pm 0.69~.
\eeq

\item
For $\Delta M_{B_s}$  we use~\cite{Buras:2000qz,Aaij:2013mpa}:
\beq
\frac{\Delta M_{B_s}^{\rm exp}}{\Delta M_{B_s}^{\rm SM}}=0.97\pm 0.2~.
\eeq
\end{itemize}

Since the quartic couplings in the Higgs potential are given by the SM gauge
couplings, the~lightest Higgs boson mass is not a free~parameter,
but rather predicted in terms of other parameters. Higher-order corrections are crucial for a precise prediction of $M_h$; see~\citeres{habilSH,awb2,PomssmRep} for reviews.

The discovery of a Higgs-like particle at ATLAS and CMS in July 2012
\cite{Aad:2012tfa,Chatrchyan:2012xdj} can be~interpreted as the discovery of the light $\cal
CP$-even Higgs boson of the~MSSM Higgs spectrum~\cite{Mh125,hifi,hifi2}. The
experimental average for the~(SM) Higgs boson mass obtained at the LHC
Run~I is given by
\cite{Aad:2015zhl}
\beq
M_H^{\rm exp}=125.1\pm 0.3~{\rm GeV}~.
\eeq
More recent Run~II measurements confirm this measurement. The
uncertainty, however is dominated by the theoretical accuracy for the
prediction of $\Mh$ in the MSSM, which was estimated to be at the level of
$3 \gev$~\cite{Degrassi:2002fi,Buchmueller:2013psa,BHHW}. It should be
noted that this estimate is only valid if the most accurate prediction
of $\Mh$ is employed.
For the following phenomenological analyses the code
{\tt FeynHiggs}~\cite{Degrassi:2002fi,BHHW,FeynHiggs} (Version 2.14.0~beta)
was used to predict the light Higgs mass. The~evaluation
of the Higgs masses with {\tt FeynHiggs} is based on the combination of a
fixed order diagrammatic calculation and a resummation of
the~(sub)leading logarithmic contributions at all orders of perturbation
theory. This~combination ensures a reliable
evaluation of $M_h$ also for large SUSY scales. Refinements in the
combination~of the fixed order log resummed calculation have been included
w.r.t.\ previous~versions~\cite{BHHW}. They resulted in a
more precise $M_h$ evaluation for high supersymmetric mass scales and also in
a downward shift of $M_h$ at the level of ${\cal O}(2~{\rm GeV})$ for
large SUSY masses. For our analyses we used two estimates for the theory
uncertainty of $3 (2) \gev$. The a total uncertainty for
$\Mh$, combined of the experimental and the theoretical uncertainty,
is then given by
\beq
M_h=125.1\pm 3.1~(2.1)~{\rm GeV}~.
\eeq

We finally briefly comment on possible Cold Dark Matter (CDM) constraints.
Since it is well known that the lightest neutralino, being the Lightest
SUSY Particle (LSP), is an
excellent candidate for CDM~\cite{EHNOS},
one can in principle demand that the lightest neutralino is
indeed the LSP and parameters leading to a different LSP could be discarded.
The current bound, favoured by a joint analysis of WMAP/Planck and other
astrophysical and cosmological data, is at
$2\,\sigma$~level given by~\cite{Komatsu:2010fb,Komatsu:2014ioa}
\beq
\Omega_{\rm CDM} h^2 = 0.1120 \pm 0.0112~.
\label{cdmexp}
\eeq
However, in the analyzed parameter space the relic abundance turns out
to be too high in comparison with \refeq{cdmexp}.
Consequently, on a more general basis a mechanism is needed in our models to
reduce the CDM abundance in the early universe.  This issue could, for
instance, be related to another problem, that of neutrino masses.
Within the FUTs this type of masses cannot be generated naturally,
although a non-zero value for neutrino masses has clearly been
established~\cite{PDG18}. However, the FUTs discussed here
can, in principle, be easily extended by introducing bilinear R-parity
violating terms that preserve finiteness and introduce the desired
neutrino masses~\cite{Valle:1998bs}.
More generally, $R$-parity violation~\cite{herbi}
would have a small impact on the collider phenomenology presented here
(apart from fact the SUSY search strategies could not rely on a
`missing energy' signature), but remove the CDM bound of
\refeq{cdmexp} completely.
Consequently, \refeq{cdmexp} was not taken into account in the analyses
presented below.

Finally, we comment on the anomalous magnetic moment of the muon, $(g-2)_\mu$
(with $a_\mu \equiv (g-2)_\mu/2$). As will be shown in the numerical
analysis, the resulting SUSY spectra are relatively large.
Consequently (despite the large values of $\tb$, see below)
the models gives only a negligible correction to the SM prediction.
The comparison of the experimental result and the SM value shows a
deviation of
$\sim 3.5\,\sig$~\cite{newBNL,g-2,Jegerlehner:2011ti,Benayoun:2012wc}.
Consequently, since the results would be very close to the SM results, the
models have the same level of difficulty with the $a_\mu$ measurement as
the SM.


\section{Numerical Analysis of the FUT}
\label{sec:num-fut}


\subsection{FUT Predictions for Future Colliders}
\label{sec:future-fut}

As was discussed in \refse{sec:fut}, the experimental bounds on the $\mb(M_Z)$
and the $\mt$ mass clearly single out model~{\bf B} with
$\mu <0$ as~the only solution compatible with these constraints, which will
simply be called {\bf FUT} below.

\begin{figure}[htb!]
\centering
\includegraphics[width=0.7\textwidth]{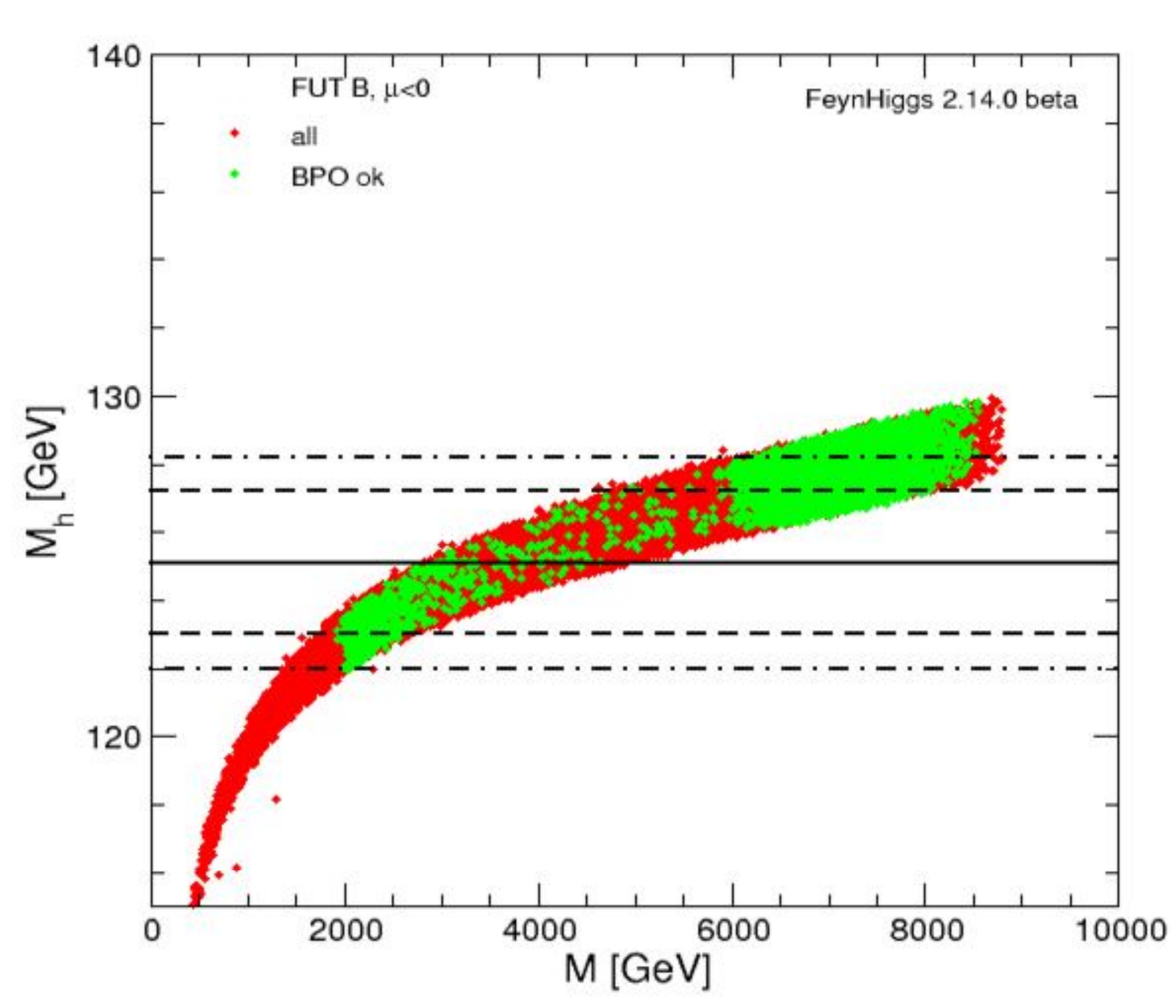}
\caption{\textit{The lightest Higgs boson mass, $M_h$, as~a
    function of $M$ for the choice $\mu<0$. The~green points are the
    ones that satisfy the $B$-physics~constraints. Taken from
    \citere{Heinemeyer:2018zpw}.}}
\label{fig:MhiggsvsM}
\end{figure}

The~prediction for $M_h$ of {\bf FUT} with $\mu<0$
is shown in \reffi{fig:MhiggsvsM} (as presented in \citere{Heinemeyer:2018zpw})
in a range for the unified gaugino mass  0.5~TeV $\lesssim$ $M$
$\lesssim$ 9~TeV. The~green
points satisfy the $B$-physics constraints as well, as discussed in
\refse{sec:pheno-constraints}. Here it should be kept in
mind that these predictions are subject to a theory uncertainty of
3~(2)~GeV~\cite{Degrassi:2002fi}.
Older analyses, including in particular less refined evaluations of the light
Higgs mass, are given in
\citeres{Heinemeyer:2012yj,Heinemeyer:2012ai,Heinemeyer:2013fga}.
However, since relatively heavy SUSY masses are favoured (see below)
these less refined evaluations cannot be considered as reliable.

\begin{figure}[htb!]
\centering
\includegraphics[width=0.48\textwidth]{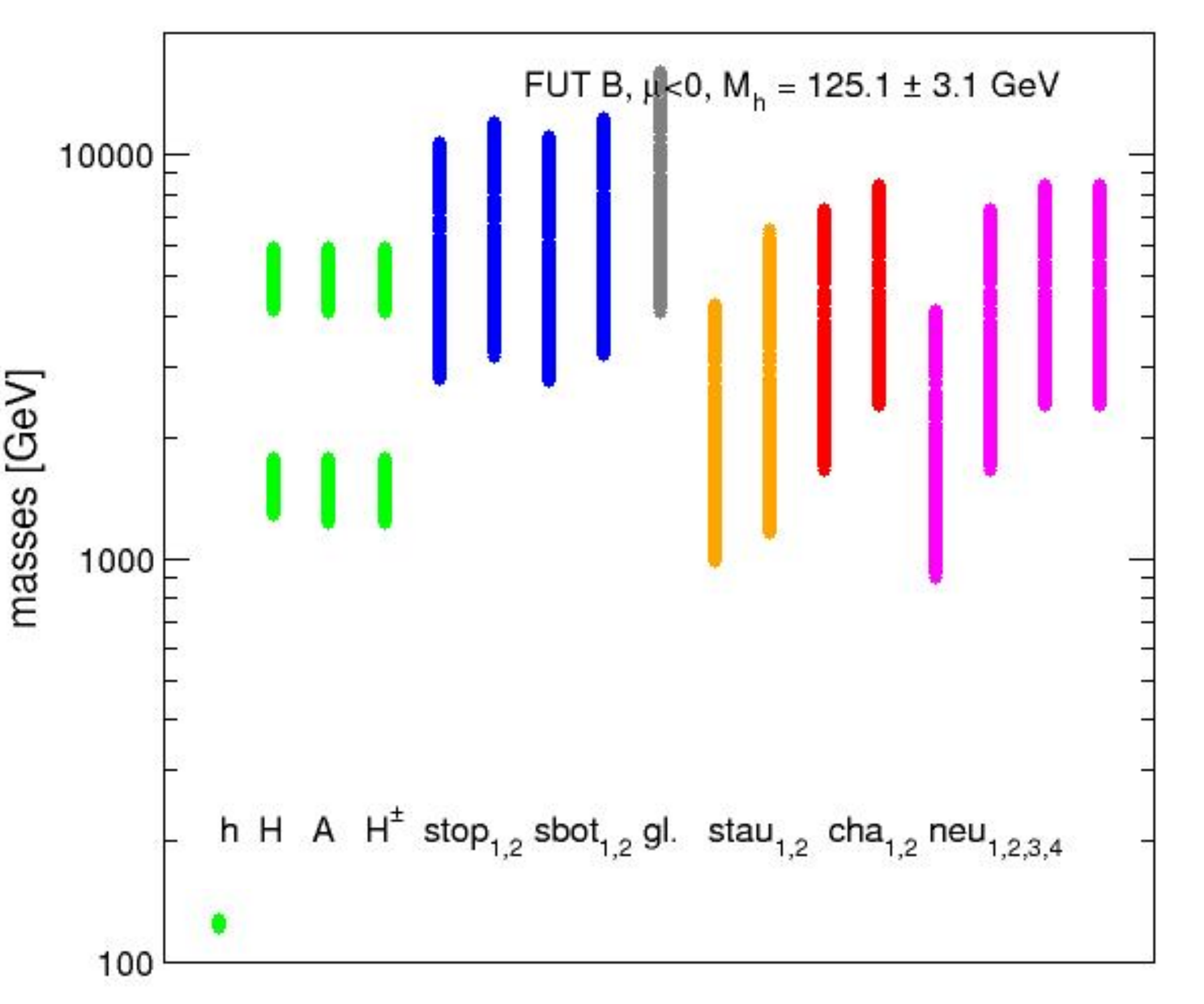}
\includegraphics[width=0.48\textwidth]{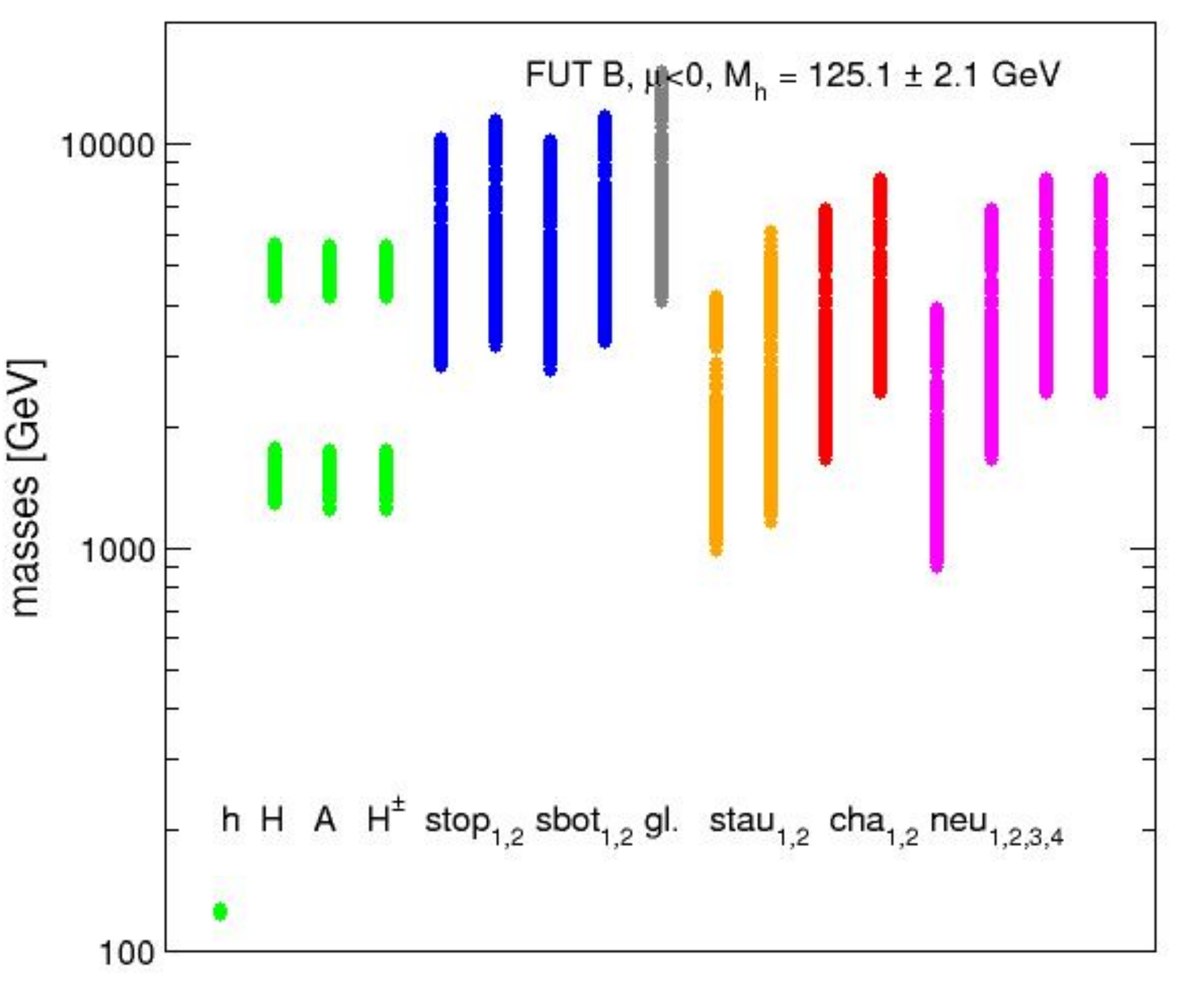}
\caption{\textit{The ({left},{right}) plots show the spectrum of
    the {\bf FUT} model after imposing the constraint
    $M_h=125.1\pm3.1(2.1)~{\rm GeV}$. The~light (green) points are the
    various Higgs boson masses; the~dark (blue) points following are the
    two scalar top and bottom masses; the~gray ones are the gluino
    masses; then come the scalar tau masses in orange (light gray);
    the~darker (red) points to the right are the two chargino masses;
    followed by the lighter shaded (pink) points indicating the
    neutralino masses. Taken from \citere{Heinemeyer:2018zpw}.}}
\label{fig:susyspectrum}
\end{figure}

The allowed values of the lightest Higgs boson mass limit  the allowed
supersymmetric masses' values, as shown in
\reffi{fig:susyspectrum}~\cite{Heinemeyer:2018zpw}. In~the left (right)
plot we impose $\Mh = 125.1\pm3.1~(2.1)~{\rm GeV}$. In
particular, very heavy coloured SUSY particles are favoured (nearly
independent of the $\Mh$ uncertainty), in~agreement with
the non-observation of those~particles at the LHC \cite{2018:59}.
The only part that can be tested at the (HL-)LHC is the lower range of
the neutral Higgs spectrum. For the $\tb$ values favoured by our
analysis, values up to $2 \tev$ are projected to be in the range of the
ATLAS/CMS searches via $pp \to H/A \to \tau^+\tau^-$~\cite{HAtautau-HL-LHC},
which would cover the lower part of the spectrum.
On the other hand, the~allowed coloured supersymmetric masses will remain
unobservable at the (HL-)LHC, the~ILC or CLIC. The lower part of the
electroweak spectrum could be accessible at CLIC with $\sqrt{s} = 3
\tev$ The coloured spectrum would~be accessible, however, at the FCC-hh
\cite{fcc-hh}, as would be the full heavy Higgs spectrum.

In \refta{tab:spectrum-fut} two example spectra of {\bf FUT} are shown,
which span~the mass range of the parameter space that is in~agreement with the
$B$-physics observables and the lightest Higgs boson mass
measurement. We show the lightest and the heaviest~spectrum (based on
$\mneu1$)  for $\de\Mh=2.1$ and $\de\Mh=3.1$.
The Higgs boson masses are denoted as $\Mh$, $\MH$, $\MA$ and
$\MHp$. $m_{\tilde{t}_{1,2}}$, $m_{\tilde{t}_{1,2}}$, $\mgl$ and
$m_{\tilde{\tau}_{1,2}}$, are the scalar top, bottom, gluino and tau
masses, respectively. $\mcha{1,2}$ and $\mneu{1,2,3,4}$ stand for
chargino and neutralino masses, respectively.
As discussed above, only the neutral Higgs spectrum of the ``light
spectrum'' is in the range of the (HL-)LHC. Scalar taus as well as the
two lighter neutralinos would be accessible at CLIC. For the ``heavy
spectrum'' only the FCC-hh could test it.

\begin{table}[htb!]
\renewcommand{\arraystretch}{1.5}
\centering
\begin{tabular}{|c|rrrrrrrrr|}
\hline
$\de\Mh=2.1$ & $\Mh$ & $\MH$ & $\MA$ & $\MHp$ & $m_{\tilde{t}_1}$ & $m_{\tilde{t}_2}$ &
  $m_{\tilde{b}_1}$& $m_{\tilde{b}_2}$ & $\mgl$ \\
\hline
lightest & 123.1 & 1533 & 1528 & 1527 & 2800 & 3161 & 2745 & 3219 & 4077 \\
heaviest & 127.2 & 4765 & 4737 & 4726 & 10328 & 11569 & 10243 & 11808 & 15268 \\

\hline

\hline
 & $m_{\tilde{\tau}_1}$ & $m_{\tilde{\tau}_2}$ &
  $\mcha1$ & $\mcha2$ & $\mneu1$ & $\mneu2$ & $\mneu3$ & $\mneu4$ & $\tb$ \\
\hline
lightest & 983 & 1163 & 1650 & 2414 & 900 & 1650 & 2410 & 2414 & 45 \\
heaviest & 4070 & 5141 & 6927 & 8237 & 3920 & 6927 & 8235 & 8237 & 46 \\

\hline
\end{tabular}

\vspace{1em}
\begin{tabular}{|c|rrrrrrrrr|}
\hline
$\de\Mh=3.1$ & $\Mh$ & $\MH$ & $\MA$ & $\MHp$ & $m_{\tilde{t}_1}$ & $m_{\tilde{t}_2}$ &
  $m_{\tilde{b}_1}$& $m_{\tilde{b}_2}$ & $\mgl$ \\
\hline
lightest & 122.8 & 1497 & 1491 & 1490 & 2795 & 3153 & 2747 & 3211 & 4070 \\
heaviest & 127.9 & 4147 & 4113 & 4103 & 10734 & 12049 & 11077 & 12296 & 16046 \\

\hline

\hline
 & $m_{\tilde{\tau}_1}$ & $m_{\tilde{\tau}_2}$ &
  $\mcha1$ & $\mcha2$ & $\mneu1$ & $\mneu2$ & $\mneu3$ & $\mneu4$ & $\tb$ \\
\hline
lightest & 1001 & 1172 & 1647 & 2399 & 899 & 647 & 2395 & 2399 & 44 \\
heaviest & 4039 & 6085 & 7300 & 8409 & 4136 & 7300 & 8406 & 8409 & 45 \\
\hline
\end{tabular}

\caption{\textit{
Two example spectra of the  {\bf FUT} .
All masses are in GeV and rounded to 1 (0.1)~GeV (for the light Higgs mass).}}
\label{tab:spectrum-fut}
\renewcommand{\arraystretch}{1.0}
\end{table}


\subsection{FUT Conclusions}
\label{sec:concl-fut}

 One can see that the
predictions of {\bf FUT} are impressive.
But one could also add some comments on the
theoretical side. The developments on~treating the problem of
divergencies include string and~non-commutative theories, as~well as $N = 4$
supersymmetric theories~\cite{Mandelstam:1982cb,Brink:1982wv}, $N = 8$
supergravity~\cite{Bern:2009kd,Kallosh:2009jb,Bern:2007hh,Bern:2006kd,Green:2006yu}
and the AdS/CFT
correspondence~\cite{Maldacena:1997re}. It is interesting that the $N = 1$ FUT
discussed here includes ideas that have survived phenomenological and
theoretical tests, as well as the ultraviolet divergence problem and
solves it in a minimal way.

In our analysis of {\bf FUT} \cite{Heinemeyer:2018zpw} we included
restrictions of third generation quark masses and $B$-physics
observables and it proved consistent with all the
phenomenological constraints. Compared to~our previous analyses
\cite{Heinemeyer:2010xt,Heinemeyer:2012yj,Heinemeyer:2013nza,Heinemeyer:2012ai,Heinemeyer:2013fga,Heinemeyer:2018roq},
the improved~evaluation of $\Mh$ prefers a heavier (Higgs) spectrum and
thus allows only a heavy supersymmetric spectrum.
The coloured spectrum easily~escapes (HL-)LHC searches, but~can likely
be tested at the FCC-hh. However, the lower~part of the EW spectrum
could be observable at CLIC.


\section{Numerical Analysis of the Reduced MSSM}
\label{sec:num-red}
In this section we analyze the particle spectrum predicted by the
reduced MSSM~\cite{Heinemeyer:2017gsv}. We first discuss the selection of
free parameters, then
apply constraints from fermion masses. Subsequently we apply the
remaining constraints and discuss the observability at current and
future colliders.

\subsection{Free Parameters of the Reduced MSSM}
\label{sec:para-red}

So far the relations among reduced parameters in terms
of the fundamental ones derived in Sect.~\ref{sec:mssm}
had a part which was RGI and a another part
originating from the corrections, which are scale dependent. In the
analysis shown here we choose the unification scale to apply the corrections
to the RGI relations.
It should be noted that we are
assuming a covering GUT, and thus unification of the three
gauge couplings, as well as a unified gaugino mass
$M$  at that scale. Also to be noted is that in the dimensionless sector of the
theory, since $Y_\tau$ cannot be reduced in favour of the fundamental
parameter $\al_3$, the mass of the $\tau$ lepton is an input parameter and
consequently $\rho_\tau$, is an independent parameter too.  At low
energies, we fix the values of $\rho_{\tau}$ and $\tan\beta$ using
the mass of the tau lepton $m_{\tau}(M_Z)=1.7462$ GeV.
For each value of
$\rho_{\tau}$ there is a corresponding value of $\tan\beta$ that
gives the appropriate $m_{\tau}(M_Z)$.  Then we use the value found
for $\tan\beta$ together with $G_{t,b}$, as obtained from the
reduction equations and their respective corrections, to determine the top and bottom quark masses.  We
require that both the bottom and top masses are within 2$\sigma$ of
their experimental value, which singles out  large $\tan\beta$
values, $\tan\beta \sim 42 - 47$.
Correspondingly, in the dimensionful sector of the theory the $\rho_{h_\tau}$
is a free parameter, since $h_\tau$ cannot be reduced in favour of the
fundamental parameter $M$ (the unified gaugino mass scale).
  $\mu$ is a free parameter, as it cannot be reduced in favour of $M_3$
as discussed above. On the other hand $m_3^2$ could be reduced, but here
it is chosen to leave it free.
However, $\mu$ and $m_3^2$ are restricted from the requirement of
EWSB, and only $\mu$ is taken as an independent parameter.
Finally, the other parameter in the Higgs-boson sector, the $\cp$-odd
Higgs-boson mass $\MA$ is evaluated from $\mu$, as well as from $m_{H_u}^2$ and
$m_{H_d}^2$, which are obtained from the reduction equations.
In total we vary the parameters $\rho_\tau$, $\rho_{h_\tau}$, $M$
  and $\mu$.


\subsection{Constraints from Fermion Masses}
\label{sec:mf-red}

\begin{figure}[htb!]
\begin{center}
\includegraphics[width=0.45\textwidth,height=8cm]{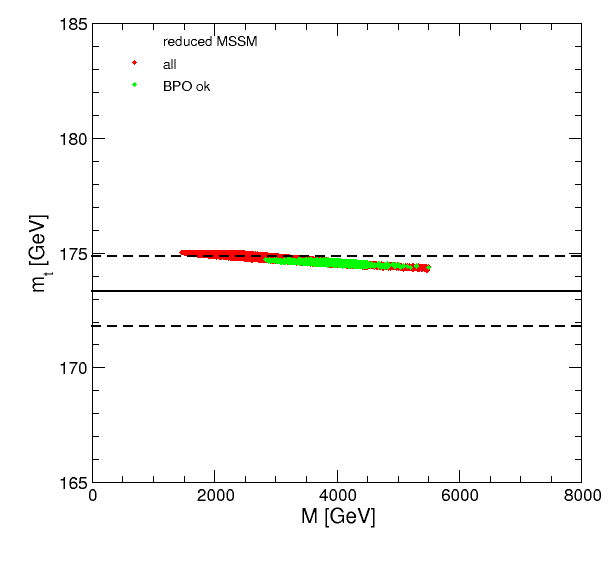}
\includegraphics[width=0.45\textwidth,height=8cm]{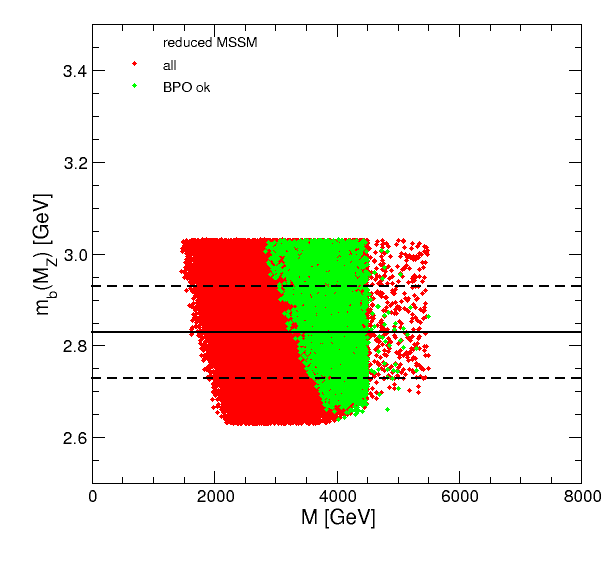}
\caption{\textit{The left (right) plot shows our results within the reduced
    MSSM for the top (bottom) quark mass. The horizontal lines indicate the
    experimental values as given in \refeq{mtmb}. Taken from
    \citere{Heinemeyer:2017gsv}.
}}
\label{fig:mf-red}
\end{center}
\vspace{-1em}
\end{figure}

The first step of the numerical analysis concerns the top and the bottom
quark masses.
As mentioned above, the variation of $\rho_\tau$ yields the
values of $\mt$ (the top pole mass) and $\mb(\MZ)$, the running
bottom quark mass at the $Z$~boson mass scale,  where
scan points which are not within $2\sigma$ of
the experimental data are neglected. This is shown in
\reffi{fig:mf-red}~\cite{Heinemeyer:2017gsv}. The experimental values
are indicated by the
horizontal lines and are taken to be~\cite{PDG14}
(the same comments on the experimental values as in Sect.~\ref{sec:mq} apply)
\be
\mt = 173.34 \pm 1.52 \gev~, \quad
\mb(\MZ) = 2.83 \pm 0.1 \gev~,
\label{mtmb}
\ee
with the uncertainties at the $2\,\sig$~level. One can see that the
scan yields many parameter points that
are in very good agreement with the experimental data.
At the same time also the flavor constraints, see
\refse{sec:pheno-constraints} are applied and shown as green dots. One
can see that they are in good agreement with the measurements of the
quark masses and give restrictions in the allowed ranges of~$M$
(the common gaugino mass at the unification scale).

\subsection{Predictions of the reduced MSSM for future colliders}
\label{sec:future-red}

\begin{figure}[htb!]
\begin{center}
\includegraphics[width=0.85\textwidth]{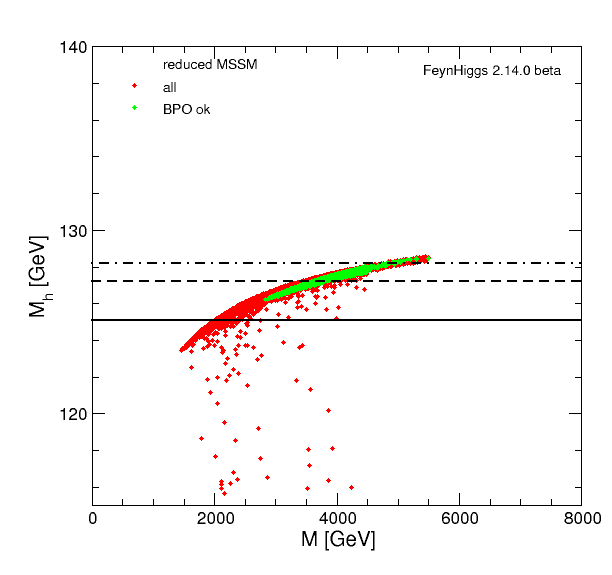}
\caption{\textit{The lightest Higgs boson mass, $\Mh$, as a function of $M$
(the common gaugino mass at the unification scale)
in the reduced MSSM. The red points is the full model prediction. The
  green points fulfill the $B$-physics constraints (see text).
Taken from \citere{Heinemeyer:2017gsv}.
}}
\label{fig:Mh-red}
\end{center}
\end{figure}

As the next step the lightest MSSM Higgs-boson mass is evaluated.
The prediction for $\Mh$ is shown in
\reffi{fig:Mh-red}~\cite{Heinemeyer:2017gsv} as a function of $M$  in the range
$1 \tev \lsim M \lsim 6 \tev$. The lightest Higgs mass ranges in
\be
\Mh \sim 124-129 \gev~ ,
\label{eq:Mhpred}
\ee
where we discard the ``spreaded'' points with possibly lower masses,
which result from a numerical instability in the Higgs-boson mass
calculation.
One should keep in mind that these predictions are subject to a theory
uncertainty of $3 (2) \gev$, see \refse{sec:pheno-constraints}.
The red points correspond to the full parameter scan, whereas the green
points are the subset that is in agreement with the $B$-physics observables
as discussed above (which do not exhibit any numerical instability). The
inclusion of the flavor observables shifts the lower bound for $\Mh$ up to
$\sim 126 \gev$.

The horizontal lines in \reffi{fig:Mh-red} show the central value of the
experimental measurement (solid), the $\pm 2.1 \gev$ uncertainty (dashed) and
the $\pm 3.1 \gev$ uncertainty (dot-dashed). The requirement to obtain a
light Higgs boson mass value in the correct range yields an upper limit
on $M$ of about $5\, (4) \tev$ for $\Mh = 125.1 \pm 2.1\, (3.1) \gev$.

\begin{figure}[htb!]
\begin{center}
\includegraphics[width=0.45\textwidth,height=8cm]{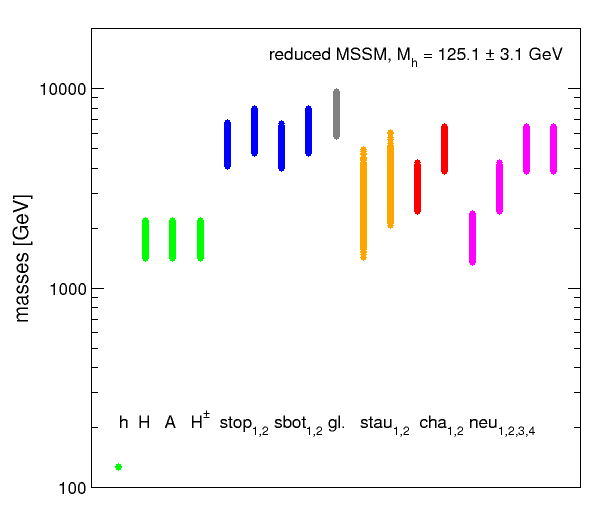}
\includegraphics[width=0.45\textwidth,height=8cm]{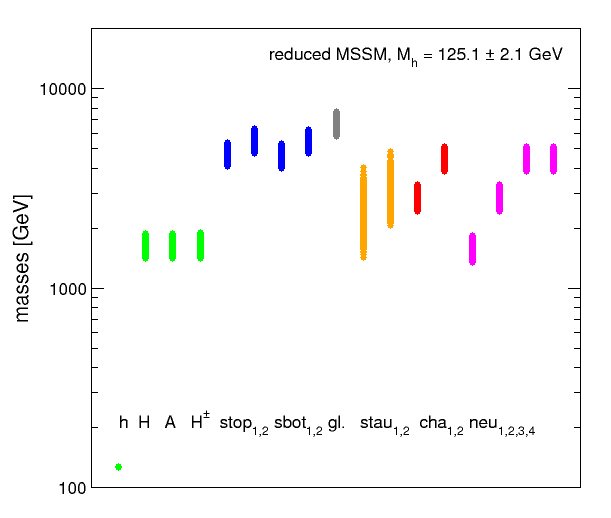}
\caption{\textit{The left (right) plot shows the spectrum of the reduced MSSM
after imposing the constraint
$\Mh = 125.1 \pm 3.1\,(2.1) \gev$.
The points shown are in agreement with the $B$-physics observables.
The light (green) points on the left are the various Higgs
boson masses. The dark (blue) points following are the
two scalar top and bottom masses, followed by the lighter
(gray) gluino mass. Next come the lighter (beige) scalar
tau masses. The darker (red) points to the right are the
two chargino masses followed by the lighter shaded (pink)
points indicating the neutralino masses.
}}
\label{fig:masses-red}
\end{center}
\end{figure}

Naturally the $\Mh$ limit also sets an upper limit on the low-energy
SUSY masses. This turns the reduced MSSM into a highly predictive and
testable theory. The full particle spectrum of the reduced MSSM
(where we restricted ourselves as before to the third generation of
sfermions)
compliant with the $B$-physics observables
is shown in \reffi{fig:masses-red}~\cite{Heinemeyer:2017gsv}.  In the
left (right) plot we impose
$\Mh = 125.1 \pm 3.1\, (2.1) \gev$.  Including
the Higgs mass constraints in general favours the somewhat higher part of the
SUSY particle mass spectra. The tighter $\Mh$ range cuts off the very
high SUSY mass scales.

The Higgs spectrum will be fully testable at the HL-LHC, which for $\tb
\gsim 40$ can explore masses up to $\sim 2 \tev$ via the channel
$pp \to H/A \to \tau^+\tau^-$~\cite{HAtautau-HL-LHC}. However, such
observations would be in agreement also with a pure 2HDM, and additional
observation of the SUSY particles will be necessary to analyze the
model.

The lighter SUSY particles are given by the electroweak spectrum, which
starts around $\sim 1.3 \tev$. They will mostly remain unobservable at the LHC
and at future $e^+e^-$ colliders such as the ILC or CLIC, with only the
very lower range mass range below $\sim 1.5 \tev$ might be observable at
CLIC (with $\sqrt{s} = 3 \tev$). The coloured mass spectrum starts at
around $\sim 4 \tev$, which will remain unobservable at the
(HL-)LHC. However, the coloured spectrum would be accessible at the
FCC-hh~\cite{fcc-hh}. This collider could definitely confirm the SUSY
spectrum of the reduced MSSM or rule out this model.

\begin{table}[t!]
\renewcommand{\arraystretch}{1.5}
\centering
\begin{tabular}{|c|rrrrrrrrr|}
\hline
& $\Mh$ & $\MH$ & $\MA$ & $\MHp$ & $\mstop1$ & $\mstop2$ &
  $\msbot1$& $\msbot2$ & $\mgl$ \\
\hline
light        & 126.2 & 1433 & 1433 & 1446 & 4052 & 4736 & 3989 & 4723 & 5789 \\
$\de\Mh=2.1$ & 127.2 & 1570 & 1570 & 1572 & 5361 & 6289 & 5282 & 6279 & 7699 \\
$\de\Mh=3.1$ & 128.1 & 1886 & 1886 & 1888 & 6762 & 7951 & 6653 & 7943 & 9683 \\
\hline
\end{tabular}

\vspace{1em}
\begin{tabular}{|c|rrrrrrrrr|}
\hline
& $\mstau1$ & $\mstau2$ &
  $\mcha1$ & $\mcha2$ & $\mneu1$ & $\mneu2$ & $\mneu3$ & $\mneu4$ & $\tb$ \\
\hline
light        & 1906 & 2066 & 2430 & 3867 & 1339 & 2430 & 3864 & 3866 & 42.6 \\
$\de\Mh=2.1$ & 1937 & 2531 & 3299 & 5166 & 1833 & 3299 & 5114 & 5116 & 43.1 \\
$\de\Mh=3.1$ & 3153 & 3490 & 4248 & 6464 & 2376 & 4248 & 6462 & 6464 & 45.2 \\
\hline
\end{tabular}
\caption{\textit{
Three example spectra of the reduced MSSM. ``light'' has the smallest $\neu1$
in our sample, ``$\de\Mh = 2.1 (3.1)$'' has the largest $\mneu1$ for
$\Mh \le 125.1 + 2.1 (3.1) \gev$.
All masses are in GeV and rounded to 1 (0.1)~GeV (for the light Higgs mass).
}}
\label{tab:spectrum-red}
\renewcommand{\arraystretch}{1.0}
\end{table}

\medskip
In \refta{tab:spectrum-red} we show three example spectra of the reduced MSSM,
which span the mass range of the parameter space that is in agreement with the
$B$-physics observables and the Higgs-boson mass measurement (using the
same notation as in \refta{tab:spectrum-fut}).
The rows labelled ``light'' correspond to
the spectrum with the smallest $\mneu1$ value (which is independent of upper
limit in $\Mh$). This point is an example for the lowest $\Mh$ values that we
can reach in our scan. As discussed above, the heavy Higgs boson spectrum starts
above $1.4 \tev$, which can be covered at the HL-LHC.
The coloured spectrum is found between $\sim 4 \tev$ and
$\sim 6 \tev$, outside the range of the (HL-)LHC. The LSP has a mass of
$\mneu1 = 1339$, which might offer the possibility of
$e^+e^- \to \neu1 \neu1 \ga$ at CLIC. All other electroweak particles are too
heavy to be produced at CLIC or the (HL-)LHC.
``$\de\Mh = 2.1 (3.1)$'' has the largest $\mneu1$ for
$\Mh \le 125.1 + 2.1 (3.1) \gev$. While, following the mass relations in the
reduced MSSM, the mass spectra are substantially heavier than in the ``light''
case, one can also observe that the smaller upper limit on $\Mh$ results in
substantially lower upper limits on the various SUSY and Higgs-boson
masses. In both cases the heavy Higgs spectrum is within the reach of
the HL-LHC, as mentioned above.
However, even in the case of $\de\Mh = 2.1 \gev$, all SUSY
particles are outside the reach of the (HL-)LHC and CLIC. On the other
hand, all spectra offer good possibilities for their discovery at the
FCC-hh~\cite{fcc-hh}, as discussed above.


\subsection{Conclusions about the Reduced MSSM}
\label{sec:concl-red}

The reduced MSSM naturally results in a light Higgs boson
in the mass range measured at the LHC.
Only the Higgs sector can be tested at the HL-LHC.
On the other hand, the rest of the SUSY spectrum will remain (likely)
unaccessible at the (HL-)LHC, ILC and
CLIC, where such a heavy spectrum also results in SM-like light Higgs
boson, in agreement with LHC measurements~\cite{HcoupLHCcomb}. In other
words, the model is naturally in full agreement with all LHC
measurements. It can be tested definitely at the FCC-hh, where large
parts of the SUSY spectrum would be in the kinematic reach.

%% file: PR2018_Chapter_7.tex
\chapter{Conclusions}

In the present review we have presented in some detail the ideas concerning the reduction
of independent parameters of various renormalizable theories, the theoretical tools which
have been developed to attack the problem and the background work on which they are based on.
Last but not least emphasis was given in presenting various models in which the reduction of
parameters has been theoretically explored and confronted with the experimental data.

The reduction of couplings principle, expressed via RGI relations among couplings, provides
us with a very interesting framework to search for more fundamental quantum field theories
in which a group of couplings are related to a primary one, thus reducing substantially the
number of free parameters of the theory, which might pave the way to search for the minimal ultimate one of Nature.

The reduction of couplings supplemented with global $N = 1$ supersymmetry, leads to theories where
the dimensionless gauge, Yukawa and the dimensionful supersymmetry breaking sectors are unified.
An admirable success of this procedure is the construction of $N = 1$ Finite Unified Theories, which solves
probably the most basic problem of field theory, namely the problem of UV divergencies, in a minimal way.

On the phenomenological side, the developed reduction of couplings machinery provides us with strict selection
rules in choosing realistic GUTs which lead to testable predictions. The celebrated success of predicting the
top-quark mass in FUTs \cite{Kapetanakis:1992vx,Mondragon:1993tw,Kubo:1994bj}
was extended to the correct prediction of the Higgs boson mass, as well as a prediction for the supersymmetric spectrum
of the MSSM \cite{Heinemeyer:2013fga,Heinemeyer:2007tz,Heinemeyer:2012yj,Heinemeyer:2013nza}.

Furthermore it is possible to apply the reduction of couplings directly in the MSSM, again decreasing substantially
the number of free parameters and making the model more predictive
\cite{Mondragon:2013aea,Mondragon:2017hki,Heinemeyer:2018zpw,Heinemeyer:2018roq,Heinemeyer:2017gsv}.
The two models selected by our analysis (FUT and reduced MSSM presented in Chapters \ref{ch:viable} and \ref{chap:FUT_LOW})
share similar features and are in natural agreement with all LHC measurements
and searches. For the reduced MSSM the heavy Higgs particles will be
accessible at the HL-LHC, while the
supersymmetric particles will likely escape the detection at the (HL-)LHC, as well as at ILC and CLIC.
In the FUT case parts of the allowed spectrum of the heavy Higgs bosons is
accessible at the HL-LHC, and parts of
the lighter scalar tau or the lighter neutralinos
might be accessible at CLIC. On the other hand, the FCC-hh will be able to test the predicted parameter
space for both models. The discovery of these particles is the next big bet on the phenomenological side.
On the theoretical side the challenge is to develop a framework in which the above successes of the field
theory models are combined with gravity.
\\
\vspace*{2cm}

\newpage
\noindent
\subsection*{Acknowledgments}

\noindent
We would like to thank a number of scientists from whom we have been
introduced in the subject and others with whom we had an open dialogue and a constant encouragement to continue.
Naturally we start by expressing our deepest gratitude to Wolfhart Zimmermann, who had agreed to be one of the
authors of the present article and to whom is dedicated. During all these years we feel we have learned a lot
from our collaborators Jisuke Kubo, Ernest Ma, Tatsuo Kobayashi and Dimitris Kapetanakis and we had the pleasure
to exchange information and ideas on the subject with Reinhard Oehme, Raymond Stora, who unfortunately are not
with us, while we are always enjoying and appreciate very much indeed the continuous exchange of ideas and the
encouragement given to us by
Genevi\`{e}ve B\'{e}langer,
Fawzi Boudjema,
C\'{e}dric Delaunay,
Abdelhak Djouadi,
Thomas Hahn,
Wolfgang Hollik,
Luis Ibanez,
Jan Kalinowski,
Dmitry Kazakov,
Kostas Kounnas,
Dieter L\"{u}st,
Carlos Mu\~noz,
Erhard Seiler,
Klaus Sibold and
Christof Wetterich.\\
The~work of S.H.\ is supported in part by the MEINCOP Spain under Contract FPA2016-78022-P, in part by the
Spanish Agencia Estatal de Investigaci\'{o}n (AEI), the EU Fondo Europeo de Desarrollo Regional (FEDER)
through the project FPA2016-78645-P, in part by the ``Spanish Red Consolider
MultiDark'' FPA2017-90566-REDC, and in part by the AEI through the grant IFT
Centro de Excelencia Severo Ochoa SEV-2016-0597.
The~work of M.M.\ is partly supported by UNAM PAPIIT through Grant IN111518.
The work of N.T.\ and G.Z.\ is partially supported by the COST actions CA15108 and CA16201.
One of us (GZ) has been supported within the Excellence Initiative funded by the German and State Governments,
at the Institute for Theoretical Physics, Heidelberg University and from the Excellence Grant Enigmass
of LAPTh. GZ thanks the ITP of Heidelberg, LAPTh of Annecy and MPI Munich for their hospitality.

\vfill